\pdfoutput=1
\documentclass[11pt]{revtex4}
\usepackage[pdftex]{graphicx}
\usepackage{amsmath}
\usepackage{amssymb}
\usepackage{amsxtra}

\begin{document}

\title{Masses and Mixing Matrices of Families of Quarks and
Leptons Within the Spin-Charge-Family-Theory, predictions beyond the tree level 
}
\author{A. Hern\'andez-Galeana${}^1$ and N.S. Manko\v c Bor\v stnik${}^2$\\
${}^1$Departamento de F\'{\i}sica,   Escuela Superior de
F\'{\i}sica y Matem\'aticas, I.P.N., \\
U. P. "Adolfo L\'opez Mateos". C. P. 07738, M\'exico, D.F.,
M\'exico.\\
${}^2$Department of Physics, FMF, University of Ljubljana,
Jadranska 19, SI-1000 Ljubljana, Slovenia}

\begin{abstract}

The {\it theory  unifying spin and charges and predicting families}, proposed by N.S.M.B.,
predicts at the low energy regime two (in the mixing matrix elements decoupled) groups of
four families.  There are two kinds of  contributions to  mass matrices in this theory.
One kind distinguishes on the tree level only among the members of one family, that is among the
$u$-quark, $d$-quark, neutrino and electron, the left and right handed, while
the other kind distinguishes only among the families. Mass matrices  for $d$-quarks and
electrons are on the tree level correspondingly strongly  correlated and so are  mass matrices
for $u$-quarks and neutrinos, up to the term, the Majorana term, which is nonzero only for right
handed neutrinos. Beyond the tree level both kinds of contributions start to contribute
coherently and it is expected that a detailed study of properties of mass matrices beyond
the tree level  explains drastic differences in masses and mixing matrices between quarks and leptons.
We report in this paper on  analysis of one loop corrections to the tree level fermion
masses and mixing matrices. Loop diagrams are mediated by the gauge bosons and  the two
kinds of scalar fields. A detailed numerical analysis of fermion masses and mixing, including
neutrinos, within this scenario is in progress and preparation.

\end{abstract}
\maketitle

\section{Introduction}
\label{introduction}
The {\it theory unifying spin and charges and predicting families} (hereafter named the
{\it spin-charge-family-theory}~\cite{norma,pikanorma,NF}), proposed by
N.S.  Manko\v c Bor\v stnik, seems promising to show the
right way beyond the {\it standard model} of fermions and bosons.
The reader is kindly asked to learn more about
this theory in the refs.~\cite{NF,norma,pikanorma} and in the references therein.
Following analyses of the ref.~\cite{NF}, we  here repeat  the parts
which are necessary  for understanding the starting assumptions and the conclusions  to which one
loop corrections beyond the tree level lead. We look at the two loop corrections and present
for the case that each group of four families
would decouple into two times two families numerical results with two loop corrections as well.
Some of this results can be found in~\cite{BledAN2010}.

The {\it spin-charge-family-theory} predicts eight massless families of quarks and leptons
before the two successive breaks --
first from $SU(2)_{I} \times SU(2)_{II} \times U(1)_{II} \times  SU(3)$   to
$SU(2)_{I} \times U(1)_{I} \times SU(3)$ and then from $SU(2)_{I} \times U(1)_{I} \times SU(3)$
to $ U(1) \times SU(3)$. Mass matrices originate in a simple starting action: They are determined
on the tree level by nonzero vacuum expectation values of  scalar (with respect to $SO(1,3)$) fields,
to which vielbeins and two kinds of  spin connection fields contribute. One kind of the spin connection
fields includes fields gauging $S^{ab}$, which are determined by the Dirac gammas ($\gamma^a$'s),
another kind gauges $\tilde{S}^{ab}$, determined by  the second kind of gammas $\tilde{\gamma}^a$'s,
used in the {\it spin-charge-family-theory}~\cite{norma,pikanorma,NF} to generate families.
Each of the two breaks is triggered by different (orthogonal) superposition of  scalar fields.
To the first break, besides vielbeins, only the spin connections of one kind contribute.
To the second break all the scalar fields contribute.

The mass matrices for eight families appear to be four times four by diagonal matrices, with
no mixing matrix elements among the upper four and the lower four families (not in comparison with the
life of the universe) also after the two breaks: The upper four families are namely
doublets with respect to two $SU(2)$ invariant subgroups (with respect to $SU(2)_{II}$,
with generators of the infinitesimal transformations $\vec{\tilde{\tau}}^{2}$, and the one
of the two $SU(2)$ subgroups of $SO(1,3)$, the subgroup $SU(2)_{R}$ with the generators of
the infinitesimal transformations $\vec{\tilde{N}}_{R}$) of the group defined by $\tilde{S}^{ab}$,
and singlets with respect two the other two invariant subgroups ($SU(2)_{I}$, with the generators
$\vec{\tilde{\tau}}^{1}$, and  the $SU(2)_{L}$, with the generators $\vec{\tilde{N}}_{L}$).
The lower four families are doublets with respect to the two subgroups, the
singlets of which are the upper four families.

There are, correspondingly, two stable families: the fifth and the observed first family.
The fifth family members are candidates to form the dark matter, the fourth family waits to be observed.

After the first of the two successive breaks (the break from
$SO(1,3) \times SU(2)_{II} \times SU(2)_{I} \times U(1)_{II} \times  SU(3)$
in both sectors, $S^{ab}$ and $\tilde{S}^{ab}$,  to
$SO(1,3) \times SU(2)_{I} \times U(1)_{I} \times SU(3)$),  which occurs,
below  $\approx 10^{13}$ GeV, the upper four families become massive. In the second break, which is the
{\it standard model}-like electroweak break, also the lower four families became massive.
The second break influences also the mass matrices of the upper four families, although the influence
is expected to be small. 

Rough estimations made so far~\cite{pikanorma,gmdn,gn} on the tree level, which took into account besides the
elementary particle data also the cosmological  data, show that the stable of the upper four families
might have masses~\cite{gn} of the order of $100$ TeV/$c^2$. (The ref.~\cite{MN}
discusses also a possibility that the masses are much smaller, of around a few TeV/$c^2$.)
For the lower four families~\cite{pikanorma,gmdn} we were not really able to predict
the masses of the fourth family members,
we only estimated for chosen masses of the fourth family members their mixing matrices.

In this paper we are studying, following suggestions from the ref.~\cite{NF},
properties of the mass matrices of  twice four families, evaluating
loops corrections to the tree level. We namely hope to see already within the one
and may be two loops corrections the explanation
for the differences in masses and mixing matrices between quarks and leptons, as well as within quarks and
within leptons.
To the loop corrections the gauge boson fields and both kinds of the scalar field contribute, as explained in
the ref.~\cite{NF}.

\section{Short review of the spin-charge-family-theory}
\label{reviewscft}

Let us here make a short review of the {\em spin-charge-family-theory}.  The simple starting action
for spinors (and gauge fields in $d=(1+13)$, ref.~\cite{NF}, Eqs.~(3,4)) manifests
at the low energy regime after several breaks of symmetries as
the effective action (see the ref.~\cite{NF},  Eq.~(5))
for eight  families of quarks and leptons ($\psi $),
left and right handed 
%
\begin{eqnarray}
{\mathcal L}_f &=&  \bar{\psi}\gamma^{n} (p_{n}- \sum_{A,i}\; g^{A}\tau^{Ai} A^{Ai}_{n}) \psi
+ \nonumber\\
               & &  \{ \sum_{s=7,8}\;  \bar{\psi} \gamma^{s} p_{0s} \; \psi \}  + \nonumber\\
               & & {\rm the \;rest},
\label{faction}
\end{eqnarray}
where $n=0,1,2,3$ and
\begin{eqnarray}
\tau^{Ai} = \sum_{a,b} \;c^{Ai}{ }_{ab} \; S^{ab},
\nonumber\\
\{\tau^{Ai}, \tau^{Bj}\}_- = i \delta^{AB} f^{Aijk} \tau^{Ak}.
\label{tau}
\end{eqnarray}
All the charge ($\tau^{Ai}$, Eq.~(\ref{tau})) and the spin ($S^{nn'}; n,n' \in\{0,1,2,3\}$) operators are
expressible with $S^{ab}$. $S^{ab}$ are  generators of spin degrees of freedom in $d=(1+13)$,
determining all the internal degree of freedom of one family members.
Index $A$ enumerates all possible spinor charges and $g^A$ is the coupling constant to a particular
gauge vector field $A^{Ai}_{n},$ as well as to a scalar field $A^{Ai}_{s},\; s > 3$.

Before the  break from $SO(1,3) \times SU(2)_{I} \times SU(2)_{II} \times U(1)_{II} \times  SU(3)\;$ to
$\;SO(1,3) \times SU(2)_{I} \times U(1)_{I} \times SU(3)$  $\vec{\tau}^{3}$  describes the colour
charge ($SU(3)$)~\footnote{
 $\vec{\tau}^{3}: = \frac{1}{2} \,\{  S^{9\;12} - S^{10\;11} ,\,
  S^{9\;11} + S^{10\;12},\, S^{9\;10} - S^{11\;12},\,
 S^{9\;14} -  S^{10\;13},\,  S^{9\;13} + S^{10\;14} ,\,
  S^{11\;14} -  S^{12\;13},
  S^{11\;13} +  S^{12\;14} ,\,
 \frac{1}{\sqrt{3}} ( S^{9\;10} + S^{11\;12} -
 2 S^{13\;14})\},
 $},
$\vec{\tau}^{1}$ the weak charge ($SU(2)_{I}$)~\footnote{
 $\vec{\tau}^{1}=\frac{1}{2} (S^{58}-  S^{67}, \,S^{57} + S^{68}, \,S^{56}-  S^{78} )
 $},
$\vec{\tau}^{2}$ the second $SU(2)_{II}$ charge~\footnote{
 $\vec{\tau}^{2}=\frac{1}{2} (S^{58}+  S^{67}, \,S^{57} - S^{68}, \,S^{56}+  S^{78} )$
	}, and $\tau^{4}$ determines the $U(1)_{II}$ charge~\footnote{ 
$\tau^{4}: = -\frac{1}{3}(S^{9\;10} + S^{11\;12} + S^{13\;14})$}.
 %
 After the  break  of $SU(2)_{II} \times U(1)_{II}$ to $ U(1)_{I}$
$A=2$ denotes the  $U(1)_{I}$ hyper charge $Y$ ($=\tau^{4} + \tau^{23}$) and
after the second break  of $SU(2)_{I} \times U(1)_{I}$ to $U(1)$
$A=2$ denotes the electromagnetic  charge $Q \; (= S^{56} + \tau^{4})$, while instead of the weak
charge $Q'\; (=\tau^{13}- \tau^{4}\, \tan^2 \theta_{1}) $ and $\tau^{1\pm}$
of the {\it standard model} manifest.

The term in the second row of Eq.~(\ref{faction}) determines mass matrices of twice four families
\begin{eqnarray}
{\mathcal L}_{m_{f}} &=&   \psi^{\dagger} \gamma^{0} M \psi \nonumber\\
                     &=&   \sum_{s=7,8}\;  \bar{\psi} \gamma^{s} p_{0s} \; \psi \,
                      =	 \, \psi^{\dagger} \, \gamma^0 \,
                      (\stackrel{78}{(+)}\, p_{0+} +  \stackrel{78}{(-)} \,p_{0-} \, \psi \,, \nonumber\\
              p_{0s} &=& f^{\sigma}{}_s p_{0\sigma} + \frac{1}{2E}\, \{ p_{\alpha}, E f^{\alpha}{}_a\}_- \;,
\; p_{0\sigma} =  p_{\sigma}  - \frac{1}{2}  S^{ab} \omega_{ab \sigma} -
                    \frac{1}{2}  \tilde{S}^{ab}   \tilde{\omega}_{ab \sigma}\,, \nonumber\\
                    \stackrel{78}{(\pm)} &=&\frac{1}{2}\,(\gamma^7 \pm i\, \gamma^8),
\, \; p_{0\pm}= p_{07} \mp i \, p_{08}\,.
\label{mfaction}
\end{eqnarray}
The main argument to take $s=7,8,$ is (so far) the required agreement with the experimental data.
The Dirac spin, described by $\gamma^a$'s, defines the  spinor representations in $d=(1+ 13)$.
The second kind of the spin~\cite{norma93,hn0203,NF},
described by $\tilde{\gamma}^a$'s ($\{\tilde{\gamma}^a, \tilde{\gamma}^b\}_{+}= 2 \, \eta^{ab}$) and
anticommuting with the Dirac $\gamma^a$ ($\{\gamma^a, \tilde{\gamma}^b\}_{+}=0$),
defines the families of spinors. One finds~\cite{NF}
\begin{eqnarray}
&& \{ \gamma^a, \gamma^b\}_{+} = 2\eta^{ab} =
\{ \tilde{\gamma}^a, \tilde{\gamma}^b\}_{+},\quad
\{ \gamma^a, \tilde{\gamma}^b\}_{+} = 0,\nonumber\\
&&S^{ab}: = (i/4) (\gamma^a \gamma^b - \gamma^b \gamma^a), \quad
\tilde{S}^{ab}: = (i/4) (\tilde{\gamma}^a \tilde{\gamma}^b
- \tilde{\gamma}^b \tilde{\gamma}^a),\quad  \{S^{ab}, \tilde{S}^{cd}\}_{-}=0.
\label{snmb:tildegclifford}
\end{eqnarray}
%
The eight massless families ($2^{(1+7)/2-1}$) manifest after the break of
$SO(1,7)$ to $SO(1,3) \times SO(4)$ (the break occurs in both sectors,
$S^{ab}$ and $\tilde{S}^{ab}$) as twice four families: Four of the families are doublets
with respect to two of the four $SU(2)$ invariant subgroups of the groups $SO(4)\times SO(1,3)$
in the $\tilde{S}^{ab}$ sector (namely, with respect to the subgroups with the infinitesimal
generators $\vec{\tilde{\tau}}^{2}$ and $\vec{\tilde{N}}_{R}$) and
singlets with respect to the remaining two $SU(2)$ invariant subgroups
(with the infinitesimal generators $\vec{\tilde{\tau}}^{1}$ and $\vec{\tilde{N}}_{RL}$),
while  the remaining four families are
singlets with respect to the first two and doublets with respect to the remaining two invariant subgroups.
At the symmetry level of $SO(1,3) \times SU(2)_{I} \times SU(2)_{II} \times U(1)_{II} \times  SU(3)$
twice four families  are  massless,  the mass matrices $M$ of any family member is equal to zero.

The break of $SU(2)_{II} \times U(1)_{II}$ to $U(1)_{I}$ in both, $S^{ab}$ and $\tilde{S}^{ab}$, sectors is
caused by the scalar fields $\vec{\tilde{A}}^{2}_{s}$ and  $\vec{\tilde{A}}^{\tilde{N}_{R}}_{s}$
\footnote{The vielbeins and the spin connections are both involved in  breaks.}, which gain
nonzero vacuum expectation values and determine the mass matrices ${\cal M}_{(o)}$  (Eq.~\ref{M0})
on the tree level.
Only families which couple to these scalar fields become massive.
 These are four families, which are  doublets with respect to the subgroups with generators of the
 infinitesimal transformations
$\vec{\tilde{\tau}}^{(2)}\; (=
  \frac{1}{2} (\tilde{S}^{58} +  \tilde{S}^{67}, \,\tilde{S}^{57} - \tilde{S}^{68}, \,
 \tilde{S}^{56} +  \tilde{S}^{78} ))$ and $\vec{\tilde{N}}_R \;
 (= \frac{1}{2}(\tilde{S}^{23}- i\tilde{S}^{01},
\tilde{S}^{31}- i\tilde{S}^{02}, \tilde{S}^{12}- i\tilde{S}^{03}))$.

The rest four families, which are singlets with respect to these  two subgroups,
remain massless until the second break of $SU(2)_{I} \times U(1)_{I}$ to $U(1)$, in which the
scalar fields $\vec{\tilde{A}}^{1}_{s}$ and  $\vec{\tilde{A}}^{\tilde{N}_{L}}_{s}$ gain
nonzero vacuum expectation values. These scalar fields couple to the rest four families through
$\vec{\tilde{\tau}}^{(1)}\; (=
  \frac{1}{2} (\tilde{S}^{58} -  \tilde{S}^{67}, \,\tilde{S}^{57} + \tilde{S}^{68}, \,
 \tilde{S}^{56} -  \tilde{S}^{78} ))$ and $\vec{\tilde{N}}_L \;
 (= \frac{1}{2}(\tilde{S}^{23}+ i\tilde{S}^{01},
\tilde{S}^{31}+ i\tilde{S}^{02}, \tilde{S}^{12}+ i\tilde{S}^{03}))$.
To this, the electroweak break,
also scalar fields in the $S^{ab}$ sector contribute. These fields - $A^{Q}_s,
A^{Q'}_s, A^{Y'}_s$ - couple  to the family members through the quantum numbers $Q,Q'$ and $Y'$,
respectively. While $\vec{\tilde{\tau}}^{(2)}\,,  \vec{\tilde{\tau}}^{(1)}\,, \vec{\tilde{N}}_{R}\,$
and $\vec{\tilde{N}}_{L}\,$ distinguish among the families, but not among the family members,
distinguish $Q,Q'$ and $Y'$ among the family members independent of the family index.

After the break of  $SU(2)_{II} \times U(1)_{II} \times SU(2)_{I} \times SU(3)\;$ into
$\; SU(2)_{I} \times U(1)_{I} \times SU(3)$ the effective Lagrange density for spinors is as follows
\begin{eqnarray}
{\mathcal L}_f &=&  \bar{\psi}\, (\gamma^{m} \, p_{0m} - \,M) \psi\,, \nonumber\\
         p_{0m}&=& p_{m} - \{g^1 \vec{\tau^{1}}\, \vec{A}^{1}_{m} + g^{Y} Y A^{Y}_{m}
         + g^3\, \vec{\tau^{3}}\, \vec{A}^{3}_{m}\nonumber\\
         &+&   g^{2} \cos \theta_2\, Y' A^{Y'}_{m}
         + \frac{g^{2}}{\sqrt{2}}\, (\tau^{2+} A^{2+}_{m} + \tau^{2-} A^{2-}_{m})  \,\}\,\nonumber\\
         \bar{\psi}\,M \, \psi &=&\bar{\psi} \gamma^s\,p_{0s} \, \psi\, ,         \nonumber\\
         p_{0s}&=& p_{s} - \{ \tilde{g}^{\tilde{N}_R} \vec{\tilde{N}}_R \vec{\tilde{A}}^{\tilde{N}_R}_{s} +
         \tilde{g}^{\tilde{Y}'}\,   \tilde{Y}' \tilde{A}^{\tilde{Y}'}_{s}
	          + \frac{\tilde{g}^{2}}{\sqrt{2}}\, (\tilde{\tau}^{2+} \,\tilde{A}^{2+}_{s}
	          + \tilde{\tau}^{2-}\,  \tilde{A}^{2-}_{s}) \,\}\,.
\label{factionII}
\end{eqnarray}
In the second row the vector gauge fields which remain massless ($\vec{A}^{1}_{m}, \,A^{Y}_{m}$ and
$\vec{A}^{3}_{m}$) and in  the third row
the massive gauge fields ($A^{Y'}_{m}$ and $A^{2\pm}_{m} $) are presented.
To the mass matrices of the upper four families $\bar{\psi}\,M \psi$ the vacuum expectation
values of the scalar fields 
$ \vec{\tilde{A}}^{\tilde{N}_R}_{s}$, $\tilde{A}^{\tilde{Y}'}_{s}$ and  $\tilde{A}^{2\pm}_{s}$,
together with the corresponding vielbeins with the scalar index,
contribute.
The new and the old gauge fields are related as follows:
$A^{23}_{m}  = A^{Y}_{m} \sin \theta_2 + A^{Y'}_{m} \cos \theta_2\, , $
$A^{4}_{m}   = A^{Y}_{m} \cos \theta_2 - A^{Y'}_{m} \sin \theta_2\, , $
$A^{2\pm}_m  = \frac{1}{\sqrt{2}}(A^{21}_m \mp  i A^{22}_m)\, ,$
with the new quantum numbers $Y= \tau^{4}+ \tau^{23},$ $Y'= \tau^{23} - \tau^{4} \tan^{2} \theta_{2},
\quad \tau^{2\pm} = \tau^{21}\pm i \tau^{22}$ and
the new coupling constants of fermions to the massive gauge fields  and the massless one become
$g^{Y}= g^{4} \cos \theta_{2}$, $g^{Y'}= g^{2} \cos \theta_{2}$,
$ \tan \theta_{2}= \frac{g^4}{g^2}$,
while $A^{2\pm}_m $ have a coupling constant $\frac{g^2}{\sqrt{2}}$.

The new and the old scalar fields are related as:
$\tilde{A}^{23}_{s}   = \tilde{A}^{\tilde{Y}}_{s} \sin \tilde{\theta}_2 + \tilde{A}^{\tilde{Y}'}_{s}
\cos \tilde{\theta}_2\, ,$
$\tilde{A}^{4}_{s}    = \tilde{A}^{\tilde{Y}}_{s} \cos \tilde{\theta}_2 - \tilde{A}^{\tilde{Y}'}_{s}
\sin \tilde{\theta}_2\,,$
$\tilde{A}^{2\pm}_s   = \frac{1}{\sqrt{2}}(\tilde{A}^{21}_s \mp  i \tilde{A}^{22}_s)\,,$
while it follows $\vec{\tilde{A}}^{2}_{s} =
2(\tilde{\omega}_{58s}, \tilde{\omega}_{57s}, \tilde{\omega}_{56s})$.
We shall make a choice in this paper of $\tilde{\theta}_2=0.$
We also have $\vec{\tilde{A}}^{\tilde{N}_R}_{s}\,=\,2(\tilde{\omega}_{23s}, \tilde{\omega}_{31s} ,
\tilde{\omega}_{12s})\,,$ and
$\vec{\tilde{N}}_R \,=\, \frac{1}{2}\,(\tilde{S}^{23}- i\tilde{S}^{01},
\tilde{S}^{31}- i\tilde{S}^{02}, \tilde{S}^{12}- i\tilde{S}^{03})$,
for $s= 7,8$. The new family quantum numbers are $\tilde{Y}= \tilde{\tau}^{4}+ \tilde{\tau}^{23}, $
$\tilde{Y'}= \tilde{\tau}^{23} - \tilde{\tau}^{4} \tan^{2} \tilde{\theta}_{2},$
$\tilde{\tau}^{2\pm} = \tilde{\tau}^{21}\pm i \tilde{\tau}^{22}$.

The reader is kindly asked
to look at the ref.~\cite{NF} for more explanations.

We present in Table~\ref{Table VII.} (from Table VIII. of  the ref.~\cite{NF})
a general shape of mass matrices of all the eight families on the tree level
after the break of $SO(2)_{II} \times U(1)_{II}$ into $U(1)_{I}$.
The lower four families 
stay massless.
The $u$-quark mass matrices (they are determined by $\vec{\tilde{A}}^{\tilde{A}}_{-}$
$=\vec{\tilde{A}}^{\tilde{A}}_{7} +  i \vec{\tilde{A}}^{\tilde{A}}_{8} $, for $\tilde{A}=2,1,\tilde{N}_R,
\tilde{N}_L$) are different than the $d$-quark ones (they are determined by $\vec{\tilde{A}}^{\tilde{A}}_{+}$
$=\vec{\tilde{A}}^{\tilde{A}}_{7} -  i \vec{\tilde{A}}^{\tilde{A}}_{8} $)
and $e$ mass matrices differ from
the $\nu$ ones, while mass matrices for quarks and leptons are identical~(ref.\cite{NF},
they are the same for $u$-quarks and neutrinos, and for  $d$ quarks and electrons. The contribution
of the scalar fields causing the Majorana right handed neutrinos (see appendix~\ref{majoranas})
is not added in this table.
 \begin{table}
 \begin{center}
\begin{tabular}{|r||c|c|c|c|c|c|c|c||}
\hline
$\Sigma_{i}$ &$ I_1 $&$ I_{2} $&$ I_{3} $&$ I_{4} $&$ II_{1} $&$ II_{2} $
 &$ II_{3} $&$ II_{4}$\\
\hline\hline
$I_{1} $ & $  \;0 \; $ & $ \;0 \; $ & $ \; 0 \;$ & $ \; 0 \;$
          & $ 0 $ & $ 0 $ & $ 0 $ & $ 0 $\\
\hline
$I_{2}$ & $ 0 $ & $ 0 $ & $ 0 $ & $ 0 $
          & $ 0 $ & $ 0 $ & $ 0 $ & $ 0 $\\
\hline
$I_{3}$ & $ 0 $ & $ 0 $ & $ 0 $ & $ 0 $
          & $ 0 $ & $ 0 $ & $ 0 $ & $ 0 $\\
\hline
$I_{4}$ & $ 0 $ & $ 0 $ & $ 0 $ & $ 0 $
          & $ 0 $ & $ 0 $ & $ 0 $ & $ 0 $\\
\hline\hline
$ II_{1} $ & $ 0 $ & $ 0 $ & $ 0 $ & $ 0 $ &
$ - \frac{1}{2}\, (\tilde{a}^{23}_{\pm} + \tilde{a}^{\tilde{N}^{3}_{R}}_{\pm})$ &
$- \tilde{a}^{\tilde{N}_{R}^{-}}_{\pm}   $&$0$&
$  - \tilde{a}^{2-}_{\pm}$ \\
\hline
$ II_{2} $ & $ 0 $ & $ 0 $ & $ 0 $ & $ 0 $ &
$- \tilde{a}^{\tilde{N}_{R}^{+}}_{\pm}   $&
$ \frac{1}{2}(-\tilde{a}^{23}_{\pm } + \tilde{a}^{\tilde{N}^{3}_{R}}_{\pm}) $ & $
-\tilde{a}^{2-}_{\pm}$ & $0$\\
\hline
$II_{3} $ & $ 0 $ & $ 0 $ & $ 0 $ & $ 0 $ &
$0$        &$  -\tilde{a}^{2+}_{\pm}$ &
$  \frac{1}{2}\,( \tilde{a}^{23}_{\pm}  - \tilde{a}^{\tilde{N}^{3}_{R}}_{\pm}) $ &
$ - \tilde{a}^{\tilde{N}_{R}^{-}}_{\pm} $ \\
\hline
$II_{4}    $ & $ 0 $ & $ 0 $ & $ 0 $ & $ 0 $ &
$- \tilde{a}^{2+}_{\pm}$   & $0$   &$- \tilde{a}^{\tilde{N}_{R}^{+}}_{\pm}   $&
$  \frac{1}{2}\,( \tilde{a}^{23}_{\pm} + \tilde{a}^{\tilde{N}^{3}_{R}}_{\pm})$ \\
\hline\hline
\end{tabular}
 \end{center}
 \caption{\label{Table VII.}  The mass matrices on the tree level (${\cal M}_{(o)}$)
 for  two groups ($\Sigma =II$ for the upper four, while $\Sigma=I$ for the lower four)
  families of quarks and
 leptons after the break of $SO(1,3)\times SU(2)_{I}  \times SU(2)_{II} \times U(1)_{II} \times SU(3)$
 to $SO(1,3) \times  SU(2)_{I} \times U(1)_{I} \times SU(3)$. The contribution comes from a particular
 superposition of  spin connection fields, the gauge fields of $\tilde{S}^{ab}$.
 $(\mp)$ distinguishes $u_{i}$ from
 $d_{i}$ and $\nu_{i}$ from $e_{i}$.
 }
\end{table}
The contributions below the tree level  change the matrix elements and  remove the degeneracy
between the $u$-quarks and neutrinos as well as between the $d$-quarks and electrons. It is expected
that they will not appreciably change the symmetry of the
matrix elements on the tree level. We shall discuss this in the next section.

To the electroweak break, when  $SU(2)_I  \times U(1)_I$ breaks
into $U(1)$,   besides the scalar fields originating
in vielbeins and in superposition  of spin connection fields of $\tilde{S}^{ab}$ (the ones, which are
orthogonal to the ones causing the first break),
also the scalar fields originating in spin connections of $S^{ab}$ contribute: $A^{Q}_{s},\,A^{Q'}_{s} $ and
$A^{Y'}_{s}$.
The three additional gauge fields and the lower four families become massive.

The new superposition of gauge fields $\vec{A}^{1}_m$ and $A^{Y'}_{m}$ manifest
(ref.~\cite{NF}
) leading to one massless $A_{m}$ ($\equiv A^{Q}_{m}$)
and three massive  gauge fields $A^{Q'}_{m}\,$ ($\equiv Z_{m}$), $ A^{\pm}_m$
($\equiv W^{\pm}_m$).

The effective Lagrange density for spinors is after the electroweak break as follows
\begin{eqnarray}
 {\mathcal L}_f &=&  \bar{\psi}\, (\gamma^{m} \, p_{0m} - \,M) \psi\, , \nonumber\\
          p_{0m}&=& p_{m} - \{ e \,Q\,A_{m} +  g^{1} \cos \theta_1 \,Q'\, Z^{Q'}_{m} +
          \frac{g^{1}}{\sqrt{2}}\, (\tau^{1+} \,W^{1+}_{m} + \tau^{1-} \,W^{1-}_{m}) + \nonumber\\
                &+& g^{2} \cos \theta_2 \,Y'\, A^{Y'}_{m} +
          \frac{g^{2}}{\sqrt{2}}\, (\tau^{2+} \,A^{2+}_{m} + \tau^{2-} \,A^{2-}_{m})\, , \nonumber\\
           \bar{\psi}\,  M \, \psi &=&  \bar{\psi}\, \gamma^{s} \, p_{0s} \psi\,\nonumber\\
          p_{0s}&=& p_{s} - \{ \tilde{g}^{\tilde{N}_R}\, \vec{\tilde{N}}_R \,\vec{\tilde{A}}^{\tilde{N}_R}_{s} +
 	                \tilde{g}^{\tilde{Y}'}  \, \tilde{Y}'\,\tilde{A}^{\tilde{Y}'}_{s}
 	       	          + \frac{\tilde{g}^{2}}{\sqrt{2}}\, (\tilde{\tau}^{2+} \,\tilde{A}^{2+}_{s}
 	          + \tilde{\tau}^{2-}\,  \tilde{A}^{2-}_{s})   \nonumber\\
 	       &+&  \tilde{g}^{\tilde{N}_L} \,\vec{\tilde{N}}_L\, \vec{\tilde{A}}^{\tilde{N}_L}_{s} +
 		  	                \tilde{g}^{\tilde{Q}'} \, \tilde{Q}'\,
 		  	                \tilde{A}^{\tilde{Q}'}_{s}
 		  	       	          + \frac{\tilde{g}^{1}}{\sqrt{2}}\, (\tilde{\tau}^{1+}
 		  	       	          \,\tilde{A}^{1+}_{s}
 	          + \tilde{\tau}^{1-}\,  \tilde{A}^{1-}_{s})\nonumber\\ 	
 	          &+&  e\, Q\, A_{s} +  g^{1}\, \cos \theta_1 \,Q'\, Z^{Q'}_{s} +
               g^{2} \cos \theta_2\, Y'\, A^{Y'}_{s}\,\}\,,
 	          \; s\in\{7,8\}\,.
\label{factionI}
\end{eqnarray}

The reader is kindly asked to learn how does the operators $(\stackrel{78}{(-)}$ transform the weak
and the hyper charge of the right handed $u_{R}$-quark  and $\nu_{R}$-lepton into those  of the
left handed ones, while $(\stackrel{78}{(+)}$ does the same for the right handed $d_{R}$-quark  and
$e_{R}$-lepton, in the ref.~\cite{NF}. One can learn there also how do the operators $\tilde{N}^{\pm}_R\,$,
$\tilde{N}^{\pm}_L\,$, $\tilde{\tau}^{2\pm}\,$ and $\tilde{\tau}^{1\pm}\,$ transform any member
of one family into the same member of another family and what transformations cause any
superposition of the operators $S^{ab}$ or of the operators $\tilde{S}^{ab}$ on
any family member of any family.
A short presentation of these properties is added in the appendix~\ref{appendixVo}.

The new fields manifest ($ A^{13}_{m} = A_{m} \sin \theta_1 + Z_{m} \cos \theta_1\,,$
 $A^{Y}_{a}  = A_{m} \cos \theta_1 -  Z_{m} \sin \theta_1\,,$
 $W^{\pm}_m  = \frac{1}{\sqrt{2}}(A^{11}_m \mp i  A^{12}_m)$), with the
 new quantum numbers ($Q  =  \tau^{13}+ Y \,= \,S^{56} +  \tau^{4},\;
 Q' \,= -Y \,\tan^2 \theta_1 + \tau^{13}, \; \tau^{1\pm}\,= \tau^{11} \pm i \,\tau^{12}\,$
 and the new coupling constants ($e =
 g^{Y} \cos \theta_1\,$ ($\equiv g^{Q}$), $\;g' = g^{1}\cos \theta_1\,$ ($\equiv g^{Q'}$)
 and $\tan \theta_1 =
 \frac{g^{Y}}{g^1} $), are in agreement with the {\it standard model}.
 Correspondingly there are new scalar field, new quantum numbers and new coupling constants.

 But there are also clearly noticeable differences  between the {\it spin-charge-family-theory}
 and the {\it standard model}, presented in Eq.~
 (\ref{factionI}),
 which should be sooner or later measurable. Like: i.) The scalar fields explaining the
 appearance i.a.) of the Higgs in the {\it standard model}, in the {\it spin-charge-family-theory}
 with weak and hyper charges  in the adjoint representation since $\gamma^s$ do what in the
 {\it standard model} the weak  and hyper charge of the Higgs do,
 and i.b.) of the Yukawa couplings, which  manifest here as new interactions (but this is the case also
 in the {\it standard model}). The scalar fields should be measured as several fields, although
 they effectively manifest as Higgs and Yukawa couplings. ii.) New gauge vector fields $A^{Y'}_{m}$,
 $A^{2\pm}_{m}$. ii.) New families predicted. iii.) New gauge scalar fields for the
 upper four families. iv.) New insight into the discrete symmetries, like charge conjugation,
 parity, charge parity (non conserved) symmetry, matter/anti-matter asymmetry and others.

 The scalar fields $\vec{\tilde{A}}^{\tilde{A}}_{s}$, $s \in \{7,8\}$, which gain in this phase transition
 a  nonzero vacuum expectation values, are: $\vec{\tilde{A}}^{\tilde{N}_{L}}_{s}$ ($=
 \frac{1}{2}\, (\tilde{\omega}_{23s} - i \, \tilde{\omega}_{01s}$,
 $\tilde{\omega}_{31s} - i \, \tilde{\omega}_{02s}$,
 $\tilde{\omega}_{12s} - i \, \tilde{\omega}_{03s}$),
 $\vec{\tilde{A}}^{1}_{s} $ (again expressible with $\tilde{\omega}_{abs}, \,s \in \{7,8,9,10\}$,
 transforming as follows $ \tilde{A}^{13}_{s} = \tilde{A}_{s} \sin \tilde{\theta}_1 +
   \tilde{Z}^{\tilde{Q}'}_{s} \cos \tilde{\theta}_1\,$ \,($\tilde{Z}^{\tilde{Q}'}_{s} \,\equiv
   \tilde{A}^{\tilde{Q}'}_{s} \,$,  $\,\tilde{A}_{s} \,\equiv
   \tilde{A}^{\tilde{Q}}_{s} $) $,
   \tilde{A}^{\tilde{Y}}_{s} = \tilde{A}_{s} \cos \tilde{\theta}_1 -
   \tilde{Z}^{\tilde{Q}'}_{s} \sin \tilde{\theta}_1\,,$ $
  \tilde{W}^{\pm}_{s} = \frac{1}{\sqrt{2}}(\tilde{A}^{11}_{s} \mp i  \tilde{A}^{12}_{s})
$ ($\,\tilde{W}^{\pm}_{s} \,\equiv
   \tilde{A}^{1 \pm}_{s} $),  and $A^{Q}_s \,(\equiv A_s) , A^{Q'}_{s}\,(\equiv Z^{Q'}_s),
   A^{Y'}_{s} $, with $Q,Q'$ and $Y'$ defined above and
with $\vec{\tilde{N}}_{(L,R)} = ( \frac{1}{2}(\tilde{S}^{23}\, (+,-) i\tilde{S}^{01},
  \tilde{S}^{31} \,(+,-) i\tilde{S}^{02}, \tilde{S}^{12}\, (+,-) i\tilde{S}^{03})$.

Let us point out again that the upper four families are singlets with respect to
$\vec{\tilde{N}}_L$ and $\vec{\tilde{\tau}}^{1} $, while the lower four families are
singlets with respect to $\vec{\tilde{N}}_R$ and $\vec{\tilde{\tau}}^{2} $.
At each break the mass matrices on the tree level ${\cal M}_{(0)}$ change.

Table~\ref{Table VIII.} represents the mass matrices for the lower
four families on the tree level. Only the contribution of the scalar fields which originate
in the gauge fields of $\tilde{S}^{ab}$ are included into the table.
The contribution from  terms like
$Q \, e \, A^{Q}_{s}, g^{Q'}\, Q' \,A^{Q'}_{s}, g^{Y'}\,Y' \,A^{Y'}_{s}$,  which are diagonal and
equal for all the families but distinguish among the members of one family, are not present.
The contribution
of the scalar fields causing the Majorana right handed neutrinos (see appendix~\ref{majoranas})
is  also not added in this table.

The notation $\tilde{a}^{\tilde{A}i}_{\pm}=$ $-\tilde{g}^{\tilde{A}}\, \tilde{A}^{\tilde{A}i}_{\pm}$ is used.

There is a mass term  within the {\it spin-charge-family-theory}, which transform
the right handed neutrino to his  charged conjugated
one, contributing to the (right handed) neutrino Majorana masses.
The Majorana terms are expected to be large and might influence strongly the neutrino masses and their
mixing matrices. The reader can find more explanation about this term in ref.~\cite{NF} and in
appendix~\ref{majoranas}. Let us add here, that it is nonzero only for the lower
four families. It needs to be studied in more details to say more.  These terms are
not yet included into Table~\ref{Table VIII.}.

 \begin{table}
 \begin{center}
\begin{tabular}{|r||c|c|c|c||}
\hline
 $I_{i}$&$ 1 $&$ 2 $&$ 3 $&$4 $\\
\hline\hline
$1 $&
$ - \frac{1}{2}\,( \tilde{a}^{13}_{\pm} + \tilde{a}^{\tilde{N}^{3}_{L}}_{\pm})$&
$\tilde{a}^{\tilde{N}_{L}^{-}}_{\pm}$&$0$&
$   \tilde{a}^{1-}_{\pm}$  \\
\hline
$2$ &  $ \tilde{a}^{\tilde{N}_{L}^{+}}_{\pm} $ &
$ \frac{1}{2}( -\tilde{a}^{13}_{\pm } + \tilde{a}^{\tilde{N}^{3}_{L}}_{\pm}) $&
$\tilde{a}^{1-}_{\pm}   $ &$0$\\
\hline
$3$ & $0$& $\tilde{a}^{1+}_{\pm}$&
$  \frac{1}{2}\,( \tilde{a}^{13}_{\pm}  - \tilde{a}^{\tilde{N}^{3}_{L}}_{\pm}) $ &
 $\tilde{a}^{\tilde{N}_{L}^{-}}_{\pm}$ \\
\hline
$4$ & $\tilde{a}^{1+}_{\pm}$& $0$&$  \tilde{a}^{\tilde{N}_{L}^{+}}_{\pm}  $ &
 $\frac{1}{2}\,( \tilde{a}^{13}_{\pm}  + \tilde{a}^{\tilde{N}^{3}_{L}}_{\pm}) $
\\
\hline\hline
\end{tabular}
 \end{center}
 \caption{\label{Table VIII.}  The mass matrices on the tree level (${\cal M}_{(o)}$)
 for the lower four  families ($\Sigma =I$) of quarks and
 leptons after the electroweak break. Only the contributions coming  from the spin connection fields,
 originating in   $\tilde{S}^{ab} $
 are presented.
 $(\mp)$ distinguishes  between the values of the $u$-quarks and $d$-quarks and
between the values of $\nu$ and $e$.
The notation $\tilde{a}^{\tilde{A}i}_{\pm}=$ $-\tilde{g}^{\tilde{A}}\, \tilde{A}^{\tilde{A}i}_{\pm}$ is used.
 The terms coming from spin connection fields originating in $S^{ss'}\,$
 are not presented here. They are the same for
 all the families,  but different for different  family members. Also possible Majorana terms are not included.
 }
\end{table}

We present in Table~\ref{Table FQN} the quantum numbers $\tilde{\tau}^{23}$,
$\tilde{N}^{3}_{R}\,$, $\tilde{\tau}^{13}$ and $\tilde{N}^{3}_{L}$ for all eight
families~\cite{NF}. The first four families are singlets with respect to
$\tilde{\tau}^{23}$ and $\tilde{N}^{3}_{R}\,$, while they are doublets
with respect to $\tilde{\tau}^{13}$ and $\tilde{N}^{3}_{L}$. The upper four families are doublets
with respect to $\tilde{\tau}^{23}$ and $\tilde{N}^{3}_{R}\,$ and are singlets
with respect to $\tilde{\tau}^{13}$ and $\tilde{N}^{3}_{L}$. The representations of families in the
technique of the ref.~\cite{holgernorma2003} are presented in appendix~\ref{appendixVo}, in
Table~\ref{Table tec}.

 \begin{table}
 \begin{center}
 \begin{tabular}{|r||r||r||r||r|||r||r||r||r||r||}
 \hline
 $\Sigma=I/i$ &$\tilde{\tau}^{23}$&$\tilde{N}^{3}_{R}$&$\tilde{\tau}^{13}$ &$\tilde{N}^{3}_{L}$&
 $\Sigma=II/i$ &$\tilde{\tau}^{23}$&$\tilde{N}^{3}_{R}$&$\tilde{\tau}^{13}$ &$\tilde{N}^{3}_{L}$\\
 \hline \hline
 $1$  & $0$& $ 0$ &  $ \frac{1}{2}$&$ \frac{1}{2}$&  $1 $   & $ \frac{1}{2}$&$ \frac{1}{2}$ & $0$& $ 0$
 \\
 \hline
 $2$ & $0$& $ 0$ &  $ \frac{1}{2}$&$-\frac{1}{2}$&  $2$   & $ \frac{1}{2}$&$-\frac{1}{2}$ & $0$& $ 0$
 \\
 \hline
 $3$& $0$& $ 0$ &  $-\frac{1}{2}$&$-\frac{1}{2}$&  $3$  & $-\frac{1}{2}$&$-\frac{1}{2}$ & $0$& $ 0$
 \\
 \hline
 $4$ & $0$& $ 0$ &  $-\frac{1}{2}$&$ \frac{1}{2}$&  $4$ & $-\frac{1}{2}$&$ \frac{1}{2}$ & $0$& $ 0$
 \\
 \hline
 \end{tabular}
 \end{center}
 \caption{\label{Table FQN} The quantum numbers $\tilde{\tau}^{23}$,
$\tilde{N}^{3}_{R}\,$, $\tilde{\tau}^{13}$ and $\tilde{N}^{3}_{L}$ for the two groups ($\Sigma=II$
for the upper four families and $\Sigma =I$ for the lower four families) of four  families are
presented~\cite{NF}.}
 \end{table}
In Table~\ref{Table MQN} we present quantum numbers of all members of a family, any one,
after the electroweak break.
 \begin{table}
 \begin{center}
 \begin{tabular}{|r||r||r||r||r|||r||r||r||r||r||}
 \hline
 $$ &$Y$&$Y'$&$Q$&$Q'$&
 $$ &$Y$&$Y'$&$Q$&$Q'$\\
 \hline \hline
 $u_R$  & $ \frac{2}{3}$& $ \frac{1}{2}\,(1-\frac{1}{3} \tan^2 \theta_2)$ &
 $ \frac{2}{3}$& $             -\frac{2}{3} \tan^2 \theta_1 $&
 $u_L$  & $ \frac{1}{6}$& $                -\frac{1}{6} \tan^2 \theta_2 $ &
 $ \frac{2}{3}$& $\frac{1}{2}(1-\frac{1}{3} \tan^2 \theta_1)$
 \\
 \hline
 $d_R$  & $-\frac{1}{3}$& $-\frac{1}{2}\,(1+\frac{1}{3} \tan^2 \theta_2)$ &
 $-\frac{1}{3}$& $              \frac{1}{3} \tan^2 \theta_1 $&
 $d_L$  & $ \frac{1}{6}$& $                -\frac{1}{6} \tan^2 \theta_2 $ &
 $-\frac{1}{3}$& $-\frac{1}{2}(1+\frac{1}{3} \tan^2 \theta_1)$
 \\
 \hline
 $\nu_R$& $0$           & $\frac{1}{2}\,(1+             \tan^2 \theta_2)$ &
 $0$&$0$&
 $\nu_L$& $-\frac{1}{2}$&$\frac{1}{2}\,                 \tan^2 \theta_2 $ &
 $0$& $ 0$
 \\
 \hline
 $e_R$ & $-1$           & $\frac{1}{2}\,(-1+             \tan^2 \theta_2)$&
         $-1$           &                                $\tan^2 \theta_1$&
 $e_L$ & $-\frac{1}{2}$ &$ \frac{1}{2}                  \tan^2 \theta_2 $ &
         $-1$& $ -\frac{1}{2}(1-                          \tan^2 \theta_1)$
 \\
 \hline
 \end{tabular}
 \end{center}
 \caption{\label{Table MQN} The quantum numbers $Y, Y',Q, Q'$ of
 the members of one family (anyone)~\cite{NF}.}
 \end{table}

When going below the tree level all the massive gauge fields and those scalar fields of both origins,
($S^{ab}$ and $\tilde{S}^{ab}$), to which the
family members couple, start to contribute.
To the lower four families mass matrices  the scalar fields,
which are  superposition of the $\omega_{st s'}$ field, that is
of $A^{Q}_{s}, A^{Q'}_{s}$ and $A^{Y'}_{s}$, contribute
already on the tree level. This was not the case for the upper four families.
Contributions of $Q A^{Q}_{s}, Q'A^{Q'}_{s}$ and $Y'A^{Y'}_{s}$ distinguish among all
the members of one family,
but are the same for a family member belonging
to different families.  They influence after the second break also the mass matrices of the upper four families.
Below the tree level all the gauge fields and dynamical scalar fields start to
contribute coherently, as dictated by Eq.~(\ref{factionI}). These contributions are expected to be large
for the lower four families, while they influence, since the scale of these two breaks are supposed
to be very different,
 only slightly the
 upper four family mass matrices.
 According to  the estimations presented in refs.~(\cite{NF,gn}) the changes
 are within a percent or much less if
the masses are large enough (of the order of hundred TeV/$c^2$ or larger).

We study in this paper properties of both groups of four families, taking the
vacuum expectation values of the scalar fields as an input. As we already explained,
in the {\it spin-charge-family-theory} the
mass matrices of the family members are within each of the two groups very much correlated.
It is the prediction of this theory~\cite{NF} that there are terms beyond the tree level, which
are responsible for the  great differences in properties of the family members
for the observed three families. It is a hope~\cite{NF} that the mass matrices can be expressed as follows
\begin{eqnarray}
\label{qprime}
M&=& 
 \sum_{k=0, k'=0, k"=0}^{\infty}\, {Q}^k \,{Q'}^{k'} \,{Y'}^{k"} \, M_{Q \,Q'\, Y'\,k k' k"}\, ,
\end{eqnarray}
where $Q,Q'$ and $Y'$ are the operators while the matrices $M_{Q' Y'\,k k'}$  do not, hopefully, depend
on  the family member, that is that they might be the same for all the members of one family. To neutrino
an additional mass matrix might be added, which is zero for all the other family members if the
Majorana contribution is taken into account.

While for the lower four families the contributions which depend on $Q,Q'$ (and $Y'$) quantum numbers
of each of a family member are expected to be large, this should not be the case
for the upper four families
(in comparison with the contributions on the tree level).

In the next section we present the loop contributions to the three level mass matrices.
The contributions originate in two kinds
of scalar fields, namely in $\tilde{\omega}_{abs}$ and in  $\omega_{stt'}$,
and in the massive gauge fields and affect both groups of four families.
First we analyse the effect of one and two loops corrections for the case, that
each of four families would decouple into twice $2 \times 2$ mass matrices, under the
assumption that the lower two families of each group of four families weakly couple to the
upper two families of the same group. This assumption seems meaningful from the point of view
of mass matrices on the tree level, presented on tables~\ref{Table VII.},~\ref{Table VIII.},
as well as from the experimental data for the lower three families. The measured values of the mixing
matrices for the observed families supports such an  assumption for quarks, but not for leptons.
We neglect accordingly for this first step the
nonzero mass matrix elements between the lower and the upper two families for each
group. We  then proceed to take into account  one loop corrections for all four
families of each of the two groups.

\section{Mass matrices beyond the tree level}
\label{beyond}

It is the purpose of this section (and also of this paper) to manifest that,
although in the {\it spin-charge-family-theory}
the matrix elements  of different family members are within
each of the two groups of four families on the tree level very much correlated,
the loop corrections lead to mass matrices, which manifest great differences in
properties of the lowest three families.

We show that the one loop corrections originating in the massive gauge fields change  masses of
families, while they leave mixing matrices unchanged. One loop corrections
originating in dynamical scalar fields change both,  masses and  mixing matrices.

Let us repeat the assumptions~\cite{NF}:
%
{\bf i.} In the break from $SU(2)_{I} \times SU(2)_{II} \times U(1)_{II}$ to
$SU(2)_{I} \times U(1)_{I}$  the superposition of the $\tilde{\omega}_{abs}$
scalar fields which are the gauge  fields of $\vec{\tilde{\tau}}^{2}$ and $\vec{\tilde{N}}_{R}$
gain non zero    vacuum expectation values, causing nonzero mass matrices for fermions.
The lower four families, which do not couple to these scalar fields,
remain massless.
{\bf ii.} In the electroweak break  the superposition of the  $\tilde{\omega}_{abs}$ scalar fields
which  are the gauge fields of  $\vec{\tilde{\tau}}^{1}$ and  $\vec{\tilde{N}}_{L}$,  and
the superposition of scalar fields $\omega_{abs}$ which are the gauge fields of  $Q$, $Q'$ and $Y'$ gain
nonzero vacuum expectation values. (While the scalar gauge fields of $Q$, $Q'$ and $Y'$
influence  masses of all the eight families, the scalar gauge fields of  $\vec{\tilde{\tau}}^{1}$
and  $\vec{\tilde{N}}_{L}$ influence only the lower four families.)
{\bf iii.} There is also a term in loop corrections of
 a very special products of superposition of $\omega_{abs}, s=5,6,9,\cdots,14$ and
$\tilde{\omega}_{abs}\,, \, s=5,6,7,8$  scalar fields, which couple only to
the right handed neutrinos and their charge conjugated states of the lower four families,
which might change drastically the properties of neutrinos of the lower four families.

Let us clarify the notation. We have before the two breaks  two times ($\Sigma \in \{II,I\}$,
$II$ denoting the upper four and $I$  the lower four families) four
massless vectors $\psi^{\alpha }_{\Sigma (L,R)}$ for each member of a
family $\alpha=\in\{u,d,\nu,e\}$.
Let $i, \, i\in\{1,2,3,4\}\,,$ denotes one of the four family members of each of the two
groups of massless families
\begin{equation}
\label{notationvecmassless}
\left(\psi^{\alpha}_{\Sigma (L,R)} \right)^{T}= \left( \psi^{\alpha}_{\Sigma\,1},\, \psi^{\alpha }_{\Sigma\,2},\,
\psi^{\alpha}_{\Sigma \,3},\, \psi^{\alpha}_{\Sigma\,4}\right)_{(L,R)}\,.
\end{equation}
Hence, we have for the lowest four families ($\Sigma=I$) and the  $u$ family member ($\alpha=u$)
\begin{equation}
\left(\psi^{u}_{I (L,R)}\right)^T = \left(u,\, c,\, t,\, u_{4}\right)_{(L,R)} \;,\nonumber
\end{equation}
$u_{4}$ to be recognized as the new, that is the fourth family member.
We then have
\begin{eqnarray}
\label{matrixelements}
\overline{\psi}^{u}_{I\:L} \:{\cal M}^{u\, I}_{(o)} \: \psi^{u}_{I\:R}
=\overline{\psi}^{u}_{I \,L\,i} \:{\cal M}^{u\,I}_{(o)\:ij}\:\psi^{u}_{I\,R\,j}\,.
\end{eqnarray}

Let $\Psi^{\alpha}_{\Sigma (L,R)}$ be the final massive four vectors for each of the
two groups of families, with all loop corrections
included
\begin{eqnarray}
\label{notatiovecnmass}
\psi^{\alpha}_{\Sigma\, (L,R)} &=& V^{\alpha}_{\Sigma} \,\Psi^{\alpha}_{\Sigma \,(L,R)} \,,\nonumber \\
V^{\alpha}_{\Sigma} &=& V^{\alpha}_{\Sigma\,(o)}\,V^{\alpha}_{\Sigma\,(1)}\,
\cdots V^{\alpha}_{\Sigma\,(k)} \cdots \,.
\end{eqnarray}
Then $\Psi^{\alpha \, (1)}_{ \Sigma \,(L,R)}$ includes one loop corrections and
$\Psi^{\alpha \, (k)}_{ \Sigma (L,R)} $ up to $(k)$ loops corrections
\begin{eqnarray}
\label{notationvonek}
V^{\alpha}_{\Sigma\,(o)}\,\Psi^{\alpha \,(o)}_{ \Sigma \,(L,R)} &=&   \psi^{\alpha}_{\Sigma\, (L,R)}
\,,\nonumber\\
V^{\alpha}_{\Sigma\,(o)}\, V^{\alpha}_{\Sigma\, (1)}\,\cdots V^{\alpha}_{\Sigma \,(k)}\,
\Psi^{\alpha \,(k)}_{ \Sigma (L,R)} &=&
\psi^{\alpha}_{\Sigma (L,R)}\,.
\end{eqnarray}

From the starting action the mass matrices on the tree level follow as presented in
Tables~(\ref{Table VII.}, \ref{Table VIII.}). Not being able (yet) to calculate these
matrix elements, we take them as parameters. Not (yet) paying  attention to the $CP$
non conservation, we assume in this paper that mass matrices are real and symmetric.

We calculate in this paper one and for a simplified version of two decoupled $2 \times 2$ families
of a four family group two loops corrections to the tree level mass matrices.
The mass matrices, originating in the vacuum expectation values
of the scalar fields which are superposition of $\tilde{\omega}_{abd}$ fields
(appearing as $\tilde{g}^{\tilde{N}_R}\, \hat{\vec{\tilde{N}}}_R \vec{\tilde{A}}^{\tilde{N}_R}_{s}\,,$
$\tilde{g}^{\tilde{Y}'}   \, \hat{\tilde{Y'}}\,\tilde{A}^{\tilde{Y'}}_{s}\,,$
$\frac{\tilde{g}^{2}}{\sqrt{2}}\, \hat{\tilde{\tau}}^{2\pm} \,\tilde{A}^{2\pm}_{s}\,, $
$\tilde{g}^{\tilde{N}_L} \hat{\vec{\tilde{N}}}_L \vec{\tilde{A}}^{\tilde{N}_L}_{s}\,,$
$\tilde{g}^{\tilde{Q'}}\,  \, \hat{\tilde{Q'}}\,   \tilde{A}^{\tilde{Q'}}_{s}\,,$
$ \frac{\tilde{g}^{1}}{\sqrt{2}}\, \hat{\tilde{\tau}}^{1\pm} \,\tilde{A}^{1\pm}_{s}\,,$ the reader can find
the application of these operators on family members in appendix~\ref{technique}),
are in this paper assumed to be real and symmetric. On the tree level they manifest as the two  by diagonal
$4\times 4$ matrices with the symmetry on the tree level presented
bellow  
\begin{equation}
\label{M0}
{\cal M}_{(o)} = \begin{pmatrix} - a_1 & e
& 0 & b\\ e & - a_2 & b & 0\\ 0 & b & a_1 & e\\
b &  0 & e & a_2
\end{pmatrix}\,,
\end{equation}
with the matrix elements $a_1 \equiv a^{\Sigma}_{ \pm \,1}$, $a_2\equiv a^{\Sigma}_{\pm 2}$,
$b \equiv b^{\Sigma}_{\pm}$ and
$e \equiv e^{\Sigma}_{\pm}$, which are different for the upper ($\Sigma=II$) than for the lower
($\Sigma=I$) four families
\begin{equation}
\label{tildea}
a_1= \frac{1}{2} (\tilde{a}^{3}_{\pm} - \tilde{a}^{\tilde{N} 3}_{\pm}) \quad , \quad
a_2= \frac{1}{2} (\tilde{a}^{3}_{\pm} + \tilde{a}^{\tilde{N} 3}_{\pm}) \quad , \quad
b=\tilde{a}^{+}_{\pm}= \tilde{a}^{-}_{\pm} \quad , \quad
e=\tilde{a}^{\tilde{N} +}_{\pm}= \tilde{a}^{\tilde{N} -}_{\pm}\,.
\end{equation}
The matrix elements for the upper   four families ($\Sigma=II$) are:
$\tilde{a}^{3}_{\pm}= \tilde{a}^{23}_{\pm},
$ $\tilde{a}^{\tilde{N} 3}_{\pm}= \tilde{a}^{\tilde{N}_{R} 3}_{\pm}$,
$\tilde{a}^{\pm}_{\pm}= \tilde{a}^{21}_{\pm} \pm i\, \tilde{a}^{22}_{\pm}$,
$\tilde{a}^{\tilde{N} \pm}_{\pm}= \tilde{a}^{\tilde{N}_{R} 1}_{\pm}
\pm  i \, \tilde{a}^{\tilde{N}_{R} 2}_{\pm}$. For the
lower four families ($\Sigma =I\,$) we must take $\tilde{a}^{3}_{\pm}= \tilde{a}^{13}_{\pm},
$ $\tilde{a}^{\tilde{N} 3}_{\pm}= \tilde{a}^{\tilde{N}_{L}3}_{\pm}$,
$\tilde{a}^{\pm}_{\pm}= \tilde{a}^{11}_{\pm} \pm i\, \tilde{a}^{12}_{\pm}$,
$\tilde{a}^{\tilde{N}\pm}_{\pm}= \tilde{a}^{\tilde{N}_{L} 1}_{\pm}
\pm  i \, \tilde{a}^{\tilde{N}_{L}2}_{\pm}$.
 ($\pm$) in the denominator distinguishes between the matrix elements for the pair ($d$
and $e$) ($+$) and the pair ($u$ and $\nu$) ($-$). $\bar{\psi} M \psi$ in Eq.~(\ref{faction}) can, namely,
be expressed as
\begin{eqnarray}
\label{Mpm}
\bar{\psi} M \psi   &=& \sum_{s=7,8}\,\bar{\psi} \gamma^{s}\, p_{0s}\,\psi  =
 \psi^{\dagger}\, \gamma^{0}\, (\stackrel{78}{(-)}\,p_{0-} + \stackrel{78}{(+)}\,p_{0+}) \psi\, ,\nonumber\\
\stackrel{78}{(\pm)}&=&  \frac{1}{2}\,(\gamma^{7} \, \pm i\,\gamma^{8} )\,, \nonumber\\
p_{0\pm}&=& (p_{07} \mp i\, p_{08})\,,\quad s \in\{7,8\}.
\end{eqnarray}
The reader is kindly asked to learn how do the operators $(\stackrel{78}{(\mp)}$, any superposition of the
operators $S^{ab}$, or any superposition of the operators $\tilde{S}^{ab}$) apply on any family member of any
family in the ref.~\cite{NF} and in the appendix~\ref{appendixVo}, where a short presentation of
these properties is made.

To the tree level contributions of the scalar $\tilde{\omega}_{ab \pm}$ fields,
diagonal matrices have to be added, the same for all the eight families and different for
each of the family member ($u,d,\nu,e$), $ a_{\mp} \equiv a^{\alpha}_{\mp}$,
which are the tree level contributions of the scalar $\omega_{sts'}$ fields
\begin{eqnarray}
\label{diagonaltreemain}
a_{\mp}&=& e\, Q\, A_{\mp} +  g^{1} \,\cos \theta_1\, Q'\, Z^{Q'}_{\mp}\, +
               g^{2}\, \cos \theta_2\, Y'\, A^{Y'}_{\mp}\,.
\end{eqnarray}
$Q, Q'$ and $Y'$ stay  for the eigenvalues of the operators $\hat{Q},\hat{Q'}$ and $\hat{Y'}$
of the right handed $\alpha $ member of any of the families~\footnote{We shall put the operator sign $\hat{O}$
on the operator $O$ only when it is needed so that we can distinguish between the operators and their eigenvalues.
}.
Therefore, the tree level mass matrices ${\cal M}^{\alpha\, \Sigma}_{(o)}$
are different for the upper ($\Sigma=II$)  than for the lower
($\Sigma=I$) four families and  they are also different for
the pairs of ($d$,  $e$) and  ($u$, $\nu$), but are the same for $u$ and  $\nu$ and for
$d$ and  $e$ before adding $ \,a^{\alpha}_{\mp} \,\,I_{8\times8} \,$, which is different
for each family member.
The matrices $M^{\alpha}$ are indeed $8\times 8$ matrices with two by diagonal $4\times4$ matrices
also after  the loops corrections included.
The parameters Eq.~(\ref{tildea}), which enter into the tree level mass matrices after the assumptions
explained at the beginning of this section,  are presented in Table~\ref{Tablepartree}.
 \begin{table}
 \begin{center}
 \begin{tabular}{|r|r||c|c|c|c||c|c|c|}
 \hline
 $\Sigma$ &$\alpha$&&&&&&&\\
 \hline \hline
 $II$&$(u,\nu)$&$\tilde{a}^{\tilde{N}_{R}3}_{-}$& $\tilde{a}^{\tilde{N}_{R}\pm}_{-}$&
 $\tilde{a}^{23}_{-}$&$\tilde{a}^{2 \pm}_{-}$& $a^{Q}_{-}$&$a^{Q'}_{-}$&$a^{Y'}_{-}$\\
 \hline
 $II$&$(d,e)$&$\tilde{a}^{\tilde{N}_{R}3}_{+}$& $\tilde{a}^{\tilde{N}_{R}\pm}_{+}$&
  $\tilde{a}^{23}_{+}$&$\tilde{a}^{2 \pm}_{+}$& $a^{Q}_{+}$&$a^{Q'}_{+}$&$a^{Y'}_{+}$\\
 \hline
 $I$&$(u,\nu)$&$\tilde{a}^{\tilde{N}_{L}3}_{-}$& $\tilde{a}^{\tilde{N}_{L}\pm}_{-}$&
  $\tilde{a}^{13}_{-}$&$\tilde{a}^{1 \pm}_{-}$& $a^{Q}_{-}$&$a^{Q'}_{-}$&$a^{Y'}_{-}$\\
 \hline
 $I$&$(d,e)$&$\tilde{a}^{\tilde{N}_{L}3}_{+}$& $\tilde{a}^{\tilde{N}_{L}\pm}_{+}$&
   $\tilde{a}^{1 3}_{+}$&$\tilde{a}^{1 \pm}_{+}$& $a^{Q}_{+}$&$a^{Q'}_{+}$&$a^{Y'}_{+}$\\
 \hline
 \end{tabular}
 \end{center}
 \caption{\label{Tablepartree} The parameters entering into the tree level mass matrices
 are presented.  The notation $\tilde{a}^{\tilde{A}i}_{\pm}=$
 $-\tilde{g}^{\tilde{A}}\, \tilde{A}^{\tilde{A}i}_{\pm}$ (staying for
 $-\tilde{g}^{\tilde{N}_{R}}\, \tilde{A}^{\tilde{N}_{R}i}_{\pm}\,$,
 $-\tilde{g}^{2}\, \tilde{A}^{2i}_{\pm}\,$,
 $-\tilde{g}^{\tilde{N}_{L}}\, \tilde{A}^{\tilde{N}_{L}i}_{\pm}\,$,
 $-\tilde{g}^{1}\, \tilde{A}^{1i}_{\pm}\,$\,),
 $a^{Q}_{\mp}= g^{Q} \,A^{Q}_{\mp}\,$ $a^{Q}_{\mp}= g^{Q'}\, A^{Q'}_{\mp}\,$,
 $a^{Y'}_{\mp}= g^{Y'} A^{Y'}_{\mp}\,$ is used.
}
 \end{table}

On the tree level we have
\begin{equation}
\label{M}
{\cal M}^{\alpha}_{(o)} = \begin{pmatrix} {\cal M}^{\alpha\,II}_{(o)} & 0\\
0&{\cal M}^{\alpha\,I}_{(o)}
\end{pmatrix}\,.
\end{equation}
Since the upper and the lower four family mass matrices appear at two
completely different scales, determined by two orthogonal sets of scalar fields, have the two
tree level mass matrices ${\cal M}^{\alpha \,\Sigma}_{(o)}$  very little in common, only the symmetries and
the contributions from Eq.~(\ref{diagonaltreemain}).

On the tree level  we have %
$\psi^{\alpha}_{\Sigma \:(L,R)}=V^{\alpha}_{\Sigma\,(o)}\:
\Psi^{\alpha \,(o)}_{\Sigma \:(L,R)}$
%
and
\begin{equation}
\label{treenotation}
          < \psi^{\alpha}_{\Sigma\:L}|\gamma^0 \, {\cal M}^{\alpha\, \Sigma}_{(o)}\,
 |\psi^{\alpha}_{ \Sigma\:R}> = < \Psi^{\alpha \,(o)}_{\Sigma\: L}|\gamma^0 \,V^{\alpha\,
 \dagger}_{\Sigma\,(o)}\,
 {\cal M}^{\alpha\, \Sigma}_{(o)}\,V^{\alpha}_{\Sigma(o)}\,|\Psi^{\alpha \,(o)}_{\Sigma\: R \,(o)}>,
\end{equation}
from where  the tree level mass eigenvalues follow
\begin{equation}
\label{treediag}
{\cal M}^{\alpha\, \Sigma}_{(o) D}= V^{\alpha\,\dagger}_{\Sigma\,(o)}\, {\cal M}^{\alpha\,\Sigma}_{(o)}
\,V^{\alpha}_{\Sigma\,(o)}
=diag(m^{\alpha\,\Sigma}_{(o)\,1}, m^{\alpha\,\Sigma}_{(o)\,2},
m^{\alpha\,\Sigma}_{(o)\,3}, m^{\alpha\,\Sigma}_{(o)\,4}).
\end{equation}
The one loop corrections leads to
$\psi^{\alpha}_{\Sigma \:(L,R)}$ $ =V^{\alpha}_{\Sigma\,(o)}\:\Psi^{\alpha\,(o)}_{\Sigma \:(L,R)}$
$ =V^{\alpha}_{\Sigma\,(o)}\,V^{\alpha}_{\Sigma\,(1)}\,$
$\Psi^{\alpha \,(1)}_{\Sigma \:(L,R)}$ and  ${\cal M}^{\alpha\,\Sigma}_{(o \,1)}$ include all the one
loop corrections evaluated
among the massless states, so that
\begin{equation}
\label{oneloopnotationinter}
< \psi^{\alpha}_{\Sigma L}|\gamma^0 \, {\cal M}^{\alpha\, \Sigma}_{(o \,1)}\,
 |\psi^{\alpha}_{ \Sigma \:R}> = < \Psi^{\alpha\,(o)}_{\Sigma\, L}|\gamma^0 \,
 V^{\alpha\,\dagger}_{\Sigma\,(o)}\,
{\cal M}^{\alpha\,\Sigma}_{(o \,1)}\,V^{\alpha}_{\Sigma\,(o)}\,|\Psi^{\alpha\,(o)}_{\Sigma R}>\,.
\end{equation}
The mass matrix including up to one loop corrections is
\begin{equation}
\label{oneloopnotation}
{\cal M}^{\alpha\, \Sigma}_{(1)}= V^{\alpha\, \dagger}_{\Sigma\,(o)}\, {\cal M}^{\alpha\,\Sigma}_{(o\,1)}
\,V^{\alpha}_{\Sigma\,(o)} +
M^{\alpha\,\Sigma}_{(o) \,D}=
V^{\alpha\, \dagger}_{\Sigma\,(o)}\,( {\cal M}^{\alpha\,\Sigma}_{(o\,1)} +
{\cal M}^{\alpha\, \Sigma}_{(o)} )\,V^{\alpha}_{\Sigma\,(o)}\,.
\end{equation}
Thus the contribution up to one loop is $ < \Psi^{\alpha\,(o)}_{\Sigma \,L}|\gamma^0 \,
V^{\alpha\,\dagger}_{\Sigma\, (o)} \,( {\cal M}^{\alpha\,\Sigma}_{(o \,1)} +
{\cal M}^{\alpha\,\Sigma}_{(o)}) )\,V^{\alpha}_{\Sigma\,(o)} \,
|\Psi^{\alpha\,(o)}_{\Sigma\, R} > $,
which can be written as 
\begin{eqnarray}
\label{oneloopV}
&&< \psi^{\alpha}_{\Sigma\, L}|\gamma^0 \, ({\cal M}^{\alpha\, \Sigma}_{(o\,1)} +
{\cal M}^{\alpha\,\Sigma}_{(o)})  \,
 |\psi^{\alpha}_{ \Sigma \:R}> =\nonumber\\
 &&< \Psi^{\alpha\,(1)}_{\Sigma \, L}|\gamma^0 \,(V^{\alpha}_{\Sigma\, (o)}\,
 V^{\alpha}_{\Sigma\,(1)})^{\dagger}\,
 ( {\cal M}^{\alpha\,\Sigma}_{(o\,1)} + {\cal M}^{\alpha\,\Sigma}_{(o)}  )
 \,V^{\alpha}_{\Sigma \,(o)}\,V^{\alpha}_{\Sigma\,(1)}
 |\Psi^{\alpha\,(1)}_{\Sigma \, R}>\,,
\end{eqnarray}
with $V^{\alpha}_{\Sigma\,(1)}$ which is obtained from
\begin{eqnarray}
\label{M1Dmain}
{\cal M}^{\alpha\,\Sigma}_{(1)\,D}= V^{\alpha\,\dagger}_{\Sigma\,(1)}\,\left[ V^{\alpha \,\dagger}_{\Sigma \,(o)}
\,( {\cal M}^{\alpha\,\Sigma}_{(o \,1)}+
{\cal M}^{\alpha\,\Sigma}_{(o)}) \,V_{\alpha\,\Sigma\, (o)} \right]\,V^{\alpha}_{\Sigma \,(1)}=\nonumber\\
diag(m^{\alpha\,\Sigma}_{(1)\,1}, m^{\alpha\,\Sigma}_{(1)\,2}, m^{\alpha\,\Sigma}_{(1)\,3},
m^{\alpha\,\Sigma}_{(1)\,4})\,,
\end{eqnarray}
with $m^{\alpha\,\Sigma}_{(1)\,i}, \, i \in \{1,2,3,4\}\,,$
the mass eigenvalues, which include one loop corrections.

Taking into account corrections up to $(k)$ loops, we have
\begin{eqnarray}
\label{MvPsi}
&&< \psi^{\alpha}_{\Sigma\, L}|\gamma^0 \,({\cal M}^{\alpha\,\Sigma}_{(o\, k)} + \cdots +
{\cal M}^{\alpha\,\Sigma}_{(o\,1)} + {\cal M}^{\alpha\,\Sigma}_{(o)})  \,
 |\psi^{\alpha}_{ \Sigma \:R}> = \nonumber\\
 &&< \Psi^{\alpha \,(k)}_{\Sigma\, L}|\gamma^0 \,(V^{\alpha}_{\Sigma\,(o)}
 \,V^{\alpha}_{\Sigma\,(1)}\,\cdots V^{\alpha}_{\Sigma\,(k)})^{\dagger}\,
 ({\cal M^{\alpha\,\Sigma}}_{(o\, k)} + \cdots \nonumber\\
 &&+ {\cal M}^{\alpha\,\Sigma}_{(o\,1)} + {\cal M}^{\alpha\,\Sigma}_{(o)})
 \,\,V^{\alpha}_{\Sigma\,(o)}\,V^{\alpha}_{\Sigma(1)}
 \,\cdots V^{\alpha}_{\Sigma\,(k)}\,|\Psi^{\alpha\,(k)}_{\Sigma\, R}>.
\end{eqnarray}
$V^{\alpha}_{\Sigma\,(k)}$ follows  from		
\begin{eqnarray}
\label{MvD}
&&{\cal M}^{\alpha\,\Sigma}_{(k)D}= V^{\alpha\,\dagger}_{\Sigma\,(k)}\,[ V^{\alpha\,\dagger}_{\Sigma\,(k-1)}
\,\dots\, \nonumber\\
&& V^{\alpha\,\dagger}_{\Sigma\,(1)}\,V^{\alpha\,\dagger}_{\Sigma\,(o)}\,
({\cal M}^{\alpha\,\Sigma}_{(o\,k)}+ \cdots + {\cal M}^{\alpha\,\Sigma}_{(o\,1)} +
{\cal M}^{\alpha\,\Sigma}_{(o)})  \,V^{\alpha}_{\Sigma\,(o)} \,
V^{\alpha}_{\Sigma\,(1)}\cdots V^{\alpha}_{\Sigma\,(k-1)} \,]\,V^{\alpha}_{\Sigma\,(k)}\nonumber\\
&& = diag(m^{\alpha\,\Sigma}_{(k)\,1},\cdots m^{\alpha\,\Sigma}_{(k)\,4}),
\end{eqnarray}
with $m^{\alpha\,\Sigma}_{(k)\,i},\, i\in\{1,2,3,4\}\,,$ the mass eigenvalues of the states,
which take into account
up to $(k)$ loops corrections.

In what follows we shall use the indices $\Sigma$ and $\alpha$ only when we explicitly calculate
mass matrices for a particular group of families and for a particular member, otherwise we shall assume that
both indices are all the time present and we shall skip both.  Eq.(\ref{treediag}) will, for example,
accordingly read
\begin{equation}
\label{V0}
{\cal M}_{(o) D}= V^{\dagger}_{(o)}\, {\cal M}_{(o)}\,V_{(o)}
=diag(m_{(o)\,1}, m_{(o)\,2},
m_{(o)\,3}, m_{(o)\,4})\,,
\end{equation}
with the indices $\Sigma$ and $\alpha$ assumed, but not written.
Similarly Eq.~(\ref{notatiovecnmass}) reads
\begin{eqnarray}
\label{notationvecmass}
\psi_{(L,R)} &=& V \, \Psi_{(L,R)} \,,\nonumber \\
V &=& V_{(o)}\,V_{(1)}\,\cdots \,V_{(k)}\, \cdots \,,
\end{eqnarray}
connecting the massless $\psi$ and the massive $\Psi$ with all the loop corrections included.
In our case, since ${\cal M}_{(o)}$ 
are in this paper assumed to be real and symmetric,
$V^{\dagger}_{(o)} = V^{T}_{(o)}$. 

Loop corrections (with the  gauge and dynamical scalar fields  contributing
coherently) are expected to cause differences in  mass matrices among the family members of
the lower four families, and will hopefully explain the  experimental data for the so far
observed three families of quarks and leptons.  The differences among
the family members of the upper four families are  expected to be small even after taking into account
loop corrections, since the contributions to the loop corrections, which distinguish among
family members, originate in the
$\omega_{sta}$ dynamical massive fields, the  scalar and vector  ones,
whose masses can only be of the order of the electroweak scale
and this is expected to be for orders of
magnitude smaller than the scale of the break of symmetry which brings masses to the
upper four families.  The only exception is $\tau^{2i} A^{2i}_m$.

The contribution which transforms the right handed neutrino into
his charged conjugate one, influences only the lower four families, because, by the
assumption, the superposition of the
$\tilde{\omega}_{abs}$ fields couple only to the lower four families~\cite{NF}.
A short explanation is presented in appendix~\ref{majoranas}.

In appendix~\ref{appendixVo} the matrix ${\cal M}_{(o)}$ (Eq.(\ref{M0}))
is diagonalized for a general choice of
matrix elements, assuming that the matrix is real and symmetric, with the symmetry on the tree level
as presented in~Eq.(\ref{M0}).
 A possible non hermiticity of the mass matrices on the tree level is neglected.
 The diagonalizing matrix is presented.

 In the ref.~\cite{gmdn} the authors,  assuming that loop corrections  (drastically) change  mass matrix
 elements as they follow on the tree level from the {\em spin-charge-family-theory}, keep the symmetries
 of mass matrices as dictated by the {\em spin-charge-family-theory} on the tree level and
 fit the mass matrix elements for the lower four families to existing experimental data for
 a particular choice of  masses of the fourth family members.

 In this paper we make one loop corrections to the tree level mass matrices,
 demonstrating that loop corrections may contribute to
 the tree level mass matrices to the experimentally acceptable direction.

We calculate loop corrections  originating in two kinds of the scalar
dynamical fields, those originating in $\tilde{\omega}_{abs}$ ($
\tilde{g}^{\tilde{Y}'} \; \tilde{Y}' \,\tilde{A}^{\tilde{Y}'}_{s}\;$,
$\frac{\tilde{g}^{2}}{\sqrt{2}}\, \tilde{\tau}^{2\pm} \,\tilde{A}^{2\pm}_{s}, \;
	        \tilde{g}^{\tilde{N}_{L,R}} \vec{\tilde{N}}_{L,R} \,\vec{\tilde{A}}^{\tilde{N}_{L,R}}_{s}\,$, $
		  	                \tilde{g}^{\tilde{Q}'} \,   \; \tilde{Q}'\;
		  	                \tilde{A}^{\tilde{Q}'}_{s}$, $
		  	       	          \, \frac{\tilde{g}^{1}}{\sqrt{2}}\, \tilde{\tau}^{1\pm}
		  	       	          \,\tilde{A}^{1\pm}_{s}\,$)
		  	       	          and those originating in $\omega_{abs}$
	          ($e\, Q \, A_{s}\,, \;g^{1}\, \cos \theta_1 \,Q'\, Z^{Q'}_{s}\,$,
              $ g^{Y'} \, Y'\, A^{Y'}_{s}$)
and in the  massive gauge fields ($
               g^{2} \cos \theta_2 \,Y'\, A^{Y'}_{m}$, $g^{1}\, \cos \theta_1 \,Q'\, Z^{Q'}_{m}\,$)
               as it follow from~Eq.(\ref{factionI}).

\begin{figure}
\centering\includegraphics[width=12cm]{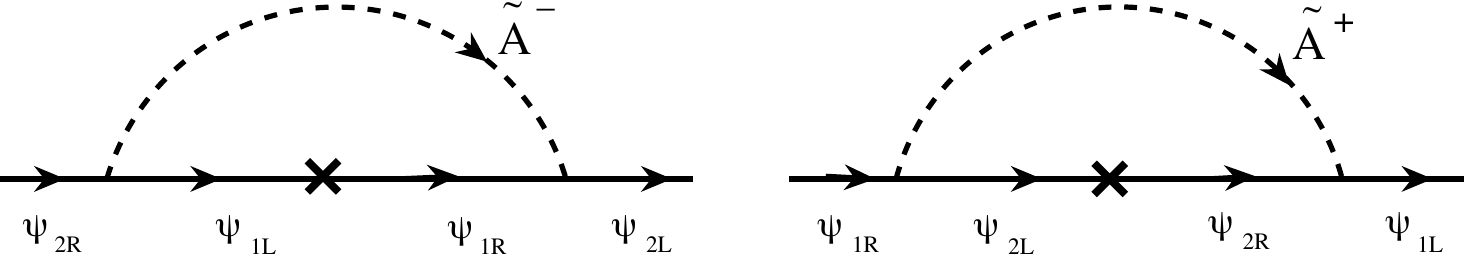}
\caption{\label{Fig1} One loop contributions originating in $\tilde{A}^{\pm}$ scalar fields,  where
$\tilde{A}^{\pm}$ stays for $\tilde{A}^{2 \pm}_{\pm}$ or $\tilde{A}^{\tilde{N}_{R} \pm}_{\pm}$ when
the upper four families are treated, while it stays for $\tilde{A}^{1 \pm}_{\pm}$ or
$\tilde{A}^{\tilde{N}_{L} \pm}_{\pm}$ when the lower four families are treated. Each of the massless states
$\psi_{i(R,L)}$ in the figure, staying instead of $\psi^{\alpha}_{\Sigma (R,L) i}$ where $\Sigma =II$ determines
the upper four families group membership and $\Sigma =I$ the lower four families group membership,
should correspondingly carry  also
the family member index $\alpha=(u,d,\nu,e)$
the group family index  $\Sigma =II,I$, where $II$ denotes the upper four
 and $I$ for the lower four families,
besides the family index $i=(1,2,3,4),$ which
distinguishes among the familes within each
of the two groups.  }
\end{figure}

\begin{figure}
\centering\includegraphics{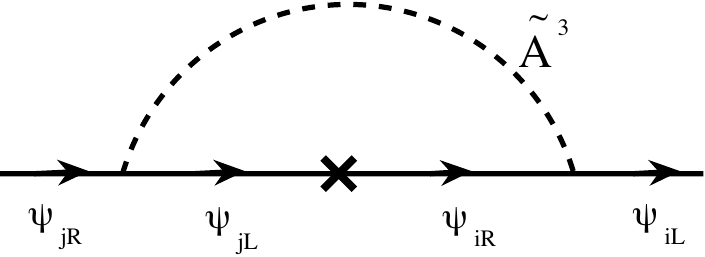}
\caption{\label{Fig2} One loop contributions  originating in $\tilde{A}^{3}$ scalar fields,  where
$\tilde{A}^{3}$ stays for $\tilde{A}^{23}_{\pm}$ or $\tilde{A}^{\tilde{N}_{R}3}_{ \pm}$ when
the upper four families are treated, while it stays for $\tilde{A}^{1 3}_{\pm}$ or
$\tilde{A}^{\tilde{N}_{L}3}_{ \pm}$ when the lower four families are treated.
The rest of comments are the same as in Fig.~\ref{Fig1}.
Each of  states carries also
the family member index $\alpha$, the $\Sigma$ index determining one of the two four families groups
and the index $i$ which distinguishes among the families within each of the two groups of four families.
} \end{figure}

\begin{figure}
\centering\includegraphics{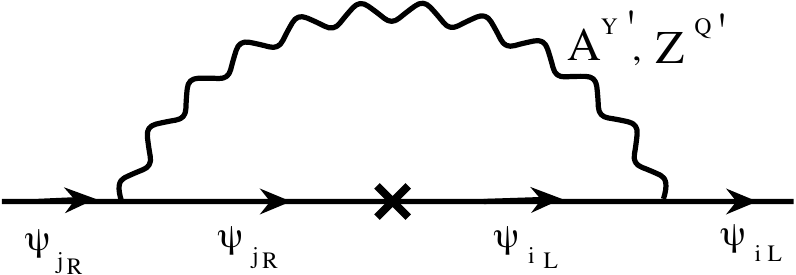}
\caption{\label{Fig3} One loop contributions from the massive gauge vector fields
$A_m^{Y^\prime}$ and $Z_m^{Q^\prime}$. Each of  states carry in addition to $i,j=(1,2,3,4)$,
which distinguishes the four members of each of the two group of four families, also the group index
$\Sigma=II,I$, for the upper and the lower four families, respectively, and
the family member index $\alpha=(u,d,\nu,e)$.}
\end{figure}

In appendix~\ref{M44massmatrix} the corresponding loop corrections are calculated in a general form,
which enables to distinguish among members of the scalar fields of both kinds and of the massive gauge fields.
The corresponding loop diagrams are presented in Figures~\ref{Fig1},\ref{Fig2},\ref{Fig3}.

Fig.~\ref{Fig1} shows the  one loop diagram for the  contribution of the terms, either
($-\gamma^0 \stackrel{78}{(\mp)}\,$$
\frac{\tilde{g}^{2}}{\sqrt{2}}\, \hat{\tilde{\tau}}^{2\pm} $ $\,\tilde{A}^{2 \pm}_{\mp}$) or
($-\gamma^0 \stackrel{78}{(\mp)}\,$$
\frac{\tilde{g}^{\tilde{N}_R}}{\sqrt{2}}\, \hat{\tilde{N}}_{R}^{\pm} $ $\,\tilde{A}^{\tilde{N}_R \pm}_{\mp}$),
both presented in~Eq.(\ref{factionI}), when the upper four families are treated. For the lower four
families the same diagram shows the one loop corrections induced by either
($-\gamma^0 \stackrel{78}{(\mp)}\,$$
\frac{\tilde{g}^{1}}{\sqrt{2}}\, \hat{\tilde{\tau}}^{1\pm} $ $\,\tilde{A}^{1 \pm}_{\mp}$) or
($-\gamma^0 \stackrel{78}{(\mp)}\,$$
\frac{\tilde{g}^{\tilde{N}_L}}{\sqrt{2}}\, \hat{\tilde{N}}_{L}^{\pm} $ $\,\tilde{A}^{\tilde{N}_L \pm}_{\mp}$),
Eq.(\ref{factionI}). These fields couple the  family members as it is presented in  Tables~(\ref{Table VII.},
\ref{Table VIII.}) and demonstrated in the diagram
\begin{equation}
\label{diagramNtau}
      \stackrel{\stackrel{\vec{\tilde{N}}_{L}}{\leftrightarrow} }{\begin{pmatrix} I_{4} & I_{3} \\
I_{1} & I_{2}   \end{pmatrix}} \updownarrow  \vec{\tilde{\tau}}^{1}   \quad ,
\quad \stackrel{\stackrel{\vec{\tilde{N}}_{R}}{\leftrightarrow} }{\begin{pmatrix} II_{4} & II_{3} \\
II_{1} & II_{2} \end{pmatrix}} \updownarrow  \vec{\tilde{\tau}}^{2}\quad.
\end{equation}
The term ($-\gamma^0 \stackrel{78}{(-)}\,
\frac{\tilde{g}^{2}}{\sqrt{2}}\, \hat{\tilde{\tau}}^{2-} \,\tilde{A}^{2-}_{-}$)
applies to $u$-quarks [$\nu$-leptons], transforming the eighth family right handed $u$-quark
[$\nu$-lepton] ($II_{4}$ in the right diagram of Eq.(\ref{diagramNtau}))
into the fifth left handed one ($II_{1}$ in  Eq.(\ref{diagramNtau}))
and the seventh family right handed $u$-quark [$\nu$-lepton] ($II_{3}$)
into the sixth family left handed one ($II_{2}$),
for example.
While the term ($-\gamma^0 \stackrel{78}{(-)}\,
\frac{\tilde{g}^{\tilde{N}_R}}{\sqrt{2}}\, \hat{\tilde{N}}_{R}^{-} \,\tilde{A}^{\tilde{N}_R -}_{-}$)
transforms the eighth family right handed $u$-quark [$\nu$-lepton]
($II_{4}$ in the right diagram of Eq.(\ref{diagramNtau}))
into the seventh left handed one  ($II_{3}$ in  Eq.(\ref{diagramNtau}))
and the sixth family right handed $u$-quark [$\nu$-lepton] ($II_{2}$ in  Eq.(\ref{diagramNtau}))
into the fifth family left handed on  ($II_{1}$) and equivalently for the lower four families.
That is, the term ($-\gamma^0 \stackrel{78}{(-)}\,
\frac{\tilde{g}^{1}}{\sqrt{2}}\, \hat{\tilde{\tau}}^{1-} \,\tilde{A}^{1-}_{-}$) transforms
the fourth family right handed $u$-quark [$\nu$-lepton] ($I_{4}$ in the left diagram of
Eq.(\ref{diagramNtau}))) into the first left handed one ($I_{1}$)
and the third family right handed $u$-quark [$\nu$-lepton] ($I_{3}$)   into the second
family left handed one ($I_{2}$)  .
The term 
 ($-\gamma^0 \stackrel{78}{(-)}\,
\frac{\tilde{g}^{\tilde{N}_L}}{\sqrt{2}}\, \hat{\tilde{N}}_{L}^{-} \,\tilde{A}^{\tilde{N}_L -}_{-}$)
transforms correspondingly the fourth family right handed $u$-quark [$\nu$-lepton] ($I_{4}$)  into the
third family left handed one ($I_{3}$)  .

Correspondingly Fig.~\ref{Fig1} represents the one loop diagrams for
the $d$-quark and $e$-leptons for either the upper of the lower four families if
($-\gamma^0 \stackrel{78}{(-)}\,\frac{\tilde{g}^{A}}{\sqrt{2}}\,
\hat{\tilde{\tau}}^{A\pm} \,\tilde{A}^{A\pm}_{-}$) (where  index $A$ denotes  $2$ or $\tilde{N}_R$ for
the upper four families, and $\hat{\tilde{\tau}}^{Ai}$ correspondingly  $\hat{\tilde{\tau}}^{2i}$ and
$\hat{\tilde{N}}_{R}^{i}$, and   $1$ or $\tilde{N}_L$ for the lower four families,
and $\hat{\tilde{\tau}}^{Ai}$ correspondingly  $\hat{\tilde{\tau}}^{1i}$ and
$\hat{\tilde{N}}_{L}^{i}$) is replaced by
($-\gamma^0 \stackrel{78}{(+)}\,\frac{\tilde{g}^{A}}{\sqrt{2}}\,
\hat{\tilde{\tau}}^{A\pm} \,\tilde{A}^{A\pm}_{+}$), Eq.~(\ref{factionI}).

In Fig.~\ref{Fig2} the terms which in the loop corrections contribute to diagonal matrix elements
of the $u$-quarks  [$\nu$-leptons] and $d$-quarks [$e$-leptons] of each of the four members of
the upper four and the lower four families are presented.  Similarly as in Fig.~\ref{Fig1},
 the terms
$(-\gamma^0 \stackrel{78}{(\mp)} \,\tilde{g}^{2}  \,
\hat{\tilde{\tau}}^{23}\, \tilde{A}^{23}_{\mp}\,)$ and $(-\gamma^0 \stackrel{78}{(\mp)}\,
\tilde{g}^{\tilde{N}_{R}} \;
\tilde{N}^{3}_{R}\, \tilde{A}^{\tilde{N}^{3}_{R}}_{\mp})$,  (Eq.~(\ref{factionI})) contribute
to  the upper four families distinguishing among families and among the family members
pairs ($u\,,\nu$),
($-$), and ($d\,,e$), ($+$), while the terms
$(-\gamma^0 \stackrel{78}{(\mp)} \,\tilde{g}^{1}  \,
\hat{\tilde{\tau}}^{13}\, \tilde{A}^{13}_{\mp}\,)$ and $(-\gamma^0 \stackrel{78}{(\mp)}\,
\tilde{g}^{\tilde{N}_{L}} \;
\hat{\tilde{N}}^{3}_{L}\, \tilde{A}^{\tilde{N}^{3}_{L}}_{\mp})$ contribute to the
mass terms of the lower four families.

The eigenvalues of the operators $\hat{\tilde{\tau}}^{2\,3}$, $\hat{\tilde{N}}_{R}^{3}$,
$\hat{\tilde{\tau}}^{1\,3}$,
and $\hat{\tilde{N}}_{L}^{3}$  are presented in Table~\ref{Table FQN}.

The same Fig.~\ref{Fig2} represents also the one loop contribution  of the dynamical scalar fields
originating in $S^{ab}$, namely of
$e\, \hat{Q}\, A_{\mp}\,, $ $g^{1}\, \cos \theta_1 \,\hat{Q'}\, Z^{Q'}_{\mp}$ and $
               g^{2} \cos \theta_2\, \hat{Y'}\, A^{Y'}_{\mp}$, ($(-)$ for $u$-quarks and $\nu$-leptons,
$(+)$ for $d$-quarks and $e$-leptons),   if these fields replace $\tilde{A}^{3}_{\mp}$. These
diagonal terms are the same for all the four families of any of the two groups, but
since the operators
$\hat{Q},\, \hat{Q'}$ and $\hat{Y'}$ have different eigenvalues on each of the family members ($u,\,d,\,\nu, \,e$),
Table~\ref{Table MQN},
these matrix elements are different for different family members.

Fig.~\ref{Fig3} represents the contribution of the massive gauge field $ A^{Y'}_{m}$, originating in the dynamical
part of the Lagrange density in
Eq.~(\ref{factionI}) ($g^{2} \cos \theta_2\, \hat{Y'}\, A^{Y'}_{m}$). Replacing $ A^{Y'}_{m}$ by $Z^{Q'}_{m}\,$ the
same figure  represents also the contribution of the term
$g^{1}\, \cos \theta_1 \,\hat{Q'}\, Z^{Q'}_{m}\,$, Eq.~(\ref{factionI}).  Both contributions distinguish
among the family members and are the same for all the eight families. The  quantum
numbers $Q'$ and $Y'$ are presented in Table~\ref{Table MQN}.

In all the loop corrections 
the strength
of  couplings ( $\tilde{g}^{(2,1)}, $ $\tilde{g}^{(\tilde{N}_R,\tilde{N}_L)} $),
the application of  the operators ($\hat{\tilde{\tau}}^{(2,1)i}, $ $\hat{\tilde{N}}_{(R,L)}^{i} $),
as well as the masses of the dynamical fields playing, as usually, an essential role,
must be taken into account.

In appendix~\ref{M44massmatrix} the explicit evaluations of all the above
discussed  loop contributions are  derived in  a general form,
that is as  functions of parameters which determine a particular contribution.

The influence of a particular contribution to the mass matrices, and accordingly also to mixing matrices,
depends strongly on whether the upper or the lower four families are concerned, on the family members involved
and on the family quantum number of states involved in the corrections. 

The final mass matrices, manifesting the Lagrange density $\psi^{\dagger} \, \gamma^{0}\,
\gamma^{s}\, p_{0s}\,\psi $ with  one loop corrections
to which the scalar dynamical and massive gauge fields contribute, have  the shape, presented in
Eq.~(\ref{oneloopV}), (\ref{M1Dmain}),
with $V_{\alpha\,\Sigma(1)}$ which is obtained from Eq.~(\ref{M1Dmain}),
%
%
and which correspondingly determines mixing matrices with the one loop corrections included,
for each of the two groups of four families ($\Sigma=II,I$)
and for each of the family member $\alpha \in \{u,d,\nu,e\}$. The graphic representation of these
loop corrections can be seen in Figs.~(\ref{Fig1}, \ref{Fig2}, \ref{Fig3}).

Matrices ${\cal M}^{\alpha\,\Sigma}_{(1)}=$ ${\cal M}^{\alpha\,\Sigma}_{(1)\, \tilde{S}} + $
${\cal M}^{\alpha\,\Sigma}_{(1)\, S}$ $+ {\cal M}^{\alpha\,\Sigma}_{(1)\, V}$ are written in
terms of the parameters presented in
Tables~(\ref{Tablepartree}, \ref{MAall}) and used in  Eqs.~[(\ref{SigmaGENscalar}),(\ref{SigmaGENgauge}),
(\ref{tildemokS}), (
and (\ref{mokV})], respectively, of   appendix~\ref{M44massmatrix}.

The tree level masses  ($m^{\alpha\,\Sigma}_{(1)\,i}, \, i \in \{1,2,3,4\}\,$) and diagonalizing
matrices ($V_{\alpha\,\Sigma\, (o)}$)  are presented in Eqs.~(\ref{characteristicequation},
\ref{Vomixingmatrix}) of appendix~\ref{appendixVo} as functions of parameters from
Table~\ref{Tablepartree}.

 To the mass matrices up to one loop~(Eqs.(\ref{oneloopnotation},\ref{oneloopMa}
 ))
 contribute~(Eq.(
 \ref{M1tildeSSV}))
 \begin{eqnarray}
 \label{Mo1}
 {\cal M}^{\alpha\,\Sigma}_{(1)}&=& \tilde{{\cal M}}^{\alpha\,\Sigma}_{(1)\tilde{S}} +
 {\cal M}^{\alpha\,\Sigma}_{(1)S}  + {\cal M}^{\alpha\,\Sigma}_{(1) V} +
 {\cal M}^{\alpha\,\Sigma}_{(o)D}\,, \nonumber\\
 &=& V^{\alpha\, \dagger}_{\Sigma (o)}\,(\tilde{{\cal M}}^{\alpha\,\Sigma}_{(o1)\tilde{S}} +
 {\cal M}^{\alpha\,\Sigma}_{(o1)S}  + {\cal M}^{\alpha\,\Sigma}_{(o1) V} +
 {\cal M}^{\alpha\,\Sigma}_{(o)} )\,V^{\alpha}_{\Sigma (o)}\,,
 \end{eqnarray}
 where $\tilde{{\cal M}}^{\alpha\,\Sigma}_{(1) \tilde{S}}$  are contributions of the scalar gauge fields
 originating in $\tilde{\omega}_{abs}$, ${\cal M}^{\alpha\,\Sigma}_{(1) S}$ are contributions from
 $\omega_{sts'}$ and ${\cal M}^{\alpha\,\Sigma}_{(1) V}$ determine one loop corrections from the
 massive boson fields.   The detailed calculations are done
 in appendices~(\ref{appendixVo},\ref{M44massmatrix}).
 From Eqs.~(\ref{tildeM1Smatrixa}, \ref{tildeM1Sdetails}) 
  one reads details for $  \tilde{{\cal M}}^{\alpha\,\Sigma}_{(1)\tilde{S}}=$
  $\,V^{\alpha\, \dagger}_{\Sigma (o)}\,$ $\tilde{{\cal M}}^{\alpha\,\Sigma}_{(o\,1) \tilde{S}}$
  $\,V^{\alpha}_{\Sigma (o)}\,$. Details about
  $  \tilde{{\cal M}}^{\alpha\,\Sigma}_{(1)S}=$
  $\,V^{\alpha\, \dagger}_{\Sigma (o)}\,$ $\tilde{{\cal M}}^{\alpha\,\Sigma}_{(o\,1)S}$
   $\,V^{\alpha}_{\Sigma (o)}\,$ are written in Eq.~(\ref{oneloopSMa})
   and details about $  \tilde{{\cal M}}^{\alpha\,\Sigma}_{(1)V}=$
  $\,V^{\alpha\, \dagger}_{\Sigma (o)}\,$ $\tilde{{\cal M}}^{\alpha\,\Sigma}_{(o\,1)V}$
   $\,V^{\alpha}_{\Sigma (o)}\,$ are written in Eq.~(\ref{oneloopVMa}).

To obtain masses and diagonalizing matrices for each family member $\alpha$ of both groups
of families $\Sigma$ the diagonalization (Eq.~(\ref{M1Dmain}))  must be performed.

\section{General properties of the mass matrices}
\label{generalproperties}

All the expressions, needed for the evaluation of masses  and  diagonalizing matrices  of each
family member $\alpha=(u\,,d\,,\nu\,,e)$ for either the
upper ($\Sigma=II$) or the lower ($\Sigma=I$) four families, on the tree level or
with the loop corrections included,
are presented in appendices~\ref{appendixVo} and~\ref{M44massmatrix}.
The final mass matrix  including one loop corrections is the sum of the three matrices presented in
Eqs.~(\ref{tildeM1Smatrixa}),~\ref{oneloopSMa},~\ref{oneloopVMa} of appendix~\ref{M44massmatrix}.
We take the mass matrix elements on the
tree level as well as the masses of the scalar and gauge fields as free parameters, fitting them to the
existing experimental data.

All the free parameters which determine the mass matrices on the tree level can be read from
Eqs.~(\ref{tildea}, \ref{diagonaltreemain}) and in Table~\ref{Tablepartree}. Since the
contributions from the scalar fields $\tilde{\omega}_{abs}$ to the tree level
mass matrix are  the same for $u$-quark and $\nu$-lepton  and the same
for $d$-quark and $e$-lepton, while they are  different for each of these two pairs ( matrix elements of
$(u\,,\nu)$ differ from those of $(d\,,e)$),
there are four free parameters due to these contributions and the additional three parameters
which originate in the scalar $\omega_{abs}$ fields, all together  therefore
 seven free parameters for each of the two pairs on the tree level.

The loop corrections originate in massive fields, that is in dynamical scalar and vector boson fields.
The scalar fields $\; \tilde{g}^{\tilde{N}_{R}}\,
\hat{\vec{\tilde{N}}}_{R} \vec{\tilde{A}}^{\tilde{N}_{R}}_{\mp}\, $ and
$\tilde{g}^{2} \,  \hat{\vec{\tilde{\tau}}}^{2}\,\vec{\tilde{A}}^{2}_{\mp}\, $
 contribute to masses of the upper four families, while
 $\; \tilde{g}^{\tilde{N}_{L}}\, \hat{\vec{\tilde{N}}}_{L}
\vec{\tilde{A}}^{\tilde{N}_{L}}_{\mp}\, $  and
$\tilde{g}^{1} \,  \hat{\vec{\tilde{\tau}}}^{1}\,\vec{\tilde{A}}^{1}_{\mp}\, $
contribute to masses of the lower four families. On the other side, the scalar fields  contributions
$e\, \hat{Q}\, A_{\mp}\,,\,  g^{Q'}\, \,\hat{Q'}\, Z^{Q'}_{\mp}\,$ and
$g^{Y'} \, \hat{Y'}\, A^{Y'}_{\mp}$, and the gauge fields  contributions $g^{Y'}\, \hat{Y'}\, A^{Y'}_{m}$ and
$g^{'} \, \hat{Q'}\, A^{Q'}_{m}$   "see" only the family member index $\alpha$ and not the
 family  index $i$.
Their masses and coupling constants are presented in Table~\ref{MAall} of
appendix~\ref{M44massmatrix}.
We use their masses as free parameters as well.

Since there is no experimental data for the upper
four families, we can try to learn from the proposed procedure by taking into account
evaluations  of properties for the
fifth family quarks~\cite{gn} more  about the mass differences of the
family members of the upper four families.

\subsection{Properties of the lower two families for each of the
two groups of four families below the tree level}
\label{2x2}

We study the influence of one loop corrections on the mass matrices and mixing matrices
of quarks and leptons for  the $2 \times 2$ case,  for
$\tilde{a}^{i+}_{\pm}= \tilde{a}^{i-}_{\pm}=0$, for $i=(2,1)$, $i=2$ for $\Sigma= II$ and
$i=1$ for $\Sigma= I$. This assumption seems acceptable as a first step  for the lower group
of four families, while, since we have almost no knowledge about the upper four families (except
rough estimations evaluated when using the {\it spin-charge-family-theory} to explain dark matter
content of our universe and the direct measurements of the dark matter~\cite{gn}), it is
questionable for the upper group of four families.

Taking the results 
from appendix~\ref{appendixVo}, Eq.~(\ref{m2x2}), which is applicable for either the
upper  or the lower group of four families and for any family member,
one recognizes
that $m^{ u \,\Sigma}_{(o)2}- m^{u \,\Sigma}_{(o)1}=m^{ \nu\,\Sigma}_{(o)2}- m^{\nu \Sigma}_{(o)1}=$
$\sqrt{(\tilde{a}^{i3}_{-})^2 + (2 \,\tilde{a}^{i+}_{-})^2}, \,i=(2,1)\,,\,$ for $\Sigma=(II,I)$, respectively,
and  $m^{ d \,\Sigma}_{(o)2}- m^{d\,\Sigma}_{(o)1}=  m^{e\,\Sigma}_{(o)2}- m^{ e\,\Sigma}_{(o)1}$ $=
\sqrt{(\tilde{a}^{i3}_{+})^2
+ (2\,\tilde{a}^{i+}_{+})^2},\, i=(2,1)\,,$ for $\Sigma=(II,I)$, respectively.
This  is in complete disagreement
with the experimental data for  $u$-quarks and neutrinos
of the lowest two of the lower group of four families, and not so bad for  $d$-quarks and electrons of the
first two families, where it almost works, as it is well known.
%
We namely have~\cite{database} ($m^{ s }- m^{d}$) ($=
(m^{ d \,I}_{(o)2}- m^{d\,I}_{(o)1})$) $=[\,(101.0 \pm 25) - (4.1-5.8)\,]\,$ MeV and
($m^{ \mu }- m^{e}$) ($= m^{e\,I}_{(o)2}- m^{ e\,I}_{(o)1}$) $=[(105.65837) - (0,5109989)\,]\,$ MeV.
It is therefore on the loop corrections to correct the disagreements.

For the lowest two families there are three matrix elements on the tree level
(Eqs.~(\ref{tildea},\ref{diagonaltreemain})),
$\,a_1\,\, (= -\frac{1}{2} (\tilde{a}^{3}_{\pm} - \tilde{a}^{\tilde{N} 3}_{\pm}) +
 a_{\mp})\,$,
$\,a_2\,\, (= -\frac{1}{2} (\tilde{a}^{3}_{\pm} + \tilde{a}^{\tilde{N} 3}_{\pm}) +
 a_{\mp})$, 
 with  $\,, a_{\mp}= e\, Q\, A_{\mp} +  g^{1} \,\cos \theta_1\, Q'\, Z^{Q'}_{\mp}\, +
               g^{2}\, \cos \theta_2\, Y'\, A^{Y'}_{\mp}\,,$
and  $\,e \,\,(=\tilde{a}^{\tilde{N}^{+}}_{\pm}= \tilde{a}^{\tilde{N}^{-}}_{\pm}\,)$,
which we shall take as free parameters. (The definition of $a_i, i=1,2,$ is now slightly
changed by taking into account contributions of Eqs.~(\ref{tildea}) and (\ref{diagonaltreemain})).

\subsection{Properties of  the two groups of four families below the tree level}
\label{4times4}

We study the influence of one and two loop corrections  
on the mass matrices and correspondingly on masses and mixing matrices
of quarks and leptons for  the lower and the upper group of  four families~(\ref{Table tec}),
taking as  an input, that
is as free parameters, the parameters from Tables~\ref{Tablepartree},~\ref{MAall}.
The loop corrections due to two kinds of scalar fields and to massive gauge fields are presented
in Figs.~(\ref{Fig1}, \ref{Fig2}, \ref{Fig3}).

The  tree level mass matrices of
each group after the electroweak break is presented in Eq.~(\ref{characteristicequation}).
The one loop contributions originating in  the scalar fields ($\; \tilde{g}^{\tilde{N}_{R}}\,
\hat{\vec{\tilde{N}}}_{R}
\vec{\tilde{A}}^{\tilde{N}_{R}}_{\mp}\,, $
$\tilde{g}^{2} \,  \hat{\vec{\tilde{\tau}}}^{2}\,\vec{\tilde{A}}^{2}_{\mp}\, $) must be added
to the  tree level mass matrices  of the upper group
of four families only, while those originating in the scalar fields
($\; \tilde{g}^{\tilde{N}_{L}}\, \hat{\vec{\tilde{N}}}_{L}
\vec{\tilde{A}}^{\tilde{N}_{L}}_{\mp}\,, $
$\tilde{g}^{1} \,  \hat{\vec{\tilde{\tau}}}^{1}\,\vec{\tilde{A}}^{1}_{\mp}\, $)
contribute to mass matrices of the lower four families only. Both are presented in
appendix~\ref{M44massmatrix}
in Eq.~(\ref{tildeM1Smatrixa}).

The contributions of the scalar fields
 ($ e\, \hat{Q}\, A_{\mp}\,,\,  g^{1}\, \cos \theta_1 \,\hat{Q'}\, Z^{Q'}_{\mp}\,,$
$g^{2} \cos \theta_2\, \hat{Y'}\, A^{Y'}_{\mp}$) are presented in Eq.(\ref{oneloopSMa}),
these ones contribute to both, the upper four and the lower four families. The contributions to the
upper four families is much weaker than to the lower four, due to much larger tree level masses
 of the upper four families. The contributions depend on the family members quantum numbers
 $Q'\,$ and $Y'\,$
        and due to $\stackrel{78}{(\mp)}\, p_{0\mp} $ distinguish also among $(u,\nu)$ and $(d,e)$ pairs.

The massive gauge vector fields $g^{1} \cos \theta_1 \,\hat{Q'}\, Z^{Q'}_{m},\,
        g^{2} \cos \theta_2 \,\hat{Y'}\, A^{Y'}_{m}$ contributions differ for different
       members of a family as well.  Their influence on
        the upper four and the lower four family members
        depends again on the three level mass matrices. These contributions  are presented in
        Eq.(\ref{oneloopVMa}).

\section{Discussion and Conclusions}
\label{discussion}

We analysed in this paper the properties of twice four families as they follow from the
{\it spin-charge-family-theory}
when loop corrections, discussed in the ref.~\cite{NF}, are taken into account.
Having experimental results only for the lowest three of the lower four families, most
discussions in this paper concern the lower group of four families.

In the {\it spin-charge-family-theory}~\cite{NF} fermions carry two kinds of spin and correspondingly
interact with the two kinds of spin connection fields. One kind of spin determines at low
energies, after several breaks of the starting symmetry, the spin and the  charges
of fermions, the second kind determines families.

After several breaks from $SO(1,13)$ to $SO(1,7)\times U(1) \times SU(3)$ and further to
$SO(1,3)\times SU(2)_{I}\times SU(2)_{II} \times U(1)_{II} \times SU(3)$  there are
eight massless families~\footnote{The massless ness of the eight families is in this paper,
following the paper~\cite{NF}, is just assumed. In the ref.~\cite{DHN}, and in the references
presented there, it is proven for a toy model that after the break there can exist
massless families of fermions.}, which after the break to
$SO(1,3)\times SU(2)_{I} \times U(1)_{I} \times SU(3)$  manifest as a massive and a massless
group of four families. Correspondingly, after this break ($SU(2)_{II} \times U(1)_{II}$ to
$ U(1)_{I}$) also vector bosons involved in this break, become massive.
This break is (assumed to be) triggered by the superposition of the scalar fields
$\tilde{S}^{ab}\, \tilde{\omega}_{abs}$,
which are triplets with respect to the two $SU(2)$ (with the generators of the infinitesimal
transformations $\hat{\vec{\tilde{N}}}_{R}$ and $\hat{\vec{\tilde{\tau}}}^{2}$).

At the electroweak break (from $SO(1,3)\times SU(2)_{I} \times U(1)_{I} \times SU(3)$
to $SO(1,3)\times  U(1) \times SU(3)$) the lowest four families become massive too,
while staying decoupled from the upper four families. The vector bosons involved in this break,
become massive.
This break is (assumed to be) triggered by the superposition of the scalar fields
$ \tilde{\omega}_{abs}$,
which are triplets with respect to the two $SU(2)$ (with the generators of the infinitesimal
transformations $\hat{\vec{\tilde{N}}}_{L}$ and $\hat{\vec{\tilde{\tau}}}^{1}$) and the superposition of
the scalar fields $\omega_{s'ts}$, which are singlets with respect to the tree $U(1)$ ($A^{Y'}_s$,
$A^{Q'}_s$, $A^{Q}_s$).

In this contribution we  report the obtained analytical forms of the
upper and lower 4x4 mass matrices taking into account all contributions from dynamical scalars and
gauge bosons up to one loop corrections. At present we are carrying out  a detailed
numerical analysis trying to fit within this scenario the known quark and lepton masses
and mixing matrices, including the neutrino properties.

\appendix

\section{Majorana mass terms}
\label{majoranas}

There are mass terms  within the {\it spin-charge-family-theory}, which transform
the right handed neutrino to his  charged conjugated
one, contributing to the right handed neutrino Majorana masses~\cite{NF}
\begin{eqnarray}
\label{CnuRdaggernuR}
&&\psi^{\dagger}\, \gamma^0 \stackrel{78}{(-)} p_{0-}\, \psi\,,\nonumber\\
&&p_{0-}=  - (\tilde{\tau}^{1 +} \,\tilde{A}^{1 +}_{-}+ \tilde{\tau}^{1 -} \,\tilde{A}^{1 -}_{-})\;
 {\cal O}^{[+]} \, \cal{A}^{O}_{[+]},\nonumber\\
&&{\cal O}^{[+]}\, =   \stackrel{78}{[+]}\,
\stackrel{56}{(-)}\,\stackrel{9\,10}{(-)}\,
\stackrel{11 \,12}{(-)}\;\;\stackrel{13\,14}{(-)}.
\end{eqnarray}
One easily checks, using the technique with the Clifford objects (see ref.~\cite{NF})
that $\gamma^0 \stackrel{78}{(-)} p_{0-}$  transforms a {\it right handed neutrino} of one of the
{\it lower four families} into the charged conjugated one, belonging to the same group of families. It does not
contribute to masses of other leptons and quarks, right or left handed. Although the operator ${\cal O}^{[+]}$
appears in a quite complicated way, that is in the higher order corrections, yet it might be helpful in
explaining the properties of neutrinos. The operator
$- (\tilde{\tau}^{1 +} \,\tilde{A}^{1 +}_{-}+ \tilde{\tau}^{1 -} \,\tilde{A}^{1 -}_{-})\;
 {\cal O}^{[+]} \, \cal{A}^{O}_{[+]}$ gives zero, when being applied on the upper
four families, since they are singlets with respect to $\tilde{\tau}^{1 \pm}$.

\section{Diagonalization of $4\times 4$ tree level mass matrix}
\label{appendixVo}

We take  mass matrices on the tree level as they follow  from
the {\em spin-charge-family-theory}, Eq.~(\ref{factionI}). The part determined by the
$\tilde{\omega}_{abs}$ fields is presented  in tables~(\ref{Table VII.}, \ref{Table VIII.}),
for the upper four and the lower four families, respectively.

After assuming that  real and symmetric matrices are good approximation for both groups of families
(this is a good enough approximation for the lower four families, if we neglect the $CP$ nonconserving
terms, while for the upper four families we have no information yet about the  discrete
$CP$ nonconserving symmetry either from studying the {\em spin-charge-family-theory}
or from the experimental point of view) the  mass matrices
presented in  tables (\ref{Table VII.}, \ref{Table VIII.}) and in Eq.~(\ref{M0}) are $4 \times 4$
matrices
%
%
\begin{equation}
\label{M0a}{\cal M}_{(o)} = \begin{pmatrix} - a_1 & e
& 0 & b\\ e & - a_2 & b & 0\\ 0 & b & a_1 & e\\
b &  0 & e & a_2
\end{pmatrix},
\end{equation}
with $a_1\,, a_2\,,b\,$ and $e$ explained in Eq~(\ref{Mpm}) of sect.~\ref{beyond}.
%
%
These matrix elements are
different for the upper four families ($a_{1}= \frac{1}{2} (\tilde{a}^{2 3}_{\pm} - $ $
\tilde{a}^{\tilde{N}_{R}^{3}}_{\pm} + a^{\alpha}_{\pm})\,$, $a_{2}= \frac{1}{2} (\tilde{a}^{2 3}_{\pm} + $ $
\tilde{a}^{\tilde{N}_{R}^{3}}_{\pm} + a^{\alpha}_{\pm})\,$, $b=\tilde{a}^{2+}_{\pm}= \tilde{a}^{2 -}_{\pm} \,$,
$e=\tilde{a}^{\tilde{N}_{R}^{+}}_{\pm}= \tilde{a}^{\tilde{N}_{R}^{-}}_{\pm}$)  and
different for the lower four families
($a_{1}= \frac{1}{2} (\tilde{a}^{1 3}_{\pm} -$ $
\tilde{a}^{\tilde{N}_{L}^{3}}_{\pm} + a^{\alpha}_{\pm})\,$, $a_{2}= \frac{1}{2} (\tilde{a}^{1 3}_{\pm} + $ $
\tilde{a}^{\tilde{N}_{L}^{3}}_{\pm} + $ $a^{\alpha}_{\pm})\,$, $b=\tilde{a}^{1+}_{\pm}= \tilde{a}^{1 -}_{\pm} \,$,
$e=\tilde{a}^{\tilde{N}_{L}^{+}}_{\pm}= \tilde{a}^{\tilde{N}_{L}^{-}}_{\pm}$)
and also different for each of the family member ($\alpha \in \{u,\,d,\,\nu,\,e\}$), distinguishing in
 between the two pairs ($d$, $e$) ($+$) and  ($u$, $\nu$) ($-$) and in the term $ a^{\alpha}_{\pm})\,$,
 with~(\ref{diagonaltreemain}) %
$a_{\mp}= e\, Q\, A_{\mp} +  g^{1} \,\cos \theta_1\, Q'\, Z^{Q'}_{\mp}\, +
               g^{2}\, \cos \theta_2\, Y'\, A^{Y'}_{\mp}\,,$
%
where $Q, Q'$ and $Y'$ stay for the quantum numbers for the right handed members of one
(anyone) family ($\alpha \in \{u,\,d,\,\nu,\,e\}$).

We present in Table~\ref{Table tec} the representation of the right handed $u_R$-quark of a particular colour
and the right handed colourless $\nu$-lepton for all the eight families with the basic massless states expressed
with the Clifford algebra objects~\cite{hn0203}.
Table is taken from  ref.~\cite{NF}.
The quantum numbers, which each of these eight families carries, are presented in Table~\ref{Table FQN}.
The same quantum family numbers carry any member of a family ($\alpha\in \{u,d,\nu,e\}$),  the
left or right handed, colourless or of any colour.
%
 \begin{table}
 \begin{center}
 \begin{tabular}{|r||c||c||c||c||}
 \hline
 $I_{R\, 1}$ & $u_{R}^{c1}$&
 $ \stackrel{03}{[+i]}\,\stackrel{12}{(+)}|\stackrel{56}{(+)}\,\stackrel{78}{[+]}||
 \stackrel{9 \;10}{(+)}\:\; \stackrel{11\;12}{[-]}\;\;\stackrel{13\;14}{[-]}$ &
 $\nu_{R}$&
 $ \stackrel{03}{[+i]}\,\stackrel{12}{(+)}|\stackrel{56}{(+)}\,\stackrel{78}{[+]}||
 \stackrel{9 \;10}{(+)}\;\;\stackrel{11\;12}{(+)}\;\;\stackrel{13\;14}{(+)}$
 \\
 \hline
  $I_{R\,2}$ & $u_{R}^{c1}$&
  $ \stackrel{03}{[+i]}\,\stackrel{12}{(+)}|\stackrel{56}{[+]}\,\stackrel{78}{(+)}||
  \stackrel{9 \;10}{(+)}\;\;\stackrel{11\;12}{[-]}\;\;\stackrel{13\;14}{[-]}$ &
  $\nu_{R}$&
  $ \stackrel{03}{(+i)}\,\stackrel{12}{[+]}|\stackrel{56}{(+)}\,\stackrel{78}{[+]}||
  \stackrel{9 \;10}{(+)}\;\;\stackrel{11\;12}{(+)}\;\;\stackrel{13\;14}{(+)}$
 \\
 \hline
 $I_{R\,3}$ & $u_{R}^{c1}$&
 $ \stackrel{03}{(+i)}\,\stackrel{12}{[+]}|\stackrel{56}{(+)}\,\stackrel{78}{[+]}||
 \stackrel{9 \;10}{(+)}\;\;\stackrel{11\;12}{[-]}\;\;\stackrel{13\;14}{[-]}$ &
 $\nu_{R}$&
 $ \stackrel{03}{(+i)}\,\stackrel{12}{[+]}|\stackrel{56}{[+]}\,\stackrel{78}{(+)}||
 \stackrel{9 \;10}{(+)}\;\;\stackrel{11\;12}{(+)}\;\;\stackrel{13\;14}{(+)}$
 \\
 \hline
 $I_{R\,4}$ & $u_{R}^{c1}$&
 $ \stackrel{03}{(+i)}\,\stackrel{12}{[+]}|\stackrel{56}{[+]}\,\stackrel{78}{(+)}||
 \stackrel{9 \;10}{(+)}\;\;\stackrel{11\;12}{[-]}\;\;\stackrel{13\;14}{[-]}$ &
 $\nu_{R}$&
 $ \stackrel{03}{[+i]}\,\stackrel{12}{(+)}|\stackrel{56}{[+]}\,\stackrel{78}{(+)}||
 \stackrel{9 \;10}{(+)}\;\;\stackrel{11\;12}{(+)}\;\;\stackrel{13\;14}{(+)}$
 \\
 \hline\hline\hline
 $II_{R\,1}$ & $u_{R}^{c1}$&
 $ \stackrel{03}{(+i)}\,\stackrel{12}{(+)}|\stackrel{56}{(+)}\,\stackrel{78}{(+)} ||
 \stackrel{9 \;10}{(+)}\;\;\stackrel{11\;12}{[-]}\;\;\stackrel{13\;14}{[-]}$ &
 $\nu_{R}$&
 $ \stackrel{03}{(+i)}\,\stackrel{12}{(+)}|\stackrel{56}{(+)}\,\stackrel{78}{(+)} ||
 \stackrel{9 \;10}{(+)}\;\;\stackrel{11\;12}{(+)}\;\;\stackrel{13\;14}{(+)}$
 \\
 \hline
 $II_{R\,2}$ & $u_{R}^{c1}$&
 $ \stackrel{03}{(+i)}\,\stackrel{12}{(+)}|\stackrel{56}{[+]}\,\stackrel{78}{[+]}||
 \stackrel{9 \;10}{(+)}\;\;\stackrel{11\;12}{[-]}\;\;\stackrel{13\;14}{[-]}$ &
 $\nu_{R}$&
 $ \stackrel{03}{(+i)}\,\stackrel{12}{(+)}|\stackrel{56}{[+]}\,\stackrel{78}{[+]}||
 \stackrel{9 \;10}{(+)}\;\;\stackrel{11\;12}{(+)}\;\;\stackrel{13\;14}{(+)}$
 \\
 \hline
 $II_{R\,3}$ & $u_{R}^{c1}$&
 $\stackrel{03}{[+i]}\,\stackrel{12}{[+]}|\stackrel{56}{(+)}\,\stackrel{78}{(+)}||
 \stackrel{9 \;10}{(+)}\;\;\stackrel{11\;12}{[-]}\;\;\stackrel{13\;14}{[-]}$ &
 $\nu_{R}$&
 $\stackrel{03}{[+i]}\,\stackrel{12}{[+]}|\stackrel{56}{(+)}\,\stackrel{78}{(+)}||
 \stackrel{9 \;10}{(+)}\;\;\stackrel{11\;12}{(+)}\;\;\stackrel{13\;14}{(+)}$
 \\
 \hline
 $II_{R\,4}$ & $u_{R}^{c1}$&
 $ \stackrel{03}{[+i]}\,\stackrel{12}{[+]}|\stackrel{56}{[+]}\,\stackrel{78}{[+]}||
 \stackrel{9 \;10}{(+)}\;\;\stackrel{11\;12}{[-]}\;\;\stackrel{13\;14}{[-]}$ &
 $\nu_{R}$&
 $ \stackrel{03}{[+i]}\,\stackrel{12}{[+]}|\stackrel{56}{[+]}\,\stackrel{78}{[+]}||
 \stackrel{9 \;10}{(+)}\;\;\stackrel{11\;12}{(+)}\;\;\stackrel{13\;14}{(+)}$
 \\
 \hline
 \end{tabular}
 \end{center}
 \caption{\label{Table tec} Eight families of the right handed $u_R$ quark with the spin $\frac{1}{2}$,
  the colour charge $\tau^{33}=1/2$, $\tau^{38}=1/(2\sqrt{3})$ and of the colourless right handed
  neutrino $\nu_R$ of the spin $\frac{1}{2}$ are presented in the left and in the right column,
  respectively.
  $S^{ab}, a,b \in \{0,1,2,3,5,6,7,8\}$ transform $u_{R}^{c1}$ of the spin $\frac{1}{2}$ and the
  chosen colour $c1$ to all the members of the same colour: to the right handed $u_{R}^{c1}$
  of the spin $-\frac{1}{2}$,
  to the left $u_{L}^{c1}$ of both spins ($\pm \frac{1}{2}$), to the right handed $d_{R}^{c1}$ of both spins
  ($\pm \frac{1}{2}$) and to the left handed $ d_{L}^{c1}$ of both spins ($\pm \frac{1}{2}$). They transform
  equivalently the right handed   neutrino $\nu_R$ of the spin $\frac{1}{2}$. }
 \end{table}

While are the diagonal matrix elements, originating in $\omega_{abs}$ scalar vacuum expectation values,
expected to cause large (and desired) changes in mass matrices for
the lower four families, their contribution to the mass matrices of the
upper four families is expected to be very small, because of the difference in the
strength of the tree level contributions from $\tilde{\omega}_{abs}$ sectors  in both groups of
four families.

The matrix of Eq.(\ref{M0a}) can be diagonalized by the orthogonal
matrix $V_{(o)}$ (Eq.~(\ref{treediag}))
\begin{equation}
V_{(o)}^T {\cal M}_{(o)}\: V_{(o)} = {\cal M}_{(o)D} =
diag(m_{(o)1}, m_{(o)2}, m_{(o)3}, m_{(o)4}) \;.
\label{treeleveldiagonalmatrix} \end{equation}
The diagonal contributions $a_{\mp} $ to mass matrices (Eq.(\ref{diagonaltreemain})),
the same for all the eight
families, do not influence the diagonalization.

Four eigenvalues $m_{(o)i}\; (i = 1, 2, 3, 4)$ of the tree level mass matrices ${\cal M}_{(o)}$,
different for each of the two groups of four families, and also  different for different family members
 (due to the diagonal contribution of Eq.~(\ref{diagonaltreemain}) and to ($\mp$), are expressible
 in terms of $\Delta_o$ and$\Lambda_{o} $
\begin{eqnarray}
\label{characteristicequation}
\eta^4 - \Delta_{o} \eta^2  + \Lambda_{(o)} &=& 0\, , \qquad
\Delta_{(o)} = a_1^2 + a_2^2 + 2 b^2 + 2 e^ 2 \, , \quad \Lambda_{(o)} = (a_1 a_2 + b^2 - e^2)^2\,, \nonumber\\
m^{\alpha}_{(o)1}&=& - \eta_1 + a^{\alpha}\,,\quad
m^{\alpha}_{(o)2} =  - \eta_2 + a^{\alpha}\,,\quad
m^{\alpha}_{(o)3} =    \eta_1 + a^{\alpha}\,,\quad
m^{\alpha}_{(o)4} =    \eta_2 + a^{\alpha}\,, \nonumber\\
\eta_{1}    &=& \frac{1}{\sqrt{2}} \sqrt{\Delta_{(o)} - R_1 R_2} \,,\quad
\eta_{2}     =  \frac{1}{\sqrt{2}} \sqrt{\Delta_{(o)} + R_1 R_2} \,, \nonumber\\
R_1         &=& \sqrt{(a_1 + a_2)^2+4 b^2} \, , \quad
R_2          =  \sqrt{(a_2 - a_1)^2+4 e^2},\nonumber\\
R_1^2 R_2^2 &=& \Lambda_{(o)}^2 - 4 \Lambda_{(o)}\,,  \quad
 \eta_1^2 + \eta_2^2 = \Delta_{(o)}\,, \nonumber\\
 \eta_1^2 \,\eta_2^2&=&\Lambda_{(o)}\,, \quad  \eta_2^2 - \eta_1^2= R_1\:R_2\,.
\end{eqnarray}
Computing the eigenvectors, we obtain the orthogonal matrix $V_{(o)}$
\begin{equation}
V_{(o)} = \begin{pmatrix}  s_1 & - s_2 & -s_3 & s_4\\
s_2 & s_1 & s_4 & s_3\\ - s_3 & - s_4 & s_1 & s_2\\ - s_4 &  s_3 & - s_2 & s_1
\end{pmatrix} \,,  \quad  \quad s_1\:s_3 = s_2\:s_4\,.
\label{Vomixingmatrix}
\end{equation}
 $s_1 , s_2 , s_3 , s_4$ are mixing angles  defined
in terms of the parameters and eigenvalues as follows
\begin{eqnarray}
s_1=\frac{1}{2}\:\sqrt{ \frac{(\eta_2 + a_2)^2 - (a_1 + \eta_1)^2}{\eta_2^2 - \eta_1^2} } \quad , \quad
s_2=\frac{1}{2}\:\sqrt{ \frac{(\eta_2 + a_1)^2 - (a_2 + \eta_1)^2}{\eta_2^2 - \eta_1^2} } \nonumber \\
s_3=\frac{1}{2}\:\sqrt{ \frac{(\eta_2 - a_2)^2 - (a_1 - \eta_1)^2}{\eta_2^2 - \eta_1^2} } \quad , \quad
s_4=\frac{1}{2}\:\sqrt{ \frac{(\eta_2 - a_1)^2 - (a_2 - \eta_1)^2}{\eta_2^2 - \eta_1^2} }\,.
\label{treelevelmixingangles}
\end{eqnarray}
%
It is easy to check the orthogonality
of $V_{(o)}$, $V_{(o)}^T V_{(o)}=I$,  and  Eq.(\ref{treeleveldiagonalmatrix}).

The matrix $V_{(o)} $, 
which is different for the upper ($\Sigma=II$)
than for the lower four families ($\Sigma= I$)  and different for the pair ($u, \nu$) than the pair ($d,e$)
transforms the massless states $\psi$ ($\psi^{\alpha}_{\Sigma}$) into the massive basis~(Eq.(\ref{notationvonek}))
$\Psi^{(o)}$ ($\Psi^{\alpha \,(o)}_{\Sigma}$)
\begin{equation}
\label{PSIoa}
V^{\alpha }_{\Sigma (o)}\,\Psi^{\alpha\,(o)}_{\Sigma} =
\psi^{\alpha}_{\Sigma}\,, \;\;\Sigma=II,I,\; \alpha \in \{u,\,d,\,\nu,\,e\}.
\end{equation}
\subsection{Some useful relationships 
}
\label{usefulformulas}

\begin{eqnarray}
\label{usefulrel}
%
%
&& s_1^2\,s_2^2 + s_3^2\,s_4^2=\frac{e^2\,((a_2+a_1)^2+2 b^2)}{R_1^2\,R_2^2} \quad , \quad
 s_1^2\,s_4^2 + s_2^2\,s_3^2=\frac{b^2\,((a_2-a_1)^2+2 e^2)}{R_1^2\,R_2^2} \quad , \quad
 s_1^2\,s_3^2=\frac{b^2\,e^2}{R_1^2\,R_2^2}\,, \nonumber\\
&& s_1\,s_2(s_1^2 - s_2^2)+ s_3\,s_4(s_4^2 - s_3^2)=\frac{e(a_2 - a_1)\, ((a_2+a_1)^2+2 b^2)}{R_1^2\,R_2^2}\,,
\nonumber\\
&& s_1\,s_4 (s_1^2 - s_4^2)+ s_2\,s_3(s_2^2 - s_3^2)=\frac{b(a_2 + a_1)\, ((a_2-a_1)^2+2 e^2)}{R_1^2\,R_2^2}\,,
\nonumber\\
&& s_1\,s_2(s_4^2 - s_3^2)+ s_3\,s_4(s_1^2 - s_2^2)=2 s_1\,s_3 (s_1\,s_4 - s_2\,s_3)=
\frac{2 e b^2 (a_2 - a_1)}{R_1^2\,R_2^2}\,,\nonumber\\
&& s_1\,s_4(s_2^2 - s_3^2)+ s_2\,s_3(s_1^2 - s_4^2)=2 s_1\,s_3 (s_1\,s_2 - s_3\,s_4)=
\frac{2 b e^2 (a_2 + a_1)}{R_1^2\,R_2^2}\,,\nonumber\\
&&s_1\,s_3( s_1^2 + s_3^2 - s_2^2 - s_4^2) = (s_1\,s_2 - s_3\,s_4 )(s_1\,s_4 - s_2\,s_3)=
\frac{b e (a_2^2 - a_1^2)}{R_1^2\,R_2^2}\,.
\end{eqnarray}
%


\subsection{$2\times 2$ matrices in the limit $b=0$ within the $4\times 4$ ones}
\label{2x2II}

We study here the limit when the off diagonal matrix elements $b$ in Eq.~(\ref{M0},\ref{M0a})
are small in comparison with the other nonzero matrix elements. We put in what follows $b=0$.
The mass matrices of Eq.~(\ref{M0a}) then simplifies into two by diagonal $2\times 2$ matrices.
 In this limit it follows, after using the relation
$\left(\frac{1}{2}\:\left[a_1+a_2 \pm \sqrt{(a_2 - a_1)^2+4 e^2} \right]
\right)^2 = \frac{1}{2}\: \left[ a_1^2+a_2^2+ 2 e^2 \pm \sqrt{(a_1+a_2)^2}\:
\sqrt{(a_2 - a_1)^2+4 e^2} \right] \;$ in
Eq.~(\ref{characteristicequation}),
\begin{eqnarray}
\label{eta12}
\eta_{1 ,2} 
                & = & \frac{1}{2}\:\left[a_1+a_2 \mp \sqrt{(a_2 - a_1)^2+4 e^2} \right]\,.
\end{eqnarray}
Now $\eta_{i},  \, i=1,2\;$  obey   relations:
$\eta_{i}^2 - (a_1+a_2)\:\eta_{i} + a_1 a_2 - e^2=0 \,, $
$\eta_{2}+\eta_{1} = a_1 + a_2\,$, 
$\eta_{2}-\eta_{1} = \sqrt{(a_2 - a_1)^2 + 4 e^2} \,$,
$\eta_{1}\:\eta_{2} = a_1 a_2 - e^2\,$.
Correspondingly one finds:
$(\eta_{2} - a_2)^2-(a_1 - \eta_{1} )^2=\,$
$(\eta_{1}+\eta_{2} - a_1 - a_2)(\eta_{2} - \eta_{1} - a_2 + a_1)$
 and
$(\eta_{2} - a_1)^2-(a_2 - \eta_{1} )^2=\,$
$(\eta_{1}+\eta_{2} - a_1 - a_2)(\eta_{2} - \eta_{1} + a_2 - a_1)$.

From the above equations it follows  that $s_3=0= s_4$ and
\begin{equation}
\label{s1s2}
s_1 = \sqrt{\frac{1}{2} \left( 1 + \frac{1}{\sqrt{1+ (\frac{2 e}{a_2 - a_1})^2}} \right)} \; , \quad
s_2 = \sqrt{\frac{1}{2} \left( 1 - \frac{1}{\sqrt{1 + (\frac{2 e}{a_2 - a_1})^2}} \right)}\,.
\end{equation}
The masses of the first two families in each group of four families are then
\begin{eqnarray}
\label{m2x2}
m^{\alpha}_{(o)1}&=& - \eta_1 + a^{\alpha}\,,\quad
m^{\alpha}_{(o)2} =  - \eta_2 + a^{\alpha}\,,\quad
m^{\alpha}_{(o)3} =    \eta_1 + a^{\alpha}\,,\quad
m^{\alpha}_{(o)4} =    \eta_2 + a^{\alpha}\,,
\end{eqnarray}
in this case of the two by two diagonal matrices $\eta_{1,2} =
\frac{1}{2}\: (a_1+a_2 \mp \sqrt{(a_2 - a_1)^2+4 e^2})$.

%
\section{
Mass matrices with one loop gauge and scalar corrections included}
\label{M44massmatrix}

According to Eq.~(\ref{factionI}) to one loop corrections to the tree level mass matrices ${\cal M}_{(o)}$
the scalar fields of the two kinds  and the massive gauge fields contribute.
As discussed in sect.~\ref{beyond}  to the loop corrections contribute:

{\bf i.)} The scalar fields
 expressible with $\tilde{\omega}_{abs}$  contribute after the electroweak break to masses
 of both groups of four families. The scalar fields
($\; \tilde{g}^{\tilde{N}_{R}}\, \vec{\tilde{N}}_{R}
\vec{\tilde{A}}^{\tilde{N}_{R}}_{\mp}\,, $
$\tilde{g}^{2} \,  \vec{\tilde{\tau}}^{2}\,\vec{\tilde{A}}^{2}_{\mp}\, $)
 contribute to masses of the upper four families, while
 ($\; \tilde{g}^{\tilde{N}_{L}}\, \vec{\tilde{N}}_{L}
\vec{\tilde{A}}^{\tilde{N}_{L}}_{\mp}\,, $
$\tilde{g}^{1} \,  \vec{\tilde{\tau}}^{1}\,\vec{\tilde{A}}^{1}_{\mp}\, $)
contribute to masses of the lower four families.
Each group of these scalar fields appear at  a
different scale. The contributions in both groups of scalar fields distinguish among the pairs
$(u\,,\nu)$ and $(d\,,e)$ due to the term 
$\stackrel{78}{(\mp)}\, p_{0\mp} $ 
in Eq.(\ref{mfaction}), which contributes to $(u\,,\nu)$ for $(-)$ and to
$(d\,,e)$ for $(+)$.

{\bf ii.)} The scalar fields
expressible with $\omega_{abs}$  ($ e\, Q\, A_{\mp}\,,\,  g^{1}\, \cos \theta_1 \,Q'\, Z^{Q'}_{\mp}\,,$
$g^{2} \cos \theta_2\, Y'\, A^{Y'}_{\mp}$). These scalar fields, which gain masses
during the electroweak break, contribute to only the diagonal matrix elements,
distinguishing  among the family members $\alpha\in\{u,d,/nu, e\}$ through the
eigenvalues of the operators $Q,\, Q' $ and $Y'$ and through the term
$\stackrel{78}{(\mp)}\, p_{0\mp} $.
The effect of their loops
contributions depends strongly on the three level mass matrices.

{\bf iii.)} The massive gauge vector fields $g^{1} \cos \theta_1 \,Q'\, Z^{Q'}_{m},\,
        g^{2} \cos \theta_2 \,Y'\, A^{Y'}_{m}$ contributions differ for different
        members of a family according to the eigenvalues of the operators $Q'\,$ and $Y'\,$
        and due to $\stackrel{78}{(\mp)}\, p_{0\mp} $.  Their influence on
        the upper four and the lower four family members
        depends on the three level mass matrices.

The one loop contributions to the tree level mass matrices are illustrated in
figures~\ref{Fig1},\ref{Fig2},\ref{Fig3}  presented in section~\ref{beyond}.

We discuss these contributions separately for both kinds of scalar fields and for
gauge bosons.

According to Eqs.(\ref{notatiovecnmass},\ref{MvPsi}) the contributions taking
into account up to $(k)$ loops corrections
read
\begin{eqnarray}
\label{MvPsia}
&& \psi^{\dagger}_{L}\,\gamma^0 \,({\cal M}_{(o\, k)} + \cdots +
{\cal M}_{(o\,1)} + {\cal M}_{(o)})  \,
 \psi_{R}  = \psi^{\dagger}_{L}\,\gamma^0 \,{\cal M}_{( k)}   \,
 \psi_{R}  \nonumber\\
 &&\Psi^{(k)\,\dagger}_{L}\,\gamma^0 \,(V_{(o)}
 \,V_{(1)}\,\cdots V_{(k)})^{\dagger}\,
 ({\cal M}_{(o\, k)} + \cdots + {\cal M}_{(o\,1)} + {\cal M}_{(o)})
 \,\,V_{(o)}\,V_{(1)}
 \,\cdots V_{(k)}\,\Psi^{(k)}_{ R},
\end{eqnarray}
where $\psi_{(L,R)} $ are the massless states and $\Psi^{(k)}_{(L, R)}$ the massive ones
when $(k)$ loops corrections are taken into account
\begin{eqnarray}
\label{notatiovecnmassa}
\psi_{(L,R)} &=& V_{(o)}\,V_{(1)}\,
\cdots V_{\Sigma\,(k)} \,\Psi^{(k)}_{(L, R)}
\end{eqnarray}
and we skipped the indices $\Sigma$ and $\alpha$, assuming that they are present and will be determined when
numerical calculations will be performed.

Accordingly we have up to one loop corrections
\begin{eqnarray}
\label{oneloopa}
&&\Psi^{(1)\,\dagger}_{L}\,\gamma^0 \,(V_{(o)} \,V_{(1)})^{\dagger}\,
 ({\cal M}_{(o\,1)} + {\cal M}_{(o)})  \,\,V_{(o)} \,V_{(1)} \,\Psi^{(1)}_{ R}=
 \psi^{\dagger}_{L}\,\gamma^0 \,
  ({\cal M}_{(o\,1)} + {\cal M}_{(o)})  \,\psi_{ R} \,,\nonumber\\
&&{\cal M}_{(o\,1)}= \tilde{{\cal M}}_{(o\,1)_{\tilde{S}}}
+ {\cal M}_{(o\,1)_S} + {\cal M}_{(o\,1)_V}\,,
\end{eqnarray}
where  ${\cal M}_{(o\,1)}$ stays for the sum of the one loop contributions  of
the scalar fields originating in $\tilde{\omega}_{ab\pm}$
($\,\vec{\tilde{A}}^{\tilde{N}_{(L,R)}}_{\mp}\,, $
$\,\vec{\tilde{A}}^{(2,1)}_{\mp}\, $),
we write them as  $\tilde{{\cal M}}_{(o\,1)_{\tilde{S}}}$, of those
originating in $\omega_{st\pm}$ ($\,A_{\mp}, \, Z^{Q'}_{\mp}\,,$ $\, A^{Y'}_{\mp}$),
we write them as  ${\cal M}_{(o\,1)_S} $ and of those originating in the massive boson fields
($ Z^{Q'}_{m},\, A^{Y'}_{m}$), we write them as ${\cal M}_{(o\,1)_G}$.
All these contributions will be calculated in the next three subsections for
the tree level mass matrices ${\cal M}_{(o)}$ from Eq.(\ref{M0}).

The mass matrix up to one loop is~(Eq.(\ref{oneloopnotation}))
\begin{equation}
\label{oneloopMa}
{\cal M}_{(1)}=
V^{\dagger}_{(o)}\,( {\cal M}_{(o\,1)} +
{\cal M}_{(o)} )\,V_{(o)}\,,
\end{equation}
with ${\cal M}_{(o\,1)}\,$  $=(\tilde{{\cal M}}_{(o\,1)_{\tilde{S}}}  + {\cal M}_{(o\,1)_S} +
{\cal M}_{(o\,1)_V})\,$ to be calculated in the subsections of this appendix and with
$V_{(1)}$ which follows from~(Eq.(\ref{MvD}))
\begin{eqnarray}
\label{MvDa}
&&{\cal M}_{(1)D}= V^{\dagger}_{(1)}\,[\, V^{\dagger}_{(o)}\,
({\cal M}_{(o\,1)}+ {\cal M}_{(o)})  \,V_{(o)}\,]\,V_{(1)}
= diag(m_{(1)\,1},\cdots m_{(1)\,4})\,.
\end{eqnarray}
Let $m_{(o)\,i}\, i=1,2,3,4$ be the diagonal mass eigenvalues from Eqs.~(\ref{V0},\ref{treeleveldiagonalmatrix})
(each carrying the quantum number of the family member $\alpha$ and the group index $\Sigma$)
\begin{equation}
\label{V0Ma}
{\cal M}_{(o) D}= V^{\dagger}_{(o)}\, {\cal M}_{(o)}\,V_{(o)}
=diag(m_{(o)\,1}, m_{(o)\,2},
m_{(o)\,3}, m_{(o)\,4})\,,
\end{equation}
from appendix~\ref{appendixVo}.

Let $M_{A}$ stays for the masses of all fields $A^{A}_{a}\,$ ($\vec{\tilde{A}}^{2}_{\mp}\, ,$
$\,\vec{\tilde{A}}^{\tilde{N}_{R}}_{\mp}\,, $  $\,\vec{\tilde{A}}^{1}_{\mp}\,, $
 $\,\vec{\tilde{A}}^{\tilde{N}_{L}}_{\mp}\,, $
 $\,A^{Q}_{\mp}\,,$ $\,   Z^{Q'}_{\mp}\,,$ $\, A^{Y'}_{\mp}$,
  $Z^{Q'}_{m},\, A^{Y'}_{m}$) contributing to loop corrections
to the tree level masses as presented in Figs.~(\ref{Fig1}, \ref{Fig2}, \ref{Fig3}) and let $g^A$
and $\tau^{A}$ stay
for the corresponding coupling constants (as presented in Table~\ref{MAall}) and the eigenvalues of the
operators $\hat{\tau}^{A}$.
\small{
 \begin{table}
 \begin{center}
 \begin{tabular}{|r|| c| c| c |c || c| c |c |c ||c | c | c ||c | c|}
 \hline
 $A^{A}_{a}$          & $\tilde{A}^{\tilde{N}_{R}(1,2)}_{\mp}$&$\tilde{A}^{\tilde{N}_{R}3}_{\mp}$&
$\tilde{A}^{2(1,2)}_{\mp}$               & $\tilde{A}^{2 \,3}_{\mp}$&
$\tilde{A}^{\tilde{N}_{L}(1,2)}_{\mp}$&$\tilde{A}^{\tilde{N}_{L}3}_{\mp}$&
$\tilde{A}^{1(1,2)}_{\mp}$& $\tilde{A}^{1 \,3}_{\mp}$&
$A^{Y'}_{\mp}$&$A^{Q'}_{\mp}$ & $A^{Q}_{\mp}$&$A^{Y'}_{m}$&$Z^{Q'}_{m}$\\
\hline
$g^A$&$\tilde{g}^{\tilde{N}_{R}}$&$\tilde{g}^{\tilde{N}_{R}}$&$\tilde{g}^{2}$&$\tilde{g}^{2}$&
$\tilde{g}^{\tilde{N}_{L}}$&$\tilde{g}^{\tilde{N}_{L}}$&$\tilde{g}^{1}$&$\tilde{g}^{1}$&
$g^{Y'}$&$g^{Q'}$&$g^{Y}$&$g^{Y'}$&$g^{Q'}$\\
 \hline
  $M_{A}$             &$M_{\tilde{N}_{R}} $&$M_{\tilde{N}_{R}3}$
                      &$M_{\tilde{2}}$    &$M_{\tilde{23}}$
                      &$M_{\tilde{N}_{L}}$&$M_{\tilde{N}_{L}3}$
                      &$M_{\tilde{1}}$    &$M_{\tilde{13}}$
                      &$M_{Y'S}$& $M_{Q'S}$& $M_{QS}$& $M_{Y'}$&$M_{Q'}$\\
 \hline
 \end{tabular}
 \end{center}
 \caption{\label{MAall} Notation for masses of dynamical scalar and vector boson fields.
 We used in Table~\ref{MAall} the notation $M_{\tilde{N}_{R}}=\,$
$M_{\tilde{N}_{R}1}=M_{\tilde{N}_{R}2}\,$, $M_{\tilde{2}}=$ $M_{\tilde{21}}=M_{\tilde{22}}\,$,
$M_{\tilde{N}_{L}}=\,$ $M_{\tilde{N}_{L}1}=M_{\tilde{N}_{L}2}\,$ and
$M_{\tilde{1}}=\,$ $M_{\tilde{11}}=M_{\tilde{12}}$.}
 \end{table}
 }
 %
Then the one loop contributions of both kinds of the scalar fields can be read from
Figs.~(\ref{Fig1}, \ref{Fig2}) leading to
\begin{eqnarray}
\Sigma^{A}_{kS} &=& m_{(o)k}\:\frac{(g^{A} \vec{\tau}^{A})^2}{16 \,\pi^2}\;
\frac{(M_{A})^2}{(M_{A})^2-(m_{(o)k})^2}\,\;\;
\ln{\frac{(M_{A})^2}{(m_{(o)k})^2}} \, .
\label{SigmaGENscalar}
\end{eqnarray}
We  must keep in mind that $\tau^{A}$ ($\equiv (Y'\,,Q'\,,Q)$, applied on the
right handed massless states, or, if taking the hermitean conjugate value of
the mass term $ (\psi^{\dagger} \, \gamma^0 \, \stackrel{78}{(\pm)}\, p_{0\pm} \, \psi)^{\dagger}$
of Eq.~(\ref{mfaction}), on the left handed massless states, which brings the same result, with the
eigenvalues  presented in Table~\ref{Table MQN}) are the same for all
the families of both groups and so are $M_{A}$ ($\equiv (M_{Y'}\,,M_{Q'}\,,M_{Q})$) and that
accordingly  contributions $\Sigma^{(Y',Q',Q)}_k$ of the scalar fields $A^{Y'}_{\mp}\,,$
$A^{Q'}_{\mp}\,$ and $A^{Q}_{\mp}$  bring different contributions  for different families
only through $m^{\alpha\,\Sigma}_{(o)k}$. The contributions of ($\vec{\tilde{A}}^{2}_{\mp}\, ,$
$\,\vec{\tilde{A}}^{\tilde{N}_{R}}_{\mp}\,, $  $\,\vec{\tilde{A}}^{1}_{\mp}\,, $
 $\,\vec{\tilde{A}}^{\tilde{N}_{L}}_{\mp}\, $) are different for different group of families and different
 members of one group, while they distinguish among the family members only through dependence
 of the  fields on masses ($m^{\alpha\, \Sigma}_{(o)k}$) and on $(\mp)$.

To evaluate the contributions from the gauge fields as presented in Fig.~\ref{Fig3} we must evaluate
\begin{eqnarray}
\Sigma^{A}_{kV} &=& m_{(o)k}\:\frac{(g^{A})^2 \, \tau^{{A}_L} \;\tau^{{A}_R }}{4 \, \pi^2}\;
\frac{(M_{A})^2}{(M_{A})^2-(m_{(o)k})^2}\,\;\;
\ln{\frac{(M_{A})^2}{(m_{(o)k})^2}} \, ,
\label{SigmaGENgauge}
\end{eqnarray}
where $\tau^{A}_L \;,\tau^{A}_R $ are the eigenvalues of the operators $Y'$ and $Q'$ applied on the
left ($\tau^{A}_L $) and on the right ($\tau^{A}_R $) handed member of a family $\alpha$ ($=\in \{u \,,
d \,, \nu \,, e\}$) in the massless basis, $M_{A}$ stays for $M_{Y'}$ and $M_{Q'}$.
In Table~\ref{CouplingalphaLR}  we present the values
$(g^{A})^2 \, \tau^{{A}_L} \;\tau^{{A}_R }\;$ for the members of a family $\alpha$=($u\,,d\,,\nu\,,e$).
Let us add that
\begin{eqnarray}
\label{relatedcouplings}
(g^{Y'})^2 \, Y^{'\alpha}_{L} \;Y^{'\alpha}_{R}&=& - g^{4} \sin \theta_{2} \; \tau^{4\,\alpha} \,
g^{2} \cos \theta_{2} \,Y^{'\alpha}_{R}\, = -(g^{Y'})^2 \, \tau^{4\,\alpha}\,Y^{'\alpha}_{R}\, ,\nonumber\\
(g^{Q'})^2 \, Q^{'\alpha}_{L} \;Q^{'\alpha}_{R} &=&- g^{1} \cos \theta_{1}\; Q^{'\alpha}_{L}\,
g^{Y} \sin \theta_{1} \,Y^{\alpha}_{R}\,= - e^2\,Q^{'\alpha}_{L} \;Q^{'\alpha}_{R}\,,
\end{eqnarray}
where it is $\tau^{4\,\alpha}=$ $\tau^{4\,\alpha}_{L}=$ $\tau^{4\,\alpha}_{R}$.
\small{
 \begin{table}
 \begin{center}
 \begin{tabular}{|c|| c| c| c |c |}
 \hline
 $\alpha$          & $u$&$d$&$\nu$&$e$\\
\hline
$\frac{(g^{Y'})^2 \,Y^{'\alpha}_{L}\,Y^{'\alpha}_{R}  }{4\,\pi^2}$&
$-\frac{\alpha}{12\:\pi}\frac{1}{\cos^2\theta_1}\left( 1-\frac{1}{3}\tan^2\theta_2  \right)$&
$ \frac{\alpha}{12\:\pi}\frac{1}{\cos^2\theta_1}\left( 1+\frac{1}{3}\tan^2\theta_2  \right)$&
$ \frac{\alpha}{4\:\pi} \frac{1}{\cos^2\theta_1}\left( 1+\tan^2\theta_2  \right) $&
$-\frac{\alpha}{4\:\pi} \frac{1}{\cos^2\theta_1}\left( 1-\tan^2\theta_2  \right) $\\
 \hline
$\frac{(g^{Q'})^2 \,Q^{'\alpha}_{L}\,Q^{'\alpha}_{R}  }{4\,\pi^2}$&
 $ -\frac{\alpha}{3\:\pi}\left( 1-\frac{1}{3}\tan^2\theta_1  \right)$&
 $ -\frac{\alpha}{6\:\pi}\left( 1+\frac{1}{3}\tan^2\theta_1  \right)$&
 $ 0$&
 $ -\frac{\alpha}{2\:\pi}\left( 1-\tan^2\theta_1 \right)$\\
 \hline
 \end{tabular}
 \end{center}
 \caption{\label{CouplingalphaLR} The couplings $\frac{(g^{Y'})^2 \,Y^{'\alpha}_{L}\,Y^{'\alpha}_{R}  }{4\,\pi^2}$
 are presented, evaluated for the members of a (any) family $\alpha = (u\,,d\,,\nu\,,e)$.
 $Y^{'\alpha}_{(L,R)}$ and $Q^{'\alpha}_{(L,R)}$ are the eigenvalues of the operators
 $\hat{Y'}$ and $\hat{Q'}$ applied on the
left ($Y^{'\alpha}_{L}\,,$ $Q^{'\alpha}_{L}$)  and on the right ($Y^{'\alpha}_{R}\,,$ $Q^{'\alpha}_{R}$)
 handed member of a family $\alpha$ ($=\in \{u \,,
d \,, \nu \,, e\}$) in the massless basis.}
 \end{table}
 }

\large{
We evaluate in the next subsections  the one loop corrections for all the three kinds of fields.
}
The corresponding mass matrices including one loop corrections  (Eq.~(\ref{oneloopMa}))
${\cal M}^{\alpha\,\Sigma}_{(1)}$ are the sum of  contributions of two kinds of  massive scalar
dynamical fields ${\cal M}^{\alpha\,\Sigma}_{(1)\,\tilde{S}}$ (Eq.~(\ref{tildeM1Smatrixa}) and
 ${\cal M}^{\alpha\,\Sigma}_{(1)\,S}$ (Eq.~(\ref{oneloopSMa})), of  massive vector boson fields
${\cal M}^{\alpha\,\Sigma}_{(1)\,V}$ (Eq.~(\ref{oneloopVMa})) and of the tree level mass matrices
$V^{\alpha \dagger}_{\Sigma\,(o)}\, {\cal M}^{\alpha\,\Sigma}_{(o)} \,V^{\alpha}_{\Sigma(o)}\,$
\begin{eqnarray}
\label{M1tildeSSV}
V^{\alpha \dagger}_{\Sigma\,(o)}\,( {\cal M}^{\alpha\,\Sigma}_{(o\,1)\,\tilde{S}} +
{\cal M}^{\alpha \,\Sigma}_{(o\,1)\,S} + {\cal M}^{\alpha\,\Sigma}_{(o\,1)\,V} +
{\cal M}^{\alpha\,\Sigma}_{(o)})
\,V^{\alpha }_{\Sigma\,(o)}\,.
\end{eqnarray}
We obtain the masses and the diagonalizing matrices, within one loop corrections,  from Eq.~(\ref{MvDa}).

\subsection{ Scalar fields -- $\vec{\tilde{A}}^{\tilde{N}_{(R,L)}}_{\mp}\,, $
$\,\vec{\tilde{A}}^{(2,1)}_{\mp}\,$ -- contributions to one loop corrections
to the mass matrices
}
\label{onelooptildeSsect}

We shall first study the one loop corrections to the tree level mass matrices from the scalar fields
originating in $\tilde{\omega}_{abs}$ ($\vec{\tilde{A}}^{\tilde{N}_{(R,L)}}_{\mp}\,, $
$\,\vec{\tilde{A}}^{(2,1)}_{\mp}\,$), which distinguish among all the families.
They also distinguish among the family members $\alpha$ through the dependence on $(\mp)$ and
through the tree masses $m^{\alpha\,\Sigma}_{(o)k}$. The tree level diagonalizing matrices
$V^{\alpha}_{\Sigma (o)}$ also depend on ($\mp$), that is they are different for the pair $(u\,,\nu)$
than for the pair $(d,e)$.
The corresponding diagrams are presented in Figs.~(\ref{Fig1},
\ref{Fig2}).

The operators  $\vec{\tilde{N}}_{R}\, \vec{\tilde{A}}^{\tilde{N}_R}$  and
$\vec{\tilde{\tau}}^{2}\, \vec{\tilde{A}}^{2}$  transform
the members of the upper four families while $\vec{\tilde{N}}_{L}\, \vec{\tilde{A}}^{\tilde{N}_L}$
and $\vec{\tilde{\tau}}^{1}\, \vec{\tilde{A}}^{1}$  transform
the members of the lower four families, both kinds of transformations are presented
in Eq.~(\ref{diagramNtaua})
\begin{equation}
\label{diagramNtaua}
\stackrel{ \stackrel{\tilde{N}^{i}_{L}}{\leftrightarrow} }{\begin{pmatrix} I_{4} & I_{3} \\
 I_{1} & I_{2} \end{pmatrix}} \updownarrow  \tilde{\tau}^{1i}   \quad ,
\quad \stackrel{ \stackrel{\tilde{N}^{i}_{R}}{\leftrightarrow} }{\begin{pmatrix} II_{4} & II_{3} \\
II_{1} & II_{2} \end{pmatrix}} \updownarrow  \tilde{\tau}^{2i}\quad.
\end{equation}
Let us repeat that the upper four families are doublets with respect to $\vec{\tilde{N}}_R$ and
doublets with respect to $\vec{\tilde{\tau}}^{2}$ and that they are singlets
with respect to $\vec{\tilde{N}}_L$ and
singlets with respect to $\vec{\tilde{\tau}}^{1}$, while the lower four families
are  doublets with respect to $\vec{\tilde{N}}_L$ and
doublets with respect to $\vec{\tilde{\tau}}^{1}$ and that they are singlets
with respect to $\vec{\tilde{N}}_R$ and
singlets with respect to $\vec{\tilde{\tau}}^{2}$. Accordingly the mass matrices $8\times 8$ stay to be
two by diagonal matrices $4\times 4$ also after the loops corrections.

Let us, to treat both groups of families formally all at once, accept the
notation.\\
i. Let the scalar fields $\vec{\tilde{A}}^{\tilde{N}_{(R,L)}}_{\mp}\, $ be denoted by
$\vec{\tilde{A}}^{\tilde{N}}_{\mp}\,, $ and
$\,\vec{\tilde{A}}^{(2,1)}_{\mp}\, $ by $\,\vec{\tilde{A}}_{\mp}\, $. \\
ii. Let masses of these
dynamical scalar fields be different for different components of $\tilde{A}^{\tilde{N}i}_{\mp}\,, $
so that $\tilde{A}^{\tilde{N}1}_{\mp}\, $ and $\tilde{A}^{\tilde{N}2}_{\mp}\, $ have equal masses,
and $\tilde{A}^{\tilde{N}3}_{\mp}\, $ a different one.
Equivalent assumption is made for the massless
of the components of $\,\vec{\tilde{A}}_{\mp}\, $.

Let   ($M_{\tilde{\tau} }\,, M_{\tilde{\tau} 3 }\,$)  represents the masses of  the dynamical
scalar fields $\vec{\tilde{A}}^{2}_{\mp}\, ,$
($M_{\tilde{2\,1}}= M_{\tilde{2\,2}}$ and $M_{\tilde{2\,3}}$ from Table~\ref{MAall})
when treating the upper four families,  as well as the masses of the scalar fields
$\vec{\tilde{A}}^{1}_{\mp}\, ,$ ($M_{\tilde{1\,1}}= M_{\tilde{1\,2}}$ and $M_{\tilde{1\,3}}$ from
Table~\ref{MAall}) when treating the lower four families.
%
Let ($M_{\tilde{N}}\,, M_{\tilde{N}3}$) represents the masses of  the scalar fields
$\vec{\tilde{A}}^{\tilde{N}_{(R\,,L)}}_{\mp}\, $
 ($M_{\tilde{N}_{(R\,,L)}1}=M_{\tilde{N}_{(R\,,L)}2}$ and $M_{\tilde{N}_{(R\,,L)}3}$ from
 Table~\ref{MAall}).
We shall distinguish between the
two groups of families when pointing out the differences and when
looking for the numerical evaluations. The masses of the two kinds of
the scalar fields  differ for many orders of magnitude.\\
iii. Correspondingly let $m_{(o)i}\,,i\in\{1,2,3,4\},$ be  the tree level masses
of  either the upper ($m^{\alpha \Sigma=II}_{(o)i}$)
or the lower four families ($m^{\alpha \Sigma=I}_{(o)i}$) for each of the family member
$\alpha=(u\,,d\,,\nu\,,e)$
and $\tilde{g}$ ($\tilde{g}^{(2,1)}$) determines
the couplings to the fields
$\vec{\tilde{A}}_{\mp}\, $ and $\tilde{g}^{\tilde{N}}$  ($\tilde{g}^{\tilde{N}_{(R,L)}}$)
the couplings to the fields $\vec{\tilde{A}}^{\tilde{N}}_{\mp}\, $ as presented in Table~\ref{MAall}.

From the diagrams in Figs.~(\ref{Fig1},\ref{Fig2}) one loop contributions of the
fields $\vec{\tilde{A}}_{\pm}$ and $\vec{\tilde{A}}^{\tilde{N}}_{\pm}\, $ follow
\begin{eqnarray}
\tilde{\Sigma}_{k\tilde{S}}^3\quad &=& m_{(o)k}\:\frac{(\tilde{g})^2}{4}\;\;\frac{1}{16\:\pi^2}\:
\frac{(M_{\tilde{\tau}3})^2}{(M_{\tilde{\tau}3})^2-(m_{(o)k})^2}\,\;\;
\ln{\frac{(M_{\tilde{\tau}3})^2}{(m_{(o)k})^2}} \, , \nonumber\\
\tilde{\Sigma}_{k\tilde{S}}^{\pm}\quad &=&         m_{(o)k}\:\:\frac{(\tilde{g})^2}{2}\;\;\frac{1}{16\:\pi^2}\:
\frac{(M_{\tilde{\tau}})^2}{(M_{\tilde{\tau}})^2-(m_{(o)k})^2}\,\;\;
\ln{\frac{(M_{\tilde{\tau}})^2}{(m_{(o)k})^2}} \,, \nonumber\\
\tilde{\Sigma}_{k\tilde{S}}^{\tilde{N}3} &=& m_{(o)k}\:\frac{(\tilde{g}^{\tilde{N}})^2}{4}\:\frac{1}{16\:\pi^2}\:
\frac{(M_{\tilde{N}3})^2}{(M_{\tilde{N}3})^2-(m_{(o)k})^2}\,
\ln{\frac{(M_{\tilde{N}3})^2}{(m_{(o)k})^2}} \, ,  \nonumber\\
\tilde{\Sigma}_{k\tilde{S}}^{\tilde{N}\pm} &=& m_{(o)k} \:\:\frac{(\tilde{g}^{\tilde{N}})^2}{2}\:\frac{1}{16\:\pi^2}\:
\frac{(\tilde{g}^{\tilde{N}})^2}{(M_{\tilde{N}})^2-(m_{(o)k})^2}\,
\ln{\frac{(M_{\tilde{N}})^2}{(m_{(o)k})^2}} \,,
\label{tildemokS}
\end{eqnarray}
with $k \in \{1,2,3,4\}\,$ and where also the eigenvalues of the operators $\vec{\tilde{\tau}}^{A}$
are already taken into account. 
From here the contributions to the  $\tilde{{\cal M}}_{(o\,1)_{\tilde{S}}}$ mass matrix
follows when by taking into account Eq.~(\ref{diagramNtaua})  the transformations
\begin{equation}
\label{tildesigmaij}
\tilde{\Sigma}_{\tilde{S}\,i j}^{(3,\tilde{N}3)}= \sum_{k=1}^{4}\, V_{(o)\,i k } \,V_{(o)\,j k}\,
\tilde{\Sigma}^{(3,\tilde{N}3)}_{k\tilde{S}}  \qquad , \quad
\tilde{\Sigma}_{\tilde{S}\, i i}^{(\pm,\tilde{N}\pm)}=\sum_{k=1}^{4}\, V_{(o)\,i k }\, V_{(o)\,i k}\,
\tilde{\Sigma}_{k\tilde{S}}^{(\pm,\tilde{N}\pm)}\,
\end{equation}
are performed and $\tilde{\Sigma}_{i j}^{(3,\tilde{N}3)}$ built into the mass matrix
$\tilde{{\cal M}}_{(o1)_{\tilde{S}}}$ of Eq.~(\ref{onelooptildeS}).
 \begin{eqnarray}
 &&\tilde{{\cal M}}_{(o1)_{\tilde{S}}} =\nonumber\\
 &&\left( \begin{smallmatrix} \tilde{\Sigma}_{11}^3 + \tilde{\Sigma}_{11}^{\tilde{N}3}+ \tilde{\Sigma}_{44}^{\pm} +
\tilde{\Sigma}_{22}^{\tilde{N}\pm} &  - \tilde{\Sigma}_{12}^{\tilde{N}3} & 0 & - \tilde{\Sigma}_{14}^{3} \\
- \tilde{\Sigma}_{12}^{\tilde{N}3} &  \tilde{\Sigma}_{22}^3 +
\tilde{\Sigma}_{22}^{\tilde{N}3}+ \tilde{\Sigma}_{33}^{\pm} + \tilde{\Sigma}_{11}^{\tilde{N}\pm} &
- \tilde{\Sigma}_{23}^{3} & 0 \\
0 &  - \tilde{\Sigma}_{23}^{3} & \tilde{\Sigma}_{33}^3 +
\tilde{\Sigma}_{33}^{\tilde{N}3}+ \tilde{\Sigma}_{22}^{\pm} + \tilde{\Sigma}_{44}^{\tilde{N}\pm} &
- \tilde{\Sigma}_{34}^{\tilde{N}3} \\
- \tilde{\Sigma}_{14}^{3} &  0 &  - \tilde{\Sigma}_{34}^{\tilde{N}3} &
\tilde{\Sigma}_{44}^3 + \tilde{\Sigma}_{44}^{\tilde{N}3}+ \tilde{\Sigma}_{11}^{\pm}
+ \tilde{\Sigma}_{33}^{\tilde{N}\pm}
\end{smallmatrix}  \right)\:.
\label{onelooptildeS}
\end{eqnarray}

The matrix $\tilde{{\cal M}}_{(o1)_{\tilde{S}}}$ (and correspondingly all the matrix elements)
should carry the  indices $\alpha$ and $\Sigma$, since $m_{(o)k}$  and $V_{(o)}$
carry the indices $\alpha$ and $\Sigma$ while
$M_{\tilde{\tau}}$, $M_{\tilde{\tau}3}$, $M_{\tilde{\tau}}$ and $M_{\tilde{\tau}3}$ carry the index
$\Sigma$. Correspondingly the matrix of Eq.~(\ref{onelooptildeS})
applies to any family member
of  either the upper or the lower  group of four families.

To obtain the mass matrix up to one loop ${\cal M}_{(1)}$ (Eqs.(\ref{oneloopnotation}, \ref{oneloopMa}))
one needs to find $
V^{\dagger}_{(o)}\,( {\cal M}_{(o\,1)} +
{\cal M}_{(o)} )\,V_{(o)}\,$ $=
V^{\dagger}_{(o)}\,\tilde{{\cal M}}_{(o\,1)_{\tilde{S}}}\,V_{(o)}\,  +
V^{\dagger}_{(o)}\,{\cal M}_{(o\,1)_S}\,V_{(o)}\, +
V^{\dagger}_{(o)}\,{\cal M}_{(o\,1)_V}\,V_{(o)}\,$.

Let us therefore calculate here $\tilde{{\cal M}}_{(1) \tilde{S}} $
($\equiv \tilde{{\cal M}}^{\alpha\,\Sigma}_{(1) \tilde{S}}$)
$= V^{\dagger}_{(o)}\,\tilde{{\cal M}}_{(o\,1)_{\tilde{S}}}\,V_{(o)}\,$.
Introducing
\begin{eqnarray}
\label{tildeM1S}
\tilde{\Sigma}_1&=&\tilde{\Sigma}_1^3 + \tilde{\Sigma}_1^{\tilde{N}3}+
\tilde{\Sigma}_4^{\pm} + \tilde{\Sigma}_2^{\tilde{N}\pm} \qquad , \qquad
\tilde{\Sigma}_2=\tilde{\Sigma}_2^3 + \tilde{\Sigma}_2^{\tilde{N}3}+
\tilde{\Sigma}_3^{\pm} + \tilde{\Sigma}_1^{\tilde{N}\pm}\quad,  \nonumber\\
\tilde{\Sigma}_3&=&\tilde{\Sigma}_3^3 + \tilde{\Sigma}_3^{\tilde{N}3}+
\tilde{\Sigma}_2^{\pm} + \tilde{\Sigma}_4^{\tilde{N}\pm} \qquad , \qquad
\tilde{\Sigma}_4=\tilde{\Sigma}_4^3 + \tilde{\Sigma}_4^{\tilde{N}3}+
\tilde{\Sigma}_1^{\pm} + \tilde{\Sigma}_3^{\tilde{N}\pm} \quad, \nonumber\\
&&\qquad \tilde{\Sigma}_{(1-2)}=\tilde{\Sigma}_1^{\tilde{N}3} -\tilde{\Sigma}_2^{\tilde{N}3}
 \quad , \quad
 \tilde{ \Sigma}_{(3-4)}=\tilde{\Sigma}_3^{\tilde{N}3} -\tilde{\Sigma}_4^{\tilde{N}3},\quad,
\nonumber\\
&& \qquad \tilde{ \Sigma}_{(1-4)}=\tilde{\Sigma}_1^{\tilde{N}3} -\tilde{\Sigma}_4^{\tilde{N}3}
\quad , \quad
\tilde{\Sigma}_{(2-3)}=\tilde{\Sigma}_2^{\tilde{N}3} -\tilde{\Sigma}_3^{\tilde{N}3}\quad,
\end{eqnarray}
we obtain for the mass matrix elements $(\tilde{{\cal M}}_{(1)\tilde{S}})_{ij} $
when taking into account
the matrix $V_{(o)}$ from Eq.(\ref{Vomixingmatrix}) and Eq.~(\ref{tildeM1S})
\begin{eqnarray}
\label{tildeM1Sdetailso}
&&(\tilde{{\cal M}}_{(1)\tilde{S}})_{11}= \tilde{\Sigma}_1 + 2(s_1^2 s_2^2 + s_3^2 s_4^2)
(\tilde{\Sigma}_2 - \tilde{\Sigma}_1 - \tilde{\Sigma}_{(1-2)}) +
2(s_1^2 s_4^2 + s_2^2 s_3^2)(\tilde{\Sigma}_4 - \tilde{\Sigma}_1 - \tilde{\Sigma}_{(1-4)}) \nonumber\\
&&+4 s_1^2 s_3^2 (\tilde{\Sigma}_3 - \tilde{\Sigma}_1 + \tilde{\Sigma}_{(3-4)} -
\tilde{\Sigma}_{(2-3)})\,,\nonumber\\
&&(\tilde{{\cal M}}_{(1)\tilde{S}})_{22}= \tilde{\Sigma}_2 + 2(s_1^2 s_2^2 + s_3^2 s_4^2)
(\tilde{\Sigma}_1 - \tilde{\Sigma}_2 + \tilde{\Sigma}_{(1-2)}) +
2(s_1^2 s_4^2 + s_2^2 s_3^2)(\tilde{\Sigma}_3 - \tilde{\Sigma}_2 - \tilde{\Sigma}_{(2-3)})\nonumber\\
&&+ 4 s_1^2 s_3^2 (\tilde{\Sigma}_4 - \tilde{\Sigma}_2 - \tilde{\Sigma}_{(3-4)} -
\tilde{\Sigma}_{(1-4)})\,,\nonumber\\
&&(\tilde{{\cal M}}_{(1)\tilde{S}})_{33}= \tilde{\Sigma}_3 + 2(s_1^2 s_2^2 + s_3^2 s_4^2)
(\tilde{\Sigma}_4 - \tilde{\Sigma}_3 - \tilde{\Sigma}_{(3-4)}) +
2(s_1^2 s_4^2 + s_2^2 s_3^2)(\tilde{\Sigma}_2 - \tilde{\Sigma}_3 + \tilde{\Sigma}_{(2-3)}) \nonumber\\
&&+ 4 s_1^2 s_3^2 (\tilde{\Sigma}_1 - \tilde{\Sigma}_3 + \tilde{\Sigma}_{(1-2)} +
\tilde{\Sigma}_{(1-4)})\,,\nonumber\\
&&(\tilde{{\cal M}}_{(1)\tilde{S}})_{44}- \tilde{\Sigma}_4 + 2(s_1^2 s_2^2 + s_3^2 s_4^2)
(\tilde{\Sigma}_3 - \tilde{\Sigma}_4 + \tilde{\Sigma}_{(3-4)}) +
2(s_1^2 s_4^2 + s_2^2 s_3^2)(\tilde{\Sigma}_1 - \tilde{\Sigma}_4 + \tilde{\Sigma}_{(1-4)}) \nonumber\\
&&+ 4 s_1^2 s_3^2 (\tilde{\Sigma}_2 - \tilde{\Sigma}_4 + \tilde{\Sigma}_{(2-3)} -
\tilde{\Sigma}_{(1-2)})\,,\nonumber
\end{eqnarray}
\begin{eqnarray}
\label{tildeM1Sdetailsoo}
&&(\tilde{{\cal M}}_{(1)\tilde{S}})_{12}= [s_1 s_2(s_1^2 - s_2^2) - s_3 s_4(s_3^2 - s_4^2) ] \,
(\tilde{\Sigma}_2 - \tilde{\Sigma}_1 - \tilde{\Sigma}_{(1-2)}) \nonumber\\
&&+[s_1 s_2(s_3^2 - s_4^2) - s_3 s_4(s_1^2 - s_2^2) ] (\tilde{\Sigma}_4 - \tilde{\Sigma}_3 - \tilde{\Sigma}_{(3-4)})
+ 2 s_1 s_3 (s_1 s_4 - s_2 s_3)(\tilde{\Sigma}_{(1-4)} - \tilde{\Sigma}_{(2-3)})\,,\nonumber\\
&&(\tilde{{\cal M}}_{(1)\tilde{S}})_{13}= (\tilde{{\cal M}}_{(1)\tilde{S}})_{24} =
 s_1 s_3 ( s_1^2 +  s_3^2 - s_2^2 -  s_4^2 ) (\tilde{\Sigma}_2 +
\tilde{\Sigma}_4 - \tilde{\Sigma}_1 - \tilde{\Sigma}_3) \nonumber\\
&&+(s_1 s_2 - s_3 s_4) ( s_1 s_4 - s_2 s_3 ) ( \tilde{\Sigma}_{(2-3)} - \tilde{\Sigma}_{(1-4)} -
\tilde{\Sigma}_{(1-2)} - \tilde{\Sigma}_{(3-4)} )\,, \nonumber\\
&&(\tilde{{\cal M}}_{(1)\tilde{S}})_{14}= [s_1 s_4(s_1^2 - s_4^2) + s_2 s_3(s_2^2 - s_3^2) ]
(\tilde{\Sigma}_1 - \tilde{\Sigma}_4 + \tilde{\Sigma}_{(1-4)}) \nonumber\\
&&+[s_1 s_4(s_2^2 - s_3^2) + s_2 s_3(s_1^2 - s_4^2) ]
(\tilde{\Sigma}_2 - \tilde{\Sigma}_3 + \tilde{\Sigma}_{(2-3)})
- 2 s_1 s_3 (s_1 s_2 - s_3 s_4)(\tilde{\Sigma}_{(1-2)} + \tilde{\Sigma}_{(3-4)})\,,\nonumber\\
&&(\tilde{{\cal M}}_{(1)\tilde{S}})_{23}= [s_1 s_4(s_1^2 - s_4^2) + s_2 s_3(s_2^2 - s_3^2) ]
(\tilde{\Sigma}_2 - \tilde{\Sigma}_3 + \tilde{\Sigma}_{(2-3)}) \nonumber\\
&&+[s_1 s_4(s_2^2 - s_3^2) + s_2 s_3(s_1^2 - s_4^2) ]
(\tilde{\Sigma}_1 - \tilde{\Sigma}_4 + \tilde{\Sigma}_{(1-4)}) +
+ 2 s_1 s_3 (s_1 s_2 - s_3 s_4) (\tilde{\Sigma}_{(1-2)} + \tilde{\Sigma}_{(3-4)})\,,\nonumber
\end{eqnarray}
\begin{eqnarray}
\label{tildeM1Sdetails}
&&(\tilde{{\cal M}}_{(1)\tilde{S}})_{34}= [s_1 s_2(s_1^2 - s_2^2) + s_3 s_4(s_4^2 - s_3^2) ]
(\tilde{\Sigma}_3 - \tilde{\Sigma}_4 + \tilde{\Sigma}_{(3-4)}) \nonumber\\
&&+[s_1 s_2(s_4^2 - s_3^2) + s_3 s_4(s_1^2 - s_2^2) ]
(\tilde{\Sigma}_2 - \tilde{\Sigma}_1 - \tilde{\Sigma}_{(1-2)}) +
- 2 s_1 s_3 (s_1 s_4 - s_2 s_3)(\tilde{\Sigma}_{(1-4)} + \tilde{\Sigma}_{(2-3)})\,,\nonumber\\
&&(\tilde{{\cal M}}_{(1)\tilde{S}})_{21}=(\tilde{{\cal M}}_{(1)\tilde{S}})_{12}\;,\;
(\tilde{{\cal M}}_{(1)\tilde{S}})_{31}=(\tilde{{\cal M}}_{(1)\tilde{S}})_{13}\;,\;
(\tilde{{\cal M}}_{(1)\tilde{S}})_{41}=(\tilde{{\cal M}}_{(1)\tilde{S}})_{14}\;,\nonumber\\
&&(\tilde{{\cal M}}_{(1)\tilde{S}})_{32}=(\tilde{{\cal M}}_{(1)\tilde{S}})_{23}\;,\;
(\tilde{{\cal M}}_{(1)\tilde{S}})_{43}=(\tilde{{\cal M}}_{(1)\tilde{S}})_{34}\;.
\end{eqnarray}
All the matrix elements $(\tilde{{\cal M}}_{(1)\tilde{S}})_{ij}$ carry
the indices $\alpha$, which distinguishes
among family members, and $\Sigma$, which distinguishes between the two groups of four families.
The matrix $\tilde{{\cal M}}^{\alpha \Sigma}_{(1)\tilde{S}}$ is accordingly
\begin{eqnarray}
\label{tildeM1Smatrixa}
\tilde{{\cal M}}^{\alpha \Sigma}_{(1)\tilde{S}}&=&
{V^{\alpha \dagger}_{\Sigma(o)}}\: \tilde{{\cal M}}^{\alpha \,\Sigma}_{(o1)\tilde{S}}
\:V^{\alpha}_{\Sigma(o)} \nonumber\\
&=&\begin{pmatrix} (\tilde{{\cal M}}_{(1)\tilde{S}})_{11}  & (\tilde{{\cal M}}_{(1)\tilde{S}})_{12}&
(\tilde{{\cal M}}_{(1)\tilde{S}})_{13} & (\tilde{{\cal M}}_{(1)\tilde{S}})_{14}\\
 (\tilde{{\cal M}}_{(1)\tilde{S}})_{12} & (\tilde{{\cal M}}_{(1)\tilde{S}})_{22} &
 (\tilde{{\cal M}}_{(1)\tilde{S}})_{23} &(\tilde{{\cal M}}_{(1)\tilde{S}})_{13}   \\
 (\tilde{{\cal M}}_{(1)\tilde{S}})_{13} & (\tilde{{\cal M}}_{(1)\tilde{S}})_{23} &
 (\tilde{{\cal M}}_{(1)\tilde{S}})_{33} & (\tilde{{\cal M}}_{(1)\tilde{S}})_{34}  \\
 (\tilde{{\cal M}}_{(1)\tilde{S}})_{14} & (\tilde{{\cal M}}_{(1)\tilde{S}})_{13} &
 (\tilde{{\cal M}}_{(1)\tilde{S}})_{34} & (\tilde{{\cal M}}_{(1)\tilde{S}})_{44}
 \end{pmatrix}^{\alpha \Sigma}_{\tilde{S}}\,.
 \end{eqnarray}
The matrix $\tilde{{\cal M}}_{(1)\tilde{S}}$  carry the indices $\alpha$ (distinguishing
among family members) and $\Sigma$ (distinguishing between the two groups of four families),
which are added to the matrix.

\subsubsection{Contributions from scalar fields $\vec{\tilde{A}}^{\tilde{A}}_{\pm}$
which couple two families}
\label{tildefieldsin2x2}

We explain in details  the contribution to loop corrections
from the scalar fields $\vec{\tilde{A}}^{\tilde{A}}_{\pm}$,
representing $\vec{\tilde{A}}^{2}_{\pm}$  and $\vec{\tilde{A}}^{\tilde{N}_{R}}_{\pm}$
in the case of the upper four families
and $\vec{\tilde{A}}^{1}_{\pm}$  and $\vec{\tilde{A}}^{\tilde{N}_{L}}_{\pm}$ for the lower four families.
We work in the massless basis. Let  these fields  act between
the families  $(i,j)$ accordingly to Eq.(\ref{diagramNtaua}). Let these two families are the two
states of  the fundamental
representation of the associated $SU(2)$  flavour symmetry (with the corresponding
infinitesimal generators of the group, which are either $\vec{\tilde{\tau}}^{1}$ or
$\vec{\tilde{N}}_{L}$ for the lower four families or $\vec{\tilde{\tau}}^{2}$ or
$\vec{\tilde{N}}_{R}$ for the upper four families). The fields $\vec{\tilde{A}}^{\tilde{A}}_{\pm}$
couple to  the families $(i,j)$ (that is to the massless states
$\psi^{\alpha}_{\Sigma(L,R)i}$, we here omit the indices $\alpha $ and
$\Sigma$) as follows
\begin{multline}
\frac{\tilde{g}^{\tilde{A}}}{2}\:\left[
\left( \overline{\psi}_{j\,(L,R)}\:\psi_{i\,(R,L)} +
\overline{\psi}_{i\,(L,R)}\:\psi_{j\,(R,L)}
\right)\:\tilde{A}^{\tilde{A}1}_{\pm} \right.
\\ + \left( i\:
\overline{\psi}_{j\,(L,R)}\:\psi_{i\,(R,L)} - i\:
\overline{\psi}_{i\,(L,R)}\:\psi_{j\,(R,L)}
\right)\:\tilde{A}^{\tilde{A}2}_{\pm}           \\
+ \left. \left( \overline{\psi}_{i\,(L,R)}\:\psi_{i\,(R,L)} -
\overline{\psi}_{j\,(L,R)}\:\psi_{j\,(R,L)}
\right)\:\tilde{A}^{\tilde{A}3}_{\pm} \right]\, .
\label{ijscalarcoupligs}
\end{multline}
%
For particular values of  the indices $\alpha \in (u\,, d\,, \nu \,, e)$ and $\Sigma \in (II, I)$, the
pair of the families $(i,j)$  is associated to the subset of tree level
mass parameters from ${\cal M}_{(o)}$ $\equiv {\cal M}^{\alpha\, \Sigma}_{(o)}$,
Eq.(\ref{M0a}). In Table~\ref{i=4,j=1,families}
these tree level matrix  elements are presented for the case 
$(i=4,j=1)$.
%
%
\begin{table} \begin{center}
 \begin{tabular}{  c | c c  }
   &  $\psi_{4R}$  &   $\psi_{1R}$ \\
\hline  $\bar{\psi}_{4L}$  & $a_2$  & $b $   \\
 $\bar{\psi}_{1L}$  & $b$  & $- a_1$
 \end{tabular} \end{center}
 \caption{\label{i=4,j=1,families}  $2\times 2$ tree level
 parameters for $i=4, j=1$ family indices  }
 \end{table}
Using  the scalar couplings of Eq.~(\ref{ijscalarcoupligs}) and the involved tree level
mass parameters we can draw the one loop diagrams of Figs.~\ref{Fig1} and~\ref{Fig2}. From
these diagrams the one loop contributions of the fields $\vec{\tilde{A}}^{\tilde{A}}_{\pm}$
follow
\begin{eqnarray} &
\bar{\psi}_{\Sigma\,i L}^\alpha \left( \tilde{\Sigma}_{\tilde{S}\,i i}^{(3,\tilde{N}3)}
+ \tilde{\Sigma}_{\tilde{S}\, j j}^{(\pm,\tilde{N}\pm)}  \right) \psi_{\Sigma\,i R}^\alpha -
\bar{\psi}_{\Sigma\,i L}^\alpha \, \tilde{\Sigma}_{\tilde{S}\,i j}^{(3,\tilde{N}3)} \,
\psi_{\Sigma \,j R}^\alpha \nonumber\\
+ & \bar{\psi}_{\Sigma\,j L}^\alpha \left( \tilde{\Sigma}_{\tilde{S}\,j j}^{(3,\tilde{N}3)}
+ \, \tilde{\Sigma}_{\tilde{S}\, i i}^{(\pm,\tilde{N}\pm)}  \right) \psi_{\Sigma \,j R}^\alpha -
\bar{\psi}_{\Sigma\,j L}^\alpha \,
\tilde{\Sigma}_{\tilde{S}\,i j}^{(3,\tilde{N}3)} \,  \psi_{\Sigma \,i R}^\alpha \; , \end{eqnarray}
with
%
$\tilde{\Sigma}_{\tilde{S}\,i j}^{(3,\tilde{N}3)}\,$ and
%
%
$\tilde{\Sigma}_{\tilde{S}\, i i}^{(\pm,\tilde{N} \pm)}$
%
%
defined in Eqs.~(\ref{tildesigmaij}, \ref{tildemokS}).


%
\subsection{ Scalar fields -- $\vec{A}^{Y'}_{\mp}\,,$
$\vec{A}^{Q'}_{\mp}\,,$ $\vec{A}^{Q}_{\mp}\,$ -- contributions to one loop corrections
to the mass matrices
}
\label{oneloopS}

The one loop corrections  of the scalar fields originating in $\omega_{sts'}$ --
$\vec{A}^{Y'}_{\mp}\,,$ $\vec{A}^{Q'}_{\mp}\,,$ $\vec{A}^{Q}_{\mp}\,$--
($ e\, Q\, A^{Q}_{\mp},\,  g^{1}\, \cos \theta_1 \,Q'\, Z^{Q'}_{\mp}\,,$
$g^{2} \cos \theta_2\, Y'\, A^{Y'}_{\mp}$) are presented in Fig.~\ref{Fig2}.
Their contributions to the mass matrix ${\cal M}_{(o1)_{S}}$ depend on a
family member  $\alpha$ through different values for each of the two pairs
$(u\,,\nu)$ and $(d\,,e)$ ($\mp$), through  the dependence of the tree level
masses ($m_{(o)i}$) on $\alpha$, and also through the eigenvalues of the operators
($\hat{Y'}\,,\hat{Q'}\,,\hat{Q}$) on different family members $\alpha$ $=(u\,,d\,,\nu\,,e)$ as
already explained and also presented in Table~\ref{Table MQN}.
Their contributions depend also on the  group ($\Sigma=(II\,,I)$)
and family indices ($i=(1\,,2\,,3\,,4)$) through ($m^{\alpha\,\Sigma}_{(o)i}$).

Let  here $Q^{\alpha},Q^{'\alpha},Y^{'\alpha}$ stay  for the eigenvalues of the corresponding
operators on the states $\psi^{\alpha}$ (the states are indeed $\psi^{\alpha}_{\Sigma \,i}$,
carrying  also the family and the group indices as presented in Eq.~(\ref{notationvecmassless})
and skipped here). And
let $M_{Q\,S}\,,M_{Q'\,S}$ and  $M_{Y'\,S}\,$ represent the  masses of the scalar dynamical fields
$A^{Q}_{\mp}\,, Z^{Q'}_{\mp}\,$ and $ A^{Y'}_{\mp}\, $, respectively.

We have  equivalent expressions to those of Eq.~(\ref{tildemokS})
\begin{eqnarray}
\Sigma_{kS}^{Y'\alpha}\quad &=& m_{(o)k}\:\frac{(g^{Y'} \,Y^{'\alpha})^2 }{16\pi^2}\;\;
\frac{(M_{Y'S})^2}{(M_{Y'S})^2-(m_{(o)k})^2}\,\;\;
\ln{\frac{(M_{Y'S})^2}{(m_{(o)k})^2}} \, , \nonumber\\
\Sigma_{kS}^{Q' \alpha}\quad &=& m_{(o)k}\:\frac{(g^{Q'} \,Q^{'\alpha})^2 }{16\pi^2}\;\:
\frac{(M_{Q'\,S})^2}{(M_{Q'\,S})^2-(m_{(o)k})^2}\,\;\;
\ln{\frac{(M_{Q'\,S})^2}{(m_{(o)k})^2}} \, , \nonumber\\
\Sigma_{kS}^{Q \alpha}\quad &=& m_{(o)k}\:\frac{(g^{Q\,} \,\,Q^{\alpha})^2 }{16\pi^2}\;\;
\frac{(M_{Q\,S})^2}{(M_{Q\,S})^2-(m_{(o)k})^2}\,\;\;
\ln{\frac{(M_{Q\,S})^2}{(m_{(o)k})^2}} \, ,
\label{mokS}
\end{eqnarray}
where, as already explained,
$m_{(o)k}$  ($m^{\alpha \, \Sigma}_{(o)k}$) are the masses, depending on the
member of a family $\alpha$ and on the group of four families ($\Sigma$), evaluated on the tree level.

Let us evaluate  $\Sigma_{i j}^{\alpha \Sigma (Y^{'\alpha},Q^{'\alpha},Q^{\alpha})}$,
similarly as in Eq.~(\ref{tildesigmaij}),
pointing out that they 
 depend on  $\Sigma=(II,I)$ through masses $(m^{\alpha\,\Sigma}_{(o)})$ and through
 $V_{(o)i k }$ ($\equiv V^{\alpha}_{\Sigma (o)\,i k } $)
\begin{equation}
\label{sigmaijo}
\Sigma_{S\,i j}^{\alpha\,\Sigma (Y^{'\alpha},Q^{'\alpha},Q^{\alpha})}=
\sum_{k=1}^{4}\, V^{\alpha}_{\Sigma (o)\,i k }
\,V^{\alpha}_{\Sigma(o)\,j k}\,
\Sigma^{\alpha\,\Sigma (Y^{'\alpha},Q^{'\alpha},Q^{\alpha})}_{kS}  \,.
\end{equation}
we  end up with the matrix ${\cal M}_{(o1)_{S}}$, which carry the indices $\alpha $ and $\Sigma$
(${\cal M}^{\alpha}_{\Sigma (o1)_{S}}$). Due to Eq.~(\ref{oneloopMa}) we need
to calculate to obtain the mass matrices up to the one loop corrections included
${\cal M}_{(1)}$ ($
V^{\dagger}_{(o)}\,( {\cal M}_{(o\,1)} +
{\cal M}_{(o)} )\,V_{(o)}\,$ $=
V^{\dagger}_{(o)}\,\tilde{{\cal M}}_{(o\,1)_{\tilde{S}}}\,V_{(o)}\,  +
V^{\dagger}_{(o)}\,{\cal M}_{(o\,1)_S}\,V_{(o)}\, +
V^{\dagger}_{(o)}\,{\cal M}_{(o\,1)_V}\,V_{(o)}\,$.

Let us calculate here therefore $V^{\alpha \dagger}_{\Sigma\,(o)}\,
\tilde{{\cal M}}^{\alpha \, \Sigma}_{(o\,1)_{S}}\,V^{\alpha}_{\Sigma(o)}\, $,
which distinguish among members of a family ($\alpha$) and between the two groups of families($\Sigma$),
using Eqs.~(\ref{oneloopMa}, \ref{mokS}, \ref{sigmaijo})
\begin{eqnarray}
\label{V1S}
\left(V^{\alpha\,\dagger}_{\Sigma(o)}\, {\cal M}^{\alpha\, \Sigma}_{(o\,1)_{S}}
\,V^{\alpha}_{\Sigma\,(o)}\,\right)_{ij}&=& \sum_{(l,k,r)=1}^{4}\,
(V^{\alpha}_{\Sigma\,(o)li} \,V^{\alpha}_{\Sigma(o)lk})\,(V^{\alpha}_{\Sigma(o)rk}\,
V^{\alpha}_{\Sigma(o)rj} )\,
(\Sigma_{kS}^{Y'\alpha}+
 \Sigma_{kS}^{Q'\alpha}+
 \Sigma_{kS}^{Q \alpha})^{\alpha \, \Sigma} \nonumber\\&=&
\delta_{ik}\,\delta_{jk}\,(
 \Sigma_{kS}^{Y'\alpha}+
 \Sigma_{kS}^{Q'\alpha}+
 \Sigma_{kS}^{Q \alpha})^{\alpha \, \Sigma}\,.
\end{eqnarray}
We have for ${\cal M}^{\alpha\, \Sigma}_{(1)_S}  $
 \begin{eqnarray}
 &&{\cal M}^{\alpha\,\Sigma}_{(o1)_{S}}
 =\left(V^{\alpha\,\dagger}_{\Sigma(o)}\, {\cal M}^{\alpha\, \Sigma}_{(o\,1)_{S}}
\,V^{\alpha}_{\Sigma\,(o)}\,\right)\\
 &&\left( \begin{matrix}
             \Sigma_{1S}^{Y'\alpha}+ \Sigma_{1S}^{Q'\alpha}+ \Sigma_{1S}^{Q \alpha} & 0 & 0 & 0 \\
 0 &         \Sigma_{2S}^{Y'\alpha}+ \Sigma_{2S}^{Q'\alpha}+ \Sigma_{2S}^{Q \alpha} & 0 & 0     \\
 0 & 0 &     \Sigma_{3S}^{Y'\alpha}+ \Sigma_{3S}^{Q'\alpha}+ \Sigma_{3S}^{Q \alpha} & 0         \\
 0 & 0 & 0 & \Sigma_{4S}^{Y'\alpha}+ \Sigma_{4S}^{Q'\alpha}+ \Sigma_{4S}^{Q \alpha}             \\
\end{matrix}  \right)^{\alpha \Sigma}_{S}\:\nonumber.
\label{oneloopSMa}
\end{eqnarray}
%

%
\subsection{Gauge bosons -- $\,A_m^{Y^\prime}$, $\,Z^{Q'}_{m}$ -- contribution to  one
loop corrections to the mass matrices $4\times 4$}
\label{gaugebosons}

We study the one loop contributions to the tree level mass matrices from the gauge fields
 $A_m^{Y^\prime}$ and $Z^{Q'}_{m}$.
According to  ref.~\cite{NF} $A_m^{Y^\prime}$ gains a mass after
the phase transition from $SU(2)_{I}\times SU(2)_{II}\times U(1)_{II}$ into
$SU(2)_{I}\times U(1)_{I}$ (and becomes a superposition of $\vec{A}^{2}_{m}$ and $A^{4}_{m}$ fields),
while $Z^{Q'}_{m}$ gains a mass after the electroweak break
(from $SU(2)_{I}\times U(1)_{I}$ into $U(1)$) (and becomes a superposition of
$\vec{A}^{1}_{m}$ and $A^{Y}_{m}$ fields). The one loop corrections of both vector fields to the
tree level mass matrices are presented in Fig.~\ref{Fig3}.
After the electroweak break  Eq.~(\ref{factionI}) determines the covariant moments
of all the eight families. These two  massive vector fields influence mass matrices
of the upper and the lower four families.

According to  ref.~\cite{NF}  before the phase transitions $\psi_R$
transform under $SU(2)_{II} \times
U(1)_{II}$, that is with respect to $\hat{\vec{\tau}}^{2}$ and $\hat{\tau}^{4}$, as
$(2, \tau^4)$, $\tau^4=\frac{1}{6}$ ($-\frac{1}{2}$)
for quarks (leptons),  while $\psi_L$ transform  under $SU(2)_{II} \times
U(1)_{II}$ as $(1,\tau^4 )$. From the kinetic term of Eq.(\ref{factionI}) the gauge couplings to
 $A_m^{Y^\prime}$ is
\begin{equation}
\label{AYpcouplings}
 \left[g^{Y'} \,\hat{Y}^{'}\,( \,\psi_{L} +  \psi_{R})^{\alpha}_{\Sigma} \right] A_m^{Y^\prime} \;,
\end{equation}
where massless states $\psi^{\alpha}_{\Sigma (L,R)} $ carry indices $\Sigma$ (distinguishing the upper, $=II$,
and the lower, $=I$, four
families), $\alpha$ ($=(u\,, d\,, \nu\,, e)$, which distinguishes a family members) and
the  family index ($i=1,2,3,4$) of each group. We further have
$g^{Y'}\, \hat{Y}^{'}\, \,\psi^{\alpha}_{L}= - g_4 \:\sin \theta_2 \: \hat{\tau}^4 \,\psi^{\alpha}_{\Sigma L}$,
$g^{Y'}\, \hat{Y}^{'}\, \,\psi^{\alpha}_{\Sigma R}=   g_2 \:\cos \theta_2 \: \hat{Y}' \,\psi^{\alpha}_{\Sigma R}$
(see Table~\ref{Table MQN}).  $\hat{\vec{\tau}}^{2}$, $\hat{\vec{\tau}}^1 $ and $\hat{\tau}^{4}$
distinguish only among family members.

 According to  ref.~\cite{NF} massless states
 $\psi_R$ transform as $(1, Y)$ under $SU(2)_{I} \times U(1)_{Y}$, that is with respect
 to $\vec{\hat{\tau}}^{1}$ and $\hat{Y}$, while $\psi_L$ transforms as
$(2,Y )$ under $SU(2)_{I} \times
U(1)_{Y}$. Accordingly, Eq.(\ref{factionI}) dictates  the following couplings of $ Z_m^{Q^\prime}$ to
fermions in a massless basis~\footnote{Before the electroweak  break the lower four families are massless
and the massive gauge field $A^{Y'}_{m}$ contribute to masses of only the upper (massive) four families.
After the electroweak break the lower four families become massive as well. Correspondingly
both massive gauge fields, $A^{Y'}_{m}$ and $A^{Q'}_{m}\equiv $ $Z^{Q'}_{m}$, contribute to masses of the
upper and the lower four families.}
\begin{equation}
\label{ZQpcouplings}
  \left[g^{Q'} \,\hat{Q}^{'}\,( \,\psi^{\alpha}_{\Sigma L} +  \psi^{\alpha}_{\Sigma R}) \right] Z_m^{Q^\prime} \;,
\end{equation}
where  massless basis $\psi_{(L,R)} $ carry indices $\Sigma$ , $\alpha$
and the  family index ($i=1,2,3,4$) of each group, and
$g^{Q'} \hat{Q}^{'}\, \,\psi^{\alpha}_{\Sigma L}=  g_1 \:\cos\theta_1 \:\hat{Q}^{\prime}\,
\psi^{\alpha}_{\Sigma L}$ and
$g^{Q'} \hat{Q}^{'}\, \,\psi^{\alpha}_{\Sigma R}= -g_Y \:\sin\theta_1 \:\hat{Q}' \,\psi^{\alpha}_{\Sigma R}$
(see Table~\ref{Table MQN}).

The internal fermion lines in the diagram  of Fig.~\ref{Fig3}   represent the massive
basis $\Psi^{(o)}$ (carrying the index $\Sigma$, $\alpha$ and $i$, $\Psi^{\alpha (o)}_{\Sigma\, i}$) and the masses
$m^{\alpha \Sigma}_{(o)i}$ are diagonal values,
eigenvalues, of the $4\times 4$ matrix, belonging to the family member $\alpha$ and the four family group
$\Sigma$, ${\cal M}^{\alpha\,\Sigma}_{(o)}$.

Let  $Y^{'\alpha}_{}(L,R), \,Q^{'\alpha}_{(L,R)}$ stay  for the eigenvalues of the corresponding
operators on the states $\psi^{\alpha}_{\Sigma\,(L,R)\,i}$ (
Eq.~(\ref{notationvecmassless})) and
let $M_{Y'}\,$ and  $M_{Q'}\,$ represent the  masses of the vector bosons
$A^{Y'}_{m}\,$ and $ Z^{Q'}_{m}\,$ ($ \equiv A^{Q'}_{m}\, $), respectively.
From the diagram in Fig.~\ref{Fig3}  then follow 
expressions, equivalent to those from Eqs.~(\ref{tildemokS}, \ref{mokS})
\begin{eqnarray}
\Sigma_{kV}^{Y'\alpha}\quad &=& m_{(o)k}\:\frac{(g^{Y'})^2 \,Y^{'\alpha}_{L}\,Y^{'\alpha}_{R}  }{4\,\pi^2}
\;\:
\frac{(M_{Y'})^2}{(M_{Y'})^2-(m_{(o)k})^2}\,\;\;
\ln{\frac{(M_{Y'})^2}{(m_{(o)k})^2}} \, , \nonumber\\
\Sigma_{kV}^{Q' \alpha}\quad &=&
m_{(o)k}\:\frac{(g^{Q'})^2 \,Q^{'\alpha}_L \, Q^{'\alpha}_R \,}{4\,\pi^2}\;\;
\frac{(M_{Q'})^2}{(M_{Q'})^2-(m_{(o)k})^2}\,\;\;
\ln{\frac{(M_{Q'})^2}{(m_{(o)k})^2}} \, ,
\label{mokV}
\end{eqnarray}
where, as already explained,
$m_{(o)k}$ are the masses of the (upper or  lower) four families on the tree level, and
 should carry the indices of the group $\Sigma$ and of the family member $\alpha$ and
 $Y^{'\alpha}_{(L,R)}$ and $Q^{'\alpha}_{(L,R)}$ are the eigenvalues of the operators
 when applied to the right ($Y^{'\alpha}_{R}$, $\,Y^{'\alpha}_{R}$) or to the
 left ($Y^{'\alpha}_{L}$, $\,Y^{'\alpha}_{L}$) handed member of a family ($\alpha$) in
 the massless basis. 

Let us evaluate  $\Sigma_{i j}^{\Sigma (Y^{'\alpha},Q^{'\alpha})}$,
similarly as in Eqs.~(\ref{tildesigmaij},\ref{sigmaijo},
pointing out that they 
 depend on  $\Sigma=(II,I)$ through masses $(m^{\alpha\,\Sigma}_{(o)})$ and through
 $V_{(o)i k }$ ($\equiv V^{\alpha}_{\Sigma (o)\,i k } $) and are different for each of the
 family member $\alpha$
\begin{equation}
\label{sigmaij}
\Sigma_{V\,i j}^{\Sigma (Y^{'\alpha},Q^{'\alpha})}=\sum_{k=1}^{4}\, V^{\alpha}_{\Sigma (o)\,i k }
\,V^{\alpha}_{\Sigma(o)\,j k}\,
\Sigma^{(Y^{'\alpha},Q^{'\alpha})}_{kV}  \,.
\end{equation}
We  end up with the matrix ${\cal M}_{(o1)_{V}}$, which carry the indices $\alpha $ and $\Sigma$
(${\cal M}^{\alpha}_{\Sigma (o1)_{V}}$). Due to Eq.~(\ref{oneloopMa}) we need
to calculate, to obtain the mass matrices up to the one loop corrections included,
${\cal M}_{(1)}$ ($
V^{\dagger}_{(o)}\,( {\cal M}_{(o\,1)} +
{\cal M}_{(o)} )\,V_{(o)}\,$ $= V^{\dagger}_{(o)}\,\tilde{{\cal M}}_{(o\,1)_{\tilde{S}}}\,V_{(o)}\,  +
V^{\dagger}_{(o)}\,{\cal M}_{(o\,1)_S}\,V_{(o)}\, +
V^{\dagger}_{(o)}\,{\cal M}_{(o\,1)_V}\,V_{(o)}\,$.

Let us calculate here  $V^{\alpha \dagger}_{\Sigma\,(o)}\,
\tilde{{\cal M}}^{\alpha \, \Sigma}_{(o\,1)_{V}}\,V^{\alpha}_{\Sigma(o)}\, $,
which distinguish among members of a family ($\alpha$) and between the two groups of families($\Sigma$),
using Eqs.~(\ref{oneloopMa}, \ref{mokS}, \ref{sigmaij})
\begin{eqnarray}
\label{V1SI}
\left(V^{\alpha\,\dagger}_{\Sigma(o)}\, {\cal M}^{\alpha\, \Sigma}_{(o\,1)_{V}}
\,V^{\alpha}_{\Sigma\,(o)}\,\right)_{ij}&=& \sum_{(l,k,r)=1}^{4}\,
(V_{(o)li} \,V_{(o)lk})\,(V_{(o)rk}\,V_{(o)rj} )\,
(\Sigma_{kV}^{Y'\alpha}+
 \Sigma_{kV}^{Q'\alpha}+
 \Sigma_{kV}^{Q \alpha}) \nonumber\\
 &=& \delta_{ik}\,\delta_{jk}\,(
 \Sigma_{kV}^{Y'\alpha}+
 \Sigma_{kV}^{Q'\alpha}+
 \Sigma_{k}^{Q \alpha})\,.
\end{eqnarray}
We have for ${\cal M}^{\alpha\, \Sigma}_{(1)_V}  $ the expression with only diagonal terms
 \begin{eqnarray}
 &&{\cal M}^{\alpha\,\Sigma}_{(1)_{V}} = V^{\alpha \dagger}_{\Sigma\,(o)}\,
\tilde{{\cal M}}^{\alpha \, \Sigma}_{(o\,1)_{V}}\,V^{\alpha}_{\Sigma(o)}\,   \nonumber\\
 &&\left( \begin{matrix}
             \Sigma_{1V}^{Y'\alpha}+ \Sigma_{1V}^{Q'\alpha} & 0 & 0 & 0 \\
 0 &         \Sigma_{2V}^{Y'\alpha}+ \Sigma_{2V}^{Q'\alpha} & 0 & 0     \\
 0 & 0 &     \Sigma_{3V}^{Y'\alpha}+ \Sigma_{3V}^{Q'\alpha} & 0         \\
 0 & 0 & 0 & \Sigma_{4V}^{Y'\alpha}+ \Sigma_{4V}^{Q'\alpha}             \\
\end{matrix}  \right)^{\alpha\,\Sigma}_{V}\:,
\label{oneloopVMa}
\end{eqnarray}
 like in Eq.~(\ref{oneloopSMa}).
The contribution to the one loop corrections originating in the massive vector boson fields
$A^{Y'}_{m}$ and $Z^{Q'}_{m}$ 
 leads to the diagonal mass matrices ${\cal M}^{\alpha\, \Sigma}_{(1)_V}$.

The mass matrices of  Eq.(\ref{oneloopVMa}) demonstrate that the one loop contributions
from $A^{Y^\prime}$ and $A^{Y^\prime}$ gauge bosons give
corrections to the tree level mass eigenvalues, but not change the off diagonal
terms.

\section{Short presentation of technique~\cite{norma,norma93,hn0203,holgernorma2003}, taken from~\cite{NF}}
\label{technique}

In this appendix a short review of the technique~\cite{{norma93,hn0203,holgernorma2003}},
initiated and developed by one of the authors when  proposing the
{\it spin-charge-family-theory}~\cite{norma,pikanorma}  assuming that
all the internal degrees of freedom of spinors, with family quantum number included, are
describable in the space of $d$-anticommuting (Grassmann) coordinates~\cite{norma93,hn0203,holgernorma2003}, if the
dimension of ordinary space is also $d$ and further developed by both  authors of the technique..
There are two kinds of operators in the Grassmann space,
fulfilling the Clifford algebra which anti commute with one another. The technique  was further
developed in the present shape together with H.B. Nielsen~\cite{norma93,hn0203,holgernorma2003} by identifying
one kind of the Clifford objects with $\gamma^s$'s and another kind with  $\tilde{\gamma}^a$'s.
In this last stage we  constructed a spinor basis as products of nilpotents and projections  formed
as odd and even objects of $\gamma^a$'s, respectively, and  chosen to be eigenstates
of a Cartan subalgebra of the Lorentz groups defined by $\gamma^a$'s and $\tilde{\gamma}^a$'s.
The technique can be used to construct a spinor basis for any dimension $d$
and any signature in an easy and transparent way. Equipped with the graphic presentation of basic states,
the technique offers an elegant way to see all the quantum numbers of states with respect to the two
Lorentz groups, as well as transformation properties of the states under any Clifford algebra object.

The objects $\gamma^a$ and $\tilde{\gamma}^a$ have properties~(\ref{snmb:tildegclifford}),
\begin{eqnarray}
\label{gammatildegamma}
&& \{ \gamma^a, \gamma^b\}_{+} = 2\eta^{ab}\,, \quad\quad
\{ \tilde{\gamma}^a, \tilde{\gamma}^b\}_{+}= 2\eta^{ab}\,, \quad,\quad
\{ \gamma^a, \tilde{\gamma}^b\}_{+} = 0\,,
\end{eqnarray}
for any $d$, even or odd.  $I$ is the unit element in the Clifford algebra.

The Clifford algebra objects $S^{ab}$ and $\tilde{S}^{ab}$ close the algebra of the Lorentz group
\begin{eqnarray}
\label{sabtildesab}
\
S^{ab}: &=& (i/4) (\gamma^a \gamma^b - \gamma^b \gamma^a)\,, \nonumber\\
\tilde{S}^{ab}: &=& (i/4) (\tilde{\gamma}^a \tilde{\gamma}^b
- \tilde{\gamma}^b \tilde{\gamma}^a)\,,\nonumber\\
 \{S^{ab}, \tilde{S}^{cd}\}_{-}&=& 0\,,\nonumber\\
\{S^{ab},S^{cd}\}_{-} &=& i(\eta^{ad} S^{bc} + \eta^{bc} S^{ad} - \eta^{ac} S^{bd} - \eta^{bd} S^{ac})\,,
\nonumber\\
\{\tilde{S}^{ab},\tilde{S}^{cd}\}_{-} &=& i(\eta^{ad} \tilde{S}^{bc} + \eta^{bc} \tilde{S}^{ad}
- \eta^{ac} \tilde{S}^{bd} - \eta^{bd} \tilde{S}^{ac})\,,
\end{eqnarray}

We assume  the ``Hermiticity'' property for $\gamma^a$'s  and $\tilde{\gamma}^a$'s
\begin{eqnarray}
\gamma^{a\dagger} = \eta^{aa} \gamma^a\,,\quad \quad \tilde{\gamma}^{a\dagger} = \eta^{aa} \tilde{\gamma}^a\,,
\label{cliffher}
\end{eqnarray}
in order that
$\gamma^a$ and $\tilde{\gamma}^a$ are compatible with (\ref{gammatildegamma}) and formally unitary,
i.e. $\gamma^{a \,\dagger} \,\gamma^a=I$ and $\tilde{\gamma}^{a\,\dagger} \tilde{\gamma}^a=I$.

One finds from Eq.(\ref{cliffher}) that $(S^{ab})^{\dagger} = \eta^{aa} \eta^{bb}S^{ab}$.

Recognizing from Eq.(\ref{sabtildesab})  that two Clifford algebra objects
$S^{ab}, S^{cd}$ with all indices different commute, and equivalently for
$\tilde{S}^{ab},\tilde{S}^{cd}$, we  select  the Cartan subalgebra of the algebra of the
two groups, which  form  equivalent representations with respect to one another
\begin{eqnarray}
S^{03}, S^{12}, S^{56}, \cdots, S^{d-1\; d}, \quad {\rm if } \quad d &=& 2n\ge 4,
\nonumber\\
S^{03}, S^{12}, \cdots, S^{d-2 \;d-1}, \quad {\rm if } \quad d &=& (2n +1) >4\,,
\nonumber\\
\tilde{S}^{03}, \tilde{S}^{12}, \tilde{S}^{56}, \cdots, \tilde{S}^{d-1\; d},
\quad {\rm if } \quad d &=& 2n\ge 4\,,
\nonumber\\
\tilde{S}^{03}, \tilde{S}^{12}, \cdots, \tilde{S}^{d-2 \;d-1},
\quad {\rm if } \quad d &=& (2n +1) >4\,.
\label{choicecartan}
\end{eqnarray}

The choice for  the Cartan subalgebra in $d <4$ is straightforward.
It is  useful  to define one of the Casimirs of the Lorentz group -
the  handedness $\Gamma$ ($\{\Gamma, S^{ab}\}_- =0$) in any $d$
\begin{eqnarray}
\Gamma^{(d)} :&=&(i)^{d/2}\; \;\;\;\;\;\prod_a \quad (\sqrt{\eta^{aa}} \gamma^a), \quad {\rm if } \quad d = 2n,
\nonumber\\
\Gamma^{(d)} :&=& (i)^{(d-1)/2}\; \prod_a \quad (\sqrt{\eta^{aa}} \gamma^a), \quad {\rm if } \quad d = 2n +1\,.
\label{hand}
\end{eqnarray}
One can proceed equivalently for $\tilde{\gamma}^a$'s.
We understand the product of $\gamma^a$'s in the ascending order with respect to
the index $a$: $\gamma^0 \gamma^1\cdots \gamma^d$.
It follows from Eq.(\ref{cliffher})
for any choice of the signature $\eta^{aa}$ that
$\Gamma^{\dagger}= \Gamma,\;
\Gamma^2 = I.$
We also find that for $d$ even the handedness  anticommutes with the Clifford algebra objects
$\gamma^a$ ($\{\gamma^a, \Gamma \}_+ = 0$) , while for $d$ odd it commutes with
$\gamma^a$ ($\{\gamma^a, \Gamma \}_- = 0$).

To make the technique simple we introduce the graphic presentation
as follows 
\begin{eqnarray}
\stackrel{ab}{(k)}:&=&
\frac{1}{2}(\gamma^a + \frac{\eta^{aa}}{ik} \gamma^b)\,,\quad \quad
\stackrel{ab}{[k]}:=
\frac{1}{2}(1+ \frac{i}{k} \gamma^a \gamma^b)\,,\nonumber\\
\stackrel{+}{\circ}:&=& \frac{1}{2} (1+\Gamma)\,,\quad \quad
\stackrel{-}{\bullet}:= \frac{1}{2}(1-\Gamma),
\label{signature}
\end{eqnarray}
where $k^2 = \eta^{aa} \eta^{bb}$.
One can easily check by taking into account the Clifford algebra relation
(Eq.\ref{gammatildegamma}) and the
definition of $S^{ab}$ and $\tilde{S}^{ab}$ (Eq.\ref{sabtildesab})
that if one multiplies from the left hand side by $S^{ab}$ or $\tilde{S}^{ab}$ the Clifford
algebra objects $\stackrel{ab}{(k)}$
and $\stackrel{ab}{[k]}$,
it follows that
\begin{eqnarray}
        S^{ab}\, \stackrel{ab}{(k)}= \frac{1}{2}\,k\, \stackrel{ab}{(k)}\,,\quad \quad
        S^{ab}\, \stackrel{ab}{[k]}= \frac{1}{2}\,k \,\stackrel{ab}{[k]}\,,\nonumber\\
\tilde{S}^{ab}\, \stackrel{ab}{(k)}= \frac{1}{2}\,k \,\stackrel{ab}{(k)}\,,\quad \quad
\tilde{S}^{ab}\, \stackrel{ab}{[k]}=-\frac{1}{2}\,k \,\stackrel{ab}{[k]}\,,
\label{grapheigen}
\end{eqnarray}
which means that we get the same objects back multiplied by the constant $\frac{1}{2}k$ in the case
of $S^{ab}$, while $\tilde{S}^{ab}$ multiply $\stackrel{ab}{(k)}$ by $k$ and $\stackrel{ab}{[k]}$
by $(-k)$ rather than $(k)$.
This also means that when
$\stackrel{ab}{(k)}$ and $\stackrel{ab}{[k]}$ act from the left hand side on  a
vacuum state $|\psi_0\rangle$ the obtained states are the eigenvectors of $S^{ab}$.
We further recognize 
that $\gamma^a$
transform  $\stackrel{ab}{(k)}$ into  $\stackrel{ab}{[-k]}$, never to $\stackrel{ab}{[k]}$,
while $\tilde{\gamma}^a$ transform  $\stackrel{ab}{(k)}$ into $\stackrel{ab}{[k]}$, never to
$\stackrel{ab}{[-k]}$
\begin{eqnarray}
&&\gamma^a \stackrel{ab}{(k)}= \eta^{aa}\stackrel{ab}{[-k]},\;
\gamma^b \stackrel{ab}{(k)}= -ik \stackrel{ab}{[-k]}, \;
\gamma^a \stackrel{ab}{[k]}= \stackrel{ab}{(-k)},\;
\gamma^b \stackrel{ab}{[k]}= -ik \eta^{aa} \stackrel{ab}{(-k)}\,,\nonumber\\
&&\tilde{\gamma^a} \stackrel{ab}{(k)} = - i\eta^{aa}\stackrel{ab}{[k]},\;
\tilde{\gamma^b} \stackrel{ab}{(k)} =  - k \stackrel{ab}{[k]}, \;
\tilde{\gamma^a} \stackrel{ab}{[k]} =  \;\;i\stackrel{ab}{(k)},\;
\tilde{\gamma^b} \stackrel{ab}{[k]} =  -k \eta^{aa} \stackrel{ab}{(k)}\,.
\label{snmb:gammatildegamma}
\end{eqnarray}
From Eq.(\ref{snmb:gammatildegamma}) it follows
\begin{eqnarray}
\label{stildestrans}
S^{ac}\stackrel{ab}{(k)}\stackrel{cd}{(k)} &=& -\frac{i}{2} \eta^{aa} \eta^{cc}
\stackrel{ab}{[-k]}\stackrel{cd}{[-k]}\,,\,\quad\quad
\tilde{S}^{ac}\stackrel{ab}{(k)}\stackrel{cd}{(k)} = \frac{i}{2} \eta^{aa} \eta^{cc}
\stackrel{ab}{[k]}\stackrel{cd}{[k]}\,,\,\nonumber\\
S^{ac}\stackrel{ab}{[k]}\stackrel{cd}{[k]} &=& \frac{i}{2}
\stackrel{ab}{(-k)}\stackrel{cd}{(-k)}\,,\,\quad\quad
\tilde{S}^{ac}\stackrel{ab}{[k]}\stackrel{cd}{[k]} = -\frac{i}{2}
\stackrel{ab}{(k)}\stackrel{cd}{(k)}\,,\,\nonumber\\
S^{ac}\stackrel{ab}{(k)}\stackrel{cd}{[k]}  &=& -\frac{i}{2} \eta^{aa}
\stackrel{ab}{[-k]}\stackrel{cd}{(-k)}\,,\,\quad\quad
\tilde{S}^{ac}\stackrel{ab}{(k)}\stackrel{cd}{[k]} = -\frac{i}{2} \eta^{aa}
\stackrel{ab}{[k]}\stackrel{cd}{(k)}\,,\,\nonumber\\
S^{ac}\stackrel{ab}{[k]}\stackrel{cd}{(k)} &=& \frac{i}{2} \eta^{cc}
\stackrel{ab}{(-k)}\stackrel{cd}{[-k]}\,,\,\quad\quad
\tilde{S}^{ac}\stackrel{ab}{[k]}\stackrel{cd}{(k)} = \frac{i}{2} \eta^{cc}
\stackrel{ab}{(k)}\stackrel{cd}{[k]}\,.
\end{eqnarray}
From Eqs.~(\ref{stildestrans}) we conclude that $\tilde{S}^{ab}$ generate the
equivalent representations with respect to $S^{ab}$ and opposite.

Let us deduce some useful relations

\begin{eqnarray}
\stackrel{ab}{(k)}\stackrel{ab}{(k)}& =& 0\,, \quad \quad \stackrel{ab}{(k)}\stackrel{ab}{(-k)}
= \eta^{aa}  \stackrel{ab}{[k]}\,, \quad \stackrel{ab}{(-k)}\stackrel{ab}{(k)}=
\eta^{aa}   \stackrel{ab}{[-k]}\,,\quad
\stackrel{ab}{(-k)} \stackrel{ab}{(-k)} = 0\,, \nonumber\\
\stackrel{ab}{[k]}\stackrel{ab}{[k]}& =& \stackrel{ab}{[k]}\,, \quad \quad
\stackrel{ab}{[k]}\stackrel{ab}{[-k]}= 0\,, \;\;\quad \quad  \quad \stackrel{ab}{[-k]}\stackrel{ab}{[k]}=0\,,
 \;\;\quad \quad \quad \quad \stackrel{ab}{[-k]}\stackrel{ab}{[-k]} = \stackrel{ab}{[-k]}\,,
 \nonumber\\
\stackrel{ab}{(k)}\stackrel{ab}{[k]}& =& 0\,,\quad \quad \quad \stackrel{ab}{[k]}\stackrel{ab}{(k)}
=  \stackrel{ab}{(k)}\,, \quad \quad \quad \stackrel{ab}{(-k)}\stackrel{ab}{[k]}=
 \stackrel{ab}{(-k)}\,,\quad \quad \quad
\stackrel{ab}{(-k)}\stackrel{ab}{[-k]} = 0\,,
\nonumber\\
\stackrel{ab}{(k)}\stackrel{ab}{[-k]}& =&  \stackrel{ab}{(k)}\,,
\quad \quad \stackrel{ab}{[k]}\stackrel{ab}{(-k)} =0,  \quad \quad
\quad \stackrel{ab}{[-k]}\stackrel{ab}{(k)}= 0\,, \quad \quad \quad \quad
\stackrel{ab}{[-k]}\stackrel{ab}{(-k)} = \stackrel{ab}{(-k)}.
\label{graphbinoms}
\end{eqnarray}
We recognize in  the first equation of the first row and the first equation of the second row
the demonstration of the nilpotent and the projector character of the Clifford algebra objects
$\stackrel{ab}{(k)}$ and $\stackrel{ab}{[k]}$, respectively.
Defining
\begin{eqnarray}
\stackrel{ab}{\tilde{(\pm i)}} =
\frac{1}{2} \, (\tilde{\gamma}^a \mp \tilde{\gamma}^b)\,, \quad
\stackrel{ab}{\tilde{(\pm 1)}} =
\frac{1}{2} \, (\tilde{\gamma}^a \pm i\tilde{\gamma}^b)\,,
\label{deftildefun}
\end{eqnarray}
one recognizes that
\begin{eqnarray}
\stackrel{ab}{\tilde{( k)}} \, \stackrel{ab}{(k)}& =& 0\,,
\quad \;
\stackrel{ab}{\tilde{(-k)}} \, \stackrel{ab}{(k)} = -i \eta^{aa}\,  \stackrel{ab}{[k]}\,,
\quad\;
\stackrel{ab}{\tilde{( k)}} \, \stackrel{ab}{[k]} = i\, \stackrel{ab}{(k)}\,,
\quad\;
\stackrel{ab}{\tilde{( k)}}\, \stackrel{ab}{[-k]} = 0\,.
\label{graphbinomsfamilies}
\end{eqnarray}
Recognizing that
\begin{eqnarray}
\stackrel{ab}{(k)}^{\dagger}=\eta^{aa}\stackrel{ab}{(-k)}\,,\quad
\stackrel{ab}{[k]}^{\dagger}= \stackrel{ab}{[k]}\,,
\label{graphherstr}
\end{eqnarray}
we define a vacuum state $|\psi_0>$ so that one finds
\begin{eqnarray}
< \;\stackrel{ab}{(k)}^{\dagger}
 \stackrel{ab}{(k)}\; > = 1\,, \nonumber\\
< \;\stackrel{ab}{[k]}^{\dagger}
 \stackrel{ab}{[k]}\; > = 1\,.
\label{graphherscal}
\end{eqnarray}

Taking into account the above equations it is easy to find a Weyl spinor irreducible representation
for $d$-dimensional space, with $d$ even or odd.

For $d$ even we simply make a starting state as a product of $d/2$, let us say, only nilpotents
$\stackrel{ab}{(k)}$, one for each $S^{ab}$ of the Cartan subalgebra  elements (Eq.(\ref{choicecartan})),
applying it on an (unimportant) vacuum state.
For $d$ odd the basic states are products
of $(d-1)/2$ nilpotents and a factor $(1\pm \Gamma)$.
Then the generators $S^{ab}$, which do not belong
to the Cartan subalgebra, being applied on the starting state from the left,
 generate all the members of one
Weyl spinor.
\begin{eqnarray}
\stackrel{0d}{(k_{0d})} \stackrel{12}{(k_{12})} \stackrel{35}{(k_{35})}\cdots \stackrel{d-1\;d-2}{(k_{d-1\;d-2})}
\psi_0 \nonumber\\
\stackrel{0d}{[-k_{0d}]} \stackrel{12}{[-k_{12}]} \stackrel{35}{(k_{35})}\cdots \stackrel{d-1\;d-2}{(k_{d-1\;d-2})}
\psi_0 \nonumber\\
\stackrel{0d}{[-k_{0d}]} \stackrel{12}{(k_{12})} \stackrel{35}{[-k_{35}]}\cdots \stackrel{d-1\;d-2}{(k_{d-1\;d-2})}
\psi_0 \nonumber\\
\vdots \nonumber\\
\stackrel{0d}{[-k_{0d}]} \stackrel{12}{(k_{12})} \stackrel{35}{(k_{35})}\cdots \stackrel{d-1\;d-2}{[-k_{d-1\;d-2}]}
\psi_0 \nonumber\\
\stackrel{od}{(k_{0d})} \stackrel{12}{[-k_{12}]} \stackrel{35}{[-k_{35}]}\cdots \stackrel{d-1\;d-2}{(k_{d-1\;d-2})}
\psi_0 \nonumber\\
\vdots
\label{graphicd}
\end{eqnarray}
All the states have the handedness $\Gamma $, since $\{ \Gamma, S^{ab}\} = 0$.
States, belonging to one multiplet  with respect to the group $SO(q,d-q)$, that is to one
irreducible representation of spinors (one Weyl spinor), can have any phase. We made a choice
of the simplest one, taking all  phases equal to one.

The above graphic representation demonstrate that for $d$ even
all the states of one irreducible Weyl representation of a definite handedness follow from a starting state,
which is, for example, a product of nilpotents $\stackrel{ab}{(k_{ab})}$, by transforming all possible pairs
of $\stackrel{ab}{(k_{ab})} \stackrel{mn}{(k_{mn})}$ into $\stackrel{ab}{[-k_{ab}]} \stackrel{mn}{[-k_{mn}]}$.
There are $S^{am}, S^{an}, S^{bm}, S^{bn}$, which do this.
The procedure gives $2^{(d/2-1)}$ states. A Clifford algebra object $\gamma^a$ being applied from the left hand side,
transforms  a
Weyl spinor of one handedness into a Weyl spinor of the opposite handedness. Both Weyl spinors form a Dirac
spinor.

For $d$ odd a Weyl spinor has besides a product of $(d-1)/2$ nilpotents or projectors also either the
factor $\stackrel{+}{\circ}:= \frac{1}{2} (1+\Gamma)$ or the factor
$\stackrel{-}{\bullet}:= \frac{1}{2}(1-\Gamma)$.
As in the case of $d$ even, all the states of one irreducible
Weyl representation of a definite handedness follow from a starting state,
which is, for example, a product of $(1 + \Gamma)$ and $(d-1)/2$ nilpotents $\stackrel{ab}{(k_{ab})}$, by
transforming all possible pairs
of $\stackrel{ab}{(k_{ab})} \stackrel{mn}{(k_{mn})}$ into $\stackrel{ab}{[-k_{ab}]} \stackrel{mn}{[-k_{mn}]}$.
But $\gamma^a$'s, being applied from the left hand side, do not change the handedness of the Weyl spinor,
since $\{ \Gamma,
\gamma^a \}_- =0$ for $d$ odd.
A Dirac and a Weyl spinor are for $d$ odd identical and a ''family''
has accordingly $2^{(d-1)/2}$ members of basic states of a definite handedness.

We shall speak about left handedness when $\Gamma= -1$ and about right
handedness when $\Gamma =1$ for either $d$ even or odd.

While $S^{ab}$ which do not belong to the Cartan subalgebra (Eq.~(\ref{choicecartan})) generate
all the states of one representation, generate $\tilde{S}^{ab}$ which do not belong to the
Cartan subalgebra(Eq.~(\ref{choicecartan})) the states of $2^{d/2-1}$ equivalent representations.

Making a choice of the Cartan subalgebra set of the algebra $S^{ab}$ and $\tilde{S}^{ab}$
\begin{eqnarray}
S^{03}, S^{12}, S^{56}, S^{78}, S^{9 \;10}, S^{11\;12}, S^{13\; 14}\,,\nonumber\\
\tilde{S}^{03}, \tilde{S}^{12}, \tilde{S}^{56}, \tilde{S}^{78}, \tilde{S}^{9 \;10},
\tilde{S}^{11\;12}, \tilde{S}^{13\; 14}\,,
\label{cartan}
\end{eqnarray}
a left handed ($\Gamma^{(1,13)} =-1$) eigen state of all the members of the
Cartan  subalgebra, representing a weak chargeless  $u_{R}$-quark with spin up, hypercharge ($2/3$)
and  colour ($1/2\,,1/(2\sqrt{3})$), for example, can be written as 
\begin{eqnarray}
&& \stackrel{03}{(+i)}\stackrel{12}{(+)}|\stackrel{56}{(+)}\stackrel{78}{(+)}
||\stackrel{9 \;10}{(+)}\stackrel{11\;12}{(-)}\stackrel{13\;14}{(-)} |\psi \rangle = \nonumber\\
&&\frac{1}{2^7}
(\gamma^0 -\gamma^3)(\gamma^1 +i \gamma^2)| (\gamma^5 + i\gamma^6)(\gamma^7 +i \gamma^8)||
\nonumber\\
&& (\gamma^9 +i\gamma^{10})(\gamma^{11} -i \gamma^{12})(\gamma^{13}-i\gamma^{14})
|\psi \rangle \,.
\label{start}
\end{eqnarray}
This state is an eigenstate of all $S^{ab}$ and $\tilde{S}^{ab}$ which are members of the Cartan
subalgebra (Eq.~(\ref{cartan})).

The operators $ \tilde{S}^{ab}$, which do not belong to the Cartan subalgebra (Eq.~(\ref{cartan})),
generate families from the starting $u_R$ quark, transforming $u_R$ quark from Eq.~(\ref{start})
to the $u_R$ of another family,  keeping all the properties with respect to $S^{ab}$ unchanged.
In particular $\tilde{S}^{01}$ applied on a right handed $u_R$-quark, weak chargeless,  with spin up,
hypercharge ($2/3$) and the colour charge ($1/2\,,1/(2\sqrt{3})$) from Eq.~(\ref{start}) generates a
state which is again  a right handed $u_{R}$-quark,  weak chargeless,  with spin up,
hypercharge ($2/3$)
and the colour charge ($1/2\,,1/(2\sqrt{3})$)
\begin{eqnarray}
\label{tildesabfam}
\tilde{S}^{01}\;
\stackrel{03}{(+i)}\stackrel{12}{(+)}| \stackrel{56}{(+)} \stackrel{78}{(+)}||
\stackrel{9 10}{(+)} \stackrel{11 12}{(-)} \stackrel{13 14}{(-)}= -\frac{i}{2}\,
&&\stackrel{03}{[\,+i]} \stackrel{12}{[\,+\,]}| \stackrel{56}{(+)} \stackrel{78}{(+)}||
\stackrel{9 10}{(+)} \stackrel{11 12}{(-)} \stackrel{13 14}{(-)}\,.
\end{eqnarray}

Below some useful relations~\cite{pikanorma} are presented
\begin{eqnarray}
\label{plusminus}
N^{\pm}_{+}         &=& N^{1}_{+} \pm i \,N^{2}_{+} =
 - \stackrel{03}{(\mp i)} \stackrel{12}{(\pm )}\,, \quad N^{\pm}_{-}= N^{1}_{-} \pm i\,N^{2}_{-} =
  \stackrel{03}{(\pm i)} \stackrel{12}{(\pm )}\,,\nonumber\\
\tilde{N}^{\pm}_{+} &=& - \stackrel{03}{\tilde{(\mp i)}} \stackrel{12}{\tilde{(\pm )}}\,, \quad
\tilde{N}^{\pm}_{-}= 
  \stackrel{03} {\tilde{(\pm i)}} \stackrel{12} {\tilde{(\pm )}}\,,\nonumber\\
\tau^{1\pm}         &=& (\mp)\, \stackrel{56}{(\pm )} \stackrel{78}{(\mp )} \,, \quad
\tau^{2\mp}=            (\mp)\, \stackrel{56}{(\mp )} \stackrel{78}{(\mp )} \,,\nonumber\\
\tilde{\tau}^{1\pm} &=& (\mp)\, \stackrel{56}{\tilde{(\pm )}} \stackrel{78}{\tilde{(\mp )}}\,,\quad
\tilde{\tau}^{2\mp}= (\mp)\, \stackrel{56}{\tilde{(\mp )}} \stackrel{78}{\tilde{(\mp )}}\,.
\end{eqnarray}

\end{document}